\documentclass[12pt]{article}

\global\arraycolsep=1pt
\usepackage{graphicx}
\usepackage{amsmath}
\usepackage{amssymb}
\usepackage {pstricks}
\let\psgrid\relax
\newcommand{\FigureSize}{0.6}
\newcommand{\inverseFigureSize}{1.66667}
\newcommand{\appPartition}{A}
\newcommand{\appMac}{B}
\newcommand{\appSpin}{C}
\newcommand{\appDunkl}{D}
\newcommand{\appNotation}{E}
\newcommand{\beq}{\begin{equation}}
\newcommand{\eeq}{\end{equation}}
\newcommand{\beqa}{\begin{eqnarray}}
\newcommand{\eeqa}{\end{eqnarray}}
\newcommand{\CR}{\nonumber \\}
\newcommand{\be}{\begin{equation}}
\newcommand{\ee}{\end{equation}}
\newcommand{\ba}{\begin{eqnarray}}
\newcommand{\ea}{\end{eqnarray}}
\newcommand{\eqn}[1]{(\ref{#1})}
\newcommand{\ds}{\displaystyle}

\newcommand{\bebox}[1]{
 \begin{equation}
  \fbox{
\rule[-.35cm]{0cm}{1cm}{~~~${\displaystyle{#1}}$~~~}}
}
\newcommand{\eebox}{\end{equation}}
\newcommand{\babox}[1]{
 \begin{equation}
  \fbox{\rule[-.5cm]{0cm}{1cm}{~~~${
   \begin{array}{rcl}  
    #1
   \end{array}
  }$~~~}}
}
\newcommand{\eabox}{\end{equation}}
\renewcommand{\theequation}{\thesection.\arabic{equation}}
\newcommand{\Section}{\setcounter{equation}{0} \section}
\renewcommand{\thefootnote}{\fnsymbol{footnote}}
\renewcommand{\a}{\alpha}
\renewcommand{\b}{\beta}
\newcommand{\vv}{{\mbox{\boldmath $v$}}}
\newcommand{\vo}{{\mbox{\boldmath $0$}}}
\newcommand{\bC}{{\mathbb C}}
\newcommand{\bN}{{\mathbb N}}
\newcommand{\bZ}{{\mathbb Z}}
\newcommand{\bu}{\bullet}    
\newcommand{\proof}{\noindent{\it Proof.\hskip10pt}} 
\newcommand{\qed}{\hfill\fbox{}}
\newcommand{\ha}{{1\over2}}
\newcommand{\Exp}[1]{\exp\left\{#1\right\}}
\newcommand{\Endomega}[3]{\omega_{#1,#2}^{#3}}
\newcommand{\tP}{P}
\newcommand{\tPi}{\Pi_0} 
\newcommand{\tY}{\widetilde Y}
\newcommand{\tZ}{\widetilde Z}

\newcommand{\nn}{{N_c}}
\newcommand{\NNi}[1]{N_{#1}}
\newcommand{\NNii}[1]{N_{{#1^\vee}}}
\newcommand{\Li}{L_1}
\newcommand{\Lii}{L_2}
\newcommand{\Q}{{\bf e}}
\newcommand{\QQ}[2]{\Q_{#1}/\Q_{#2}}
\newcommand{\QpQp}[2]{\Q'_{#1}/\Q'_{#2}}
\newcommand{\QQp}[2]{\Q_{#1}/\Q'_{#2}}
\newcommand{\QpQ}[2]{\Q'_{#1}/\Q_{#2}}
\newcommand{\Ciio}[5]{C_{#1#2}{}^{#3}(#4,#5)}
\newcommand{\Cooi}[5]{C^{#1#2}{}_{#3}(#4,#5)}
\newcommand{\Cioo}[5]{C_{#1}{}^{#2#3}(#4,#5)}
\newcommand{\Coii}[5]{C^{#1}{}_{#2#3}(#4,#5)}
\newcommand{\Ziiio}[7]{Z_{#1#2#3}{}^{#4}(#5;#6,#7)}
\newcommand{\Ziooo}[7]{Z_{#1}{}^{#2#3#4}(#5;#6,#7)}
\newcommand{\Ziioo}[7]{Z_{#1#2}{}^{#3#4}(#5;#6,#7)}
\newcommand{\Zioio}[7]{Z_{#1}{}^{#2}{}_{#3}{}^{#4}(#5;#6,#7)}
\newcommand{\Znoniioo}[4]{Z_{#1#2}{}^{#3#4}}
\newcommand{\Znonioio}[4]{Z_{#1}{}^{#2}{}_{#3}{}^{#4}}
\newcommand{\Zoiio}[7]{Z^{#1}{}_{#2#3}{}^{#4}(#5;#6,#7)}
\newcommand{\Ya}[2]{{ #1_{#2} }}
\newcommand{\Yav}[2]{{ #1_{#2}{}^\vee }}
\newcommand{\Yai}[3]{{ #1_{#2,#3} }}
\newcommand{\Nek}[5]{N_{#1#2}\left(#3;#4,#5\right)}
\newcommand{\fNek}[9]{N^{#3,#4}_{#5#6}(#7,#1,#2;#8,#9)} 
  
\newcommand{\fnek}[8]{n^{#3,#4}_{#5#6}(#1,#2;#7,#8)} 
\newcommand{\neknon}[6]{n^{#1#2}_{#3#4}}            
\newcommand{\fnekrasov}[6]{n_{#3#4}(#1,#2;#5,#6)}    
\newcommand{\NYNY}{ N^{\{\lambda_{\alpha}\}}_{\alpha,\beta}\left(\epsilon_1,\epsilon_2;a_\alpha\right) }
\newcommand{\NY}[5]{N_{#1#2}\left(#5;#4,#3\right)}
\newcommand{\GV}[5]{G_{#1}^{#2}(#5;#4,#3)}
\newcommand{\Zk}[4]{Z_{#1}     (#4;#3,#2)}
\newcommand{\Zm}[2]{Z^{\mathrm {#1}}_{#2}}
\newcommand{\tZm}[2]{\widetilde Z^{{\rm #1}}_{#2}}
\newcommand{\ZQ}[6]{Z^{\mathrm {#1}}_{#2}\left(#5,#6;#3,#4\right)}
\newcommand{\ZlaQ}[4]{Z^{#1}_{#2}\left(#3,#4\right)}
\newcommand{\fla}[3]{f_{#1}\left(#2,#3\right)}
\newdimen\Sx 
\newdimen\Sy
\newdimen\Ex 
\newdimen\Ey
\newdimen\Mx 
\newdimen\My
\newdimen\Tx 
\newdimen\Ty

\newdimen\Fx 
\newdimen\Fy
\newdimen\Bx 
\newdimen\By
\newdimen\Rx 
\newdimen\Ry
\newdimen\Lx 
\newdimen\Ly
\newdimen\Ux 
\newdimen\Uy
\newdimen\Dx 
\newdimen\Dy

\newdimen\txtlen
\newdimen\txthgt

\newcommand{\WhiteTriangle}[4]{
\psline(0,0)(#1,#2)
\psline(0,0)(#3,#4)
\psline[linewidth=\Lwd pt](#1,#2)(#3,#4)
}
\newcommand{\AAx}{0.4131714875}
\newcommand{\AAy}{0.1711412338}
\newcommand{\BBx}{0.4131714875}
\newcommand{\BBy}{0.1711412338}
\newcommand{\Lwd}{0.5}
\newcommand{\triangleW}{
\WhiteTriangle{\AAx}{-\AAy}{\AAx}{\AAy}
}
\newcommand{\triangleSW}{
\WhiteTriangle{ \BBx}{\BBy}{\BBy}{\BBx}
}
\newcommand{\triangleS}{
\WhiteTriangle{\AAy}{\AAx}{-\AAy}{\AAx}
}
\newcommand{\triangleSE}{
\WhiteTriangle{-\BBy}{\BBx}{-\BBx}{\BBy}
}
\newcommand{\triangleE}{
\WhiteTriangle{-\AAx}{\AAy}{-\AAx}{-\AAy}
}
\newcommand{\triangleNE}{
\WhiteTriangle{-\BBx}{-\BBy}{-\BBy}{-\BBx}
}
\newcommand{\triangleN}{
\WhiteTriangle{-\AAy}{-\AAx}{\AAy}{-\AAx}
}
\newcommand{\triangleNW}{
\WhiteTriangle{\BBy}{-\BBx}{\BBx}{-\BBy}
}

\makeatletter
\def\CArrow{\@ifnextchar[{\@CArrow}{\@CArrow[]}} 
\def\@CArrow[#1]#2#3#4#5#6{{
\advance \Sx by #2\p@
\advance \Sy by #3\p@
\advance \Ex by #4\p@
\advance \Ey by #5\p@
\advance \Mx by 0.5\Sx
\advance \Mx by 0.5\Ex
\advance \My by 0.5\Sy
\advance \My by 0.5\Ey
\advance \Tx by 0.4\Sx
\advance \Tx by 0.6\Ex
\advance \Ty by 0.4\Sy
\advance \Ty by 0.6\Ey
\setbox0\hbox{#6}\advance\txtlen by 0.035146\wd0%
\setbox0\vbox{#6}\advance\txthgt by 0.035146\ht0%
\txtlen=\inverseFigureSize\txtlen
\txthgt=\inverseFigureSize\txthgt
\psline
(\expandafter\Rval\the\Mx,\expandafter\Rval\the\My)
(\expandafter\Rval\the\Ex,\expandafter\Rval\the\Ey)%
\@tfor\opti :=#1\do{
\if\opti O 
                \psline[arrowsize=5pt]
                {->}(\expandafter\Rval\the\Sx,\expandafter\Rval\the\Sy)
                (\expandafter\Rval\the\Tx,\expandafter\Rval\the\Ty)%
\fi
\if\opti W 
                \psline
                (\expandafter\Rval\the\Sx,\expandafter\Rval\the\Sy)
                (\expandafter\Rval\the\Tx,\expandafter\Rval\the\Ty)%
              \rput(\expandafter\Rval\the\Tx,\expandafter\Rval\the\Ty){\triangleW}%
\fi
\if\opti w 
                \psline
                (\expandafter\Rval\the\Sx,\expandafter\Rval\the\Sy)
                (\expandafter\Rval\the\Tx,\expandafter\Rval\the\Ty)%
              \rput(\expandafter\Rval\the\Tx,\expandafter\Rval\the\Ty){\triangleSW}%
\fi
\if\opti S 
                \psline
                (\expandafter\Rval\the\Sx,\expandafter\Rval\the\Sy)
                (\expandafter\Rval\the\Tx,\expandafter\Rval\the\Ty)%
              \rput(\expandafter\Rval\the\Tx,\expandafter\Rval\the\Ty){\triangleS}%
\fi
\if\opti s 
                \psline
                (\expandafter\Rval\the\Sx,\expandafter\Rval\the\Sy)
                (\expandafter\Rval\the\Tx,\expandafter\Rval\the\Ty)%
              \rput(\expandafter\Rval\the\Tx,\expandafter\Rval\the\Ty){\triangleSE}%
\fi
\if\opti E 
                \psline
                (\expandafter\Rval\the\Sx,\expandafter\Rval\the\Sy)
                (\expandafter\Rval\the\Tx,\expandafter\Rval\the\Ty)%
              \rput(\expandafter\Rval\the\Tx,\expandafter\Rval\the\Ty){\triangleE}%
\fi
\if\opti e 
                \psline
                (\expandafter\Rval\the\Sx,\expandafter\Rval\the\Sy)
                (\expandafter\Rval\the\Tx,\expandafter\Rval\the\Ty)%
              \rput(\expandafter\Rval\the\Tx,\expandafter\Rval\the\Ty){\triangleNE}%
\fi
\if\opti N 
                \psline
                (\expandafter\Rval\the\Sx,\expandafter\Rval\the\Sy)
                (\expandafter\Rval\the\Tx,\expandafter\Rval\the\Ty)%
              \rput(\expandafter\Rval\the\Tx,\expandafter\Rval\the\Ty){\triangleN}%
\fi
\if\opti n 
                \psline
                (\expandafter\Rval\the\Sx,\expandafter\Rval\the\Sy)
                (\expandafter\Rval\the\Tx,\expandafter\Rval\the\Ty)%
              \rput(\expandafter\Rval\the\Tx,\expandafter\Rval\the\Ty){\triangleNW}%
\fi
\if\opti F 
        \advance \Fx by  -0.2\Sx
        \advance \Fx by   1.2\Ex
        \advance \Fy by -0.2\Sy
        \advance \Fy by  1.2\Ey
        \rput(\expandafter\Rval\the\Fx,\expandafter\Rval\the\Fy){#6}
        \advance \Fx by -\Fx
        \advance \Fy by -\Fy
\fi
\if\opti B 
        \advance \Bx by   1.2\Sx
        \advance \Bx by  -0.2\Ex
        \advance \By by  1.2\Sy
        \advance \By by -0.2\Ey
        \rput(\expandafter\Rval\the\Bx,\expandafter\Rval\the\By){#6}
        \advance \Bx by -\Bx
        \advance \By by -\By
\fi
\if\opti R 
        \advance \Rx by \Mx
        \advance \Rx by 0.5\txtlen
        \advance \Rx by \txthgt
        \advance \Ry by \My
        \rput(\expandafter\Rval\the\Rx,\expandafter\Rval\the\Ry){#6}
        \advance \Rx by -\Rx
        \advance \Ry by -\Ry
\fi
\if\opti L 
        \advance \Lx by \Mx
        \advance \Lx by -0.5\txtlen
        \advance \Lx by -\txthgt
        \advance \Ly by \My
        \rput(\expandafter\Rval\the\Lx,\expandafter\Rval\the\Ly){#6}
        \advance \Lx by -\Lx
        \advance \Ly by -\Ly
\fi
\if\opti U 
        \advance \Ux by \Mx
        \advance \Uy by \My
        \advance \Uy by 0.5\txthgt
        \advance \Uy by \txthgt
        \rput(\expandafter\Rval\the\Ux,\expandafter\Rval\the\Uy){#6}
        \advance \Ux by -\Ux
        \advance \Uy by -\Uy
\fi
\if\opti D 
        \advance \Dx by \Mx
        \advance \Dy by \My
        \advance \Dy by -0.5\txthgt
        \advance \Dy by -\txthgt
        \rput(\expandafter\Rval\the\Dx,\expandafter\Rval\the\Dy){#6}
        \advance \Dx by -\Dx
        \advance \Dy by -\Dy
\fi
}
\advance \Sx by -\Sx
\advance \Sy by -\Sy
\advance \Ex by -\Ex
\advance \Ey by -\Ey
\advance \Mx by -\Mx
\advance \My by -\My
\advance \Tx by -\Tx
\advance \Ty by -\Ty
\advance\txtlen by -\txtlen
\advance\txthgt by -\txthgt
}}
{\catcode`\p=12\catcode`\t=12\gdef\Rval#1pt{#1}}
\makeatother

\makeatletter
\def\LPut{\@ifnextchar[{\@LPut}{\@LPut[]}} 
\def\@LPut[#1]#2#3#4#5#6{{
\advance \Sx by #2\p@
\advance \Sy by #3\p@
\advance \Ex by #4\p@
\advance \Ey by #5\p@
\advance \Mx by 0.5\Sx
\advance \Mx by 0.5\Ex
\advance \My by 0.5\Sy
\advance \My by 0.5\Ey
\setbox0\hbox{#6}\advance\txtlen by 0.035146\wd0%
\setbox0\vbox{#6}\advance\txthgt by 0.035146\ht0%
\txtlen=\inverseFigureSize\txtlen
\txthgt=\inverseFigureSize\txthgt
\@tfor\opti :=#1\do{
\if\opti F 
        \advance \Fx by  -0.2\Sx
        \advance \Fx by   1.2\Ex
        \advance \Fy by -0.2\Sy
        \advance \Fy by  1.2\Ey
        \rput(\expandafter\Rval\the\Fx,\expandafter\Rval\the\Fy){#6}
        \advance \Fx by -\Fx
        \advance \Fy by -\Fy
\fi
\if\opti B 
        \advance \Bx by   1.2\Sx
        \advance \Bx by  -0.2\Ex
        \advance \By by  1.2\Sy
        \advance \By by -0.2\Ey
        \rput(\expandafter\Rval\the\Bx,\expandafter\Rval\the\By){#6}
        \advance \Bx by -\Bx
        \advance \By by -\By
\fi
\if\opti R 
        \advance \Rx by \Mx
        \advance \Rx by 0.5\txtlen
        \advance \Rx by \txthgt
        \advance \Ry by \My
        \rput(\expandafter\Rval\the\Rx,\expandafter\Rval\the\Ry){#6}
        \advance \Rx by -\Rx
        \advance \Ry by -\Ry
\fi
\if\opti L 
        \advance \Lx by \Mx
        \advance \Lx by -0.5\txtlen
        \advance \Lx by -\txthgt
        \advance \Ly by \My
        \rput(\expandafter\Rval\the\Lx,\expandafter\Rval\the\Ly){#6}
        \advance \Lx by -\Lx
        \advance \Ly by -\Ly
\fi
\if\opti U 
        \advance \Ux by \Mx
        \advance \Uy by \My
        \advance \Uy by 0.5\txthgt
        \advance \Uy by \txthgt
        \rput(\expandafter\Rval\the\Ux,\expandafter\Rval\the\Uy){#6}
        \advance \Ux by -\Ux
        \advance \Uy by -\Uy
\fi
\if\opti D 
        \advance \Dx by \Mx
        \advance \Dy by \My
        \advance \Dy by -0.5\txthgt
        \advance \Dy by -\txthgt
        \rput(\expandafter\Rval\the\Dx,\expandafter\Rval\the\Dy){#6}
        \advance \Dx by -\Dx
        \advance \Dy by -\Dy
\fi
}
\advance \Sx by -\Sx
\advance \Sy by -\Sy
\advance \Ex by -\Ex
\advance \Ey by -\Ey
\advance \Mx by -\Mx
\advance \My by -\My
\advance\txtlen by -\txtlen
\advance\txthgt by -\txthgt
}}
{\catcode`\p=12\catcode`\t=12\gdef\Rval#1pt{#1}}
\makeatother

\newcommand\FigLadder{%
\begin{figure}[h]
\psset{unit=\FigureSize cm}
\begin{center}
\begin{pspicture}(-3,-2)(7,5)
\psgrid
\scriptsize{%
\CArrow[OL]{-3}4 {-1}3{$(N-1,-1)$~~~}\CArrow[OU]{3.5}3 {-1}3{$R_1$}\CArrow[OR]{3.5}3 {5.5}5{$(m+1,1)$}
\CArrow[OL]{-1}3 02{$ $}          \CArrow[OU]32 02{$R_2$}       \CArrow[OR]32 {3.5}3{$ $}
\psline(0,2)(0,1.5)                                           \psline(3,1.5)(3,2)
\psline[linestyle=dashed,dash=2pt 1pt](0  ,0.5)(0  ,1.5)
\psline[linestyle=dashed,dash=2pt 1pt](1.5,0.5)(1.5,1.5)
\psline[linestyle=dashed,dash=2pt 1pt](3  ,0.5)(3  ,1.5)
\psline(0,0.5)(0,0)                                          \psline(3,0)(3,0.5)
\CArrow[OL]00 {-2}{-2}{$(-1,-1)$} \CArrow[OD]30 00{$R_N$}      \CArrow[OR]7{-2} 30{~~~$(m+1-N,1)$}%
}
\end{pspicture}
\end{center}
\caption{
Ladder diagram for $SU(N)$ gauge theory. 
There are $N+1$ possible toric diagrams ($m= 0, \cdots N$).
}
\end{figure}
}

\newcommand\FigDivisor{%
\begin{figure}[h]
\psset{unit=\FigureSize cm}
\begin{center}
\begin{pspicture}(-2,-1)(8,5)
\psgrid
\scriptsize{%
\rput(4,1.5){${\bf F}_{N-2k+m}$}
\CArrow[OL]35 33{$(N-k,-1)$}  \CArrow[OU]53 33{$(-1,0)$}      \CArrow[OR]53 65{$(m-k+2,1)$}
\CArrow[OL]33 00{$(N-k-1,-1)$}\CArrow[OD]{6.5}0 00{$(-1,0)$}\CArrow[OR]{6.5}0 53{$(m-k+1,1)$}
\CArrow[OL]00 {-2}{-1}{$(N-k-2,-1)$}                        \CArrow[OR]8{-1} {6.5}0{$(m-k,1)$}%
}
\end{pspicture}
\end{center}
\caption{
Subdiagram of the $k$-th divisor of $SU(N)_m$ geometry ($1 \leq k \leq N-1$)
}
\end{figure}
}

\newcommand\FigTV{%
\begin{figure}[h]
\psset{unit=\FigureSize cm}
\begin{center}
\begin{pspicture}(-12,0)(12,5)
\psgrid
\rput(-3,2){
\rput(0,-2){$\Cioo{\mu}{\lambda}{\nu}qt$}
\CArrow[OB]{-1.732}{-1} 00{$\mu$}
\CArrow[NF]00 02{$\lambda$}
\CArrow[OF]00 {1.732}{-1} {$\nu$}%
}
\rput(9,2){
\rput(0,-2){$\Coii{\mu}{\lambda}{\nu}qt$}
\CArrow[OF]00 {-1.732}{-1}{$\mu$}
\CArrow[SB]02 00{$\lambda$}
\CArrow[OB]{1.732}{-1} 00{$\nu$}%
}
\rput(3,2){
\rput(0,-2){$\Cooi{\mu}{\lambda}{\nu}qt$}
\CArrow[OF]00 {-1.732}{-1}{$\mu$}
\CArrow[NF]00 02{$\lambda$}
\CArrow[OB]{1.732}{-1} 00{$\nu$}%
}
\rput(-9,2){
\rput(0,-2){$\Ciio{\mu}{\lambda}{\nu}qt$}
\CArrow[OB]{-1.732}{-1} 00{$\mu$}
\CArrow[SB]02 00{$\lambda$}
\CArrow[OF]00 {1.732}{-1} {$\nu$}%
}
\end{pspicture}
\end{center}
\caption{
Refined topological vertex:
the represantation for the preferred direction,
i.e. the middle index $\lambda$,
is indicated by the white arrow.
}
\end{figure}
}

\newcommand\FigFeynman{%
\begin{figure}[h]
\psset{unit=\FigureSize cm}
\begin{center}
\begin{pspicture}(-2,-3)(15,3)
\psgrid
\rput(0,0){
\psline[arrowsize=5pt]{->}(0,0)(-1.732,-1)
\psline[arrowsize=5pt]{->}(0,0)(0,2)
\psline[arrowsize=5pt]{->}(0,0)(1.732,-1)%
\LPut[F]00 {-1.732}{-1}{$\vv_k$}
\LPut[SF]00 02{$\vv_j$}
\LPut[F]00 {1.732}{-1} {$\vv_i$}%
\rput(0,-3){{vertex}}
}
\rput(9,0){
\psline[arrowsize=5pt]{->}(-1, 1.732)(0,0)\psline[arrowsize=5pt]{->}(0,0)(4,0)\psline[arrowsize=5pt]{->}(4,0)(5, 1.732)
\psline[arrowsize=5pt]{->}(-1,-1.732)(0,0)                                    \psline[arrowsize=5pt]{->}(4,0)(5,-1.732)%
\LPut[B]{-1}{ 1.732} 00{$\vv_i$}\LPut[D]00 40{$\vv_k$}             \LPut[F]40 5{ 1.732}{$\vv_{j'}$}
\LPut[B]{-1}{-1.732} 00{$\vv_j$}\LPut[U]00 40{$\Ya\lambda k$, $Q_k$}\LPut[F]40 5{-1.732}{$\vv_{i'}$}%
\rput(2,-3){{edge}}
}
\end{pspicture}
\end{center}
\caption{Gluing rules}
\end{figure}
}

\newcommand\FigFourpoint{%
\begin{figure}[h]
\psset{unit=\FigureSize cm}
\begin{center}
\begin{pspicture}(-1,-1.5)(26,7)
\psgrid
\rput(0,0){
\rput(2,-2){$\Ziiio\mu{\Ya\lambda 1}{\Ya\lambda 2}\nu Qqt $}%
\LPut[L]24 22{$Q$}%
                         \CArrow[OB]06 24{$\mu$}\CArrow[WB]44 24{$\Ya\lambda 1$}
                         \CArrow[OR]24 22{$\eta$}
\CArrow[OF]22 00{$\nu$}\CArrow[WB]42 22{$\Ya\lambda 2$}%
}
\rput(7,0){
\rput(2,-2){$\Ziioo{\mu}{\lambda^1}{\lambda^2}{\nu}Qqt$}%
\LPut[L]24 22{$Q$}%
                         \CArrow[OB]06 24{$\mu$}\CArrow[WB]44 24{$\Ya\lambda 1$}
                         \CArrow[OR]24 22{$\eta$}
\CArrow[WF]22 02{$\Ya\lambda 2$}\CArrow[OF]22 40{$\nu$}%
}
\rput(14,0){
\rput(2,-2){$\Zioio{\mu}{\lambda^1}{\lambda^2}{\nu}Qqt$}
\LPut[R]24 22{$Q$}%
\CArrow[WF]24 04{$\Ya\lambda 1$}\CArrow[OB]46 24{$\mu$}
                         \CArrow[OL]24 22{$\eta$}
\CArrow[OF]22 00{$\nu$}\CArrow[WB]42 22{$\Ya\lambda 2$}%
}
\rput(21,0){
\rput(2,-2){$\Ziooo\mu{\Ya\lambda 1}{\Ya\lambda 2}\nu Qqt $}%
\LPut[R]24 22{$Q$}%
\CArrow[WF]24 04{$\Ya\lambda 1$}\CArrow[OB]46 24{$\mu$}
                         \CArrow[OL]24 22{$\eta$}
\CArrow[WF]22 02{$\Ya\lambda 2$}\CArrow[OF]22 40{$\nu$}%
}
\end{pspicture}
\end{center}
\caption{Four-point function: 
the framing indices of the internal lines are
$1$, $0$, $0$ and $-1$ from the left.}
\end{figure}
}

\newcommand\FigSlice{%
\begin{figure}[h]
\psset{unit=\FigureSize cm}
\begin{center}
\begin{pspicture}(-1,-1)(16,7)
\psgrid
\rput(0,0){
\rput(2,-1.5){$\Ziioo{\bullet}{\mu}{\nu}{\bullet}Qqt$}%
\LPut[L]44 22{$Q$}%
                         \CArrow[OB]46 44{$\bu$}\CArrow[WB]64 44{$\mu$}
                         \CArrow[OR]44 22{$\eta$}
\CArrow[WF]22 02{$\nu$}\CArrow[OF]22 20{$\bu$}%
}
\rput(6.25,-1.5){$\sim$}%
\rput(9,0){
\rput(2,-1.5){$\Zoiio\bu\nu\bu\mu Qqt$}%
\LPut[L]44 22{$Q$}%
                         \CArrow[OF]44 46{$\mu$}\CArrow[OB]64 44{$\bu$}
                         \CArrow[eR]22 44{$\eta$}
\CArrow[OF]22 02{$\bu$}\CArrow[OB]20 22{$\nu$}%
}
\end{pspicture}
\end{center}
\caption{Changing the preferred direction}
\end{figure}
}

\newcommand\FigFlop{%
\begin{figure}[h]
\psset{unit=\FigureSize cm}
\begin{center}
\begin{pspicture}(-1,-1)(16,7)
\psgrid
\rput(0,0){
\rput(2,-1.5){$\Zioio{\mu}{\lambda^1}{\lambda^2}{\nu}Qqt$}%
\LPut[R]24 42{$Q$}%
\CArrow[WF]24 04{$\lambda^1$}\CArrow[OB]26 24{$\mu$}
                         \CArrow[OL]24 42{$\eta$}
                         \CArrow[OF]42 40{$\nu$}\CArrow[WB]62 42{$\lambda^2$}%
}
\rput(7.25,-1.5){$\sim$}%
\rput(9,0){%
\rput(4,-1.5){$\Ziioo{\mu}{\lambda^2}{\lambda^1}{\nu}{Q^{-1}}qt$}%
\LPut[L]44 22{$Q^{-1}$}%
                         \CArrow[OB]46 44{$\mu$}\CArrow[WB]64 44{$\lambda^2$}
                         \CArrow[OR]44 22{$\eta$}
\CArrow[WF]22 02{$\lambda^1$}\CArrow[OF]22 20{$\nu$}%
}
\end{pspicture}
\end{center}
\caption{Flop invariance}
\end{figure}
}

\newcommand\FigFlopSlice{%
\begin{figure}[h]
\psset{unit=\FigureSize cm}
\begin{center}
\begin{pspicture}(-1,-1)(25,7)
\psgrid
\rput(0,0){
\rput(2,-1.5){$\Zioio{\bu}{\nu}{\mu}{\bu}Qqt$}%
\LPut[R]24 42{$Q$}%
\CArrow[WF]24 04{$\nu$}\CArrow[OB]26 24{$\bu$}
                         \CArrow[OL]24 42{$\eta$}
                         \CArrow[OF]42 40{$\bu$}\CArrow[WB]62 42{$\mu$}%
}
\rput(7.25,-1.5){$\sim$}%
\rput(9,0){%
\rput(4,-1.5){$\Ziioo{\bu}{\mu}{\nu}{\bu}{Q^{-1}}qt$}%
\LPut[L]44 22{$Q^{-1}$}%
                         \CArrow[OB]46 44{$\bu$}\CArrow[WB]64 44{$\mu$}
                         \CArrow[OR]44 22{$\eta$}
\CArrow[WF]22 02{$\nu$}\CArrow[OF]22 20{$\bu$}%
}
\rput(16.25,-1.5){$\sim$}%
\rput(18,0){
\rput(2,-1.5){$\Zoiio{\bu}{\nu}\bu{\mu}{Q^{-1}}qt$}%
\LPut[R]24 42{$Q^{-1}$}%
\CArrow[OB]04 24{$\nu$}\CArrow[OF]24 26{$\bu$}
                         \CArrow[sL]24 42{$\eta$}
                         \CArrow[OB]40 42{$\bu$}\CArrow[OF]42 62{$\mu$}%
}
\end{pspicture}
\end{center}
\caption{Changing the preferred direction II}
\end{figure}
}

\newcommand\FigTrace{%
\begin{figure}[h]
\psset{unit=\FigureSize cm}
\begin{center}
\begin{pspicture}(-1,-1)(17,7)
\psgrid
\rput(0,0){%
\rput(2,-1.5){\small{$Z_2$}}%
\LPut[R]44 22{$\Lambda$}%
\LPut[R]22 20{$Q$}%
\CArrow[EF]44 64{$\bu$}\CArrow[OB]46 44{$\nu$}
                       \CArrow[OL]44 22{$\lambda$}
                       \CArrow[OF]22 20{$\nu$}\CArrow[EB]02 22{$\bu$}%
}
\rput(10,0){%
\rput(3,-1.5){\small{$Z_4$}}%
\CArrow[sB]06 24{$\bu$} \CArrow[OU]44 24{$\Ya\mu 4$, $Q_4$} \CArrow[wB]66 44{$\bu$}%
\CArrow[OL]24 22{$\Ya\mu 1$, $Q_1$} \CArrow[OD]22 42{$\Ya\mu 2$, $Q_2$} \CArrow[OR]42 44{$\Ya\mu 3$, $Q_3$}%
\CArrow[eB]00 22{$\bu$}                                      \CArrow[nB]60 42{$\bu$}%
}
\end{pspicture}
\end{center}
\caption{Examples for the trace formula: 
the framing indices for the internal lines of $Z_4$ are all $1$.}
\end{figure}
}

\newcommand\FigUi{%
\begin{figure}[h]
\psset{unit=\FigureSize cm}
\begin{center}
\begin{pspicture}(-1,-1)(23,6)
\psgrid
\rput(0,1){%
\rput(3,-2){\small{$U(1)$ partition function}}%
\LPut[U]22 42{$\Lambda$}%
\CArrow[OB]04 22{$\bu$}\CArrow[WD]42 22{$\lambda$}\CArrow[OB]44 42{$\bu$}
\CArrow[OF]22 20{$\bu$}                       \CArrow[OF]42 60{$\bu$}%
}
\rput(8,1){%
\rput(3,-2){\small{$\chi_y$ genus}}%
\LPut[U]22 42{$\Lambda$}%
\LPut[R]44 42{$Q$}%
\CArrow[OB]04 22{$\bu$}\CArrow[WD]42 22{$\lambda$}\CArrow[OB]44 42{$\nu$}
\CArrow[OF]22 20{$\nu$}                       \CArrow[OF]42 60{$\bu$}%
}
\rput(16,1){%
\rput(3,-2){\small{Elliptic genus}}%
\LPut[U]22 42{$\Lambda$}%
\LPut[R]44 42{$Q_2$}%
\LPut[L]04 22{$Q_1$}%
\CArrow[OB]04 22{$\mu$}\CArrow[WD]42 22{$\lambda$}\CArrow[OB]44 42{$\nu$}
\CArrow[OF]22 20{$\nu$}                       \CArrow[OF]42 60{$\mu$}%
}
\end{pspicture}
\end{center}
\caption{$U(1)$ partition function, $\chi_y$ genus and Elliptic genus }
\end{figure}
}

\newcommand\FigSUii{%
\begin{figure}[h]
\psset{unit=\FigureSize cm}
\begin{center}
\begin{pspicture}(-1,0)(7,7)
\psgrid
                         \LPut[U]44 24{$\Lambda Q_{1,2}$}
\LPut[L]24 22{} \LPut[D]42 22{$\Lambda Q_{1,2}$} \LPut[R]42 44{}%
\CArrow[OB]06 24{$\bu$} \CArrow[WD]44 24{$\Ya\lambda 1$} \CArrow[OB]66 44{$\bu$}
\CArrow[OL]24 22{$Q_{1,2}$, $\mu$} \CArrow[WU]42 22{$\Ya\lambda 2$} \CArrow[OR]44 42{$\nu$, $Q_{1,2}$}
\CArrow[OF]22 00{$\bu$}                                      \CArrow[OF]42 60{$\bu$}%
\end{pspicture}
\end{center}
\caption{
$SU(2)$ partition function:
the framing indices for the bottom and the right internal line are $-1$ and those for the top and the left internal lines are one.}
\end{figure}
}

\newcommand\FigSUN{%
\begin{figure}[h]
\psset{unit=\FigureSize cm}
\begin{center}
\begin{pspicture}(0,0)(28,13)
\psgrid
\rput(6,6){%
\CArrow[OB]06 14{$\bu$}  \CArrow[WU]64 14{$\Ya\lambda 1$, $\Lambda_{1,m}$}\CArrow[OB]96 64{$\bu$}
\CArrow[OR]14 12{$\!\Ya\mu 1$}\CArrow[WU]42 12{~~~$\Ya\lambda 2$, $\Lambda_{2,m}$}\CArrow[OL]64 42{$\Ya\nu 1\!$}
\CArrow[OR]12 00{$\!\Ya\mu 2$}                            \CArrow[OL]42 30{$\Ya\nu 2\!$}%
\LPut[L]14 12{$Q_{1,2}$}\LPut[R]42 64{$Q_{1,2}$}
\LPut[L]12 00{$Q_{2,3}$}\LPut[R]30 42{$Q_{2,3}$}%
}
\rput(0,0){%
\psline[linestyle=dashed,dash=3pt 2pt](6,6)(5,4)%
\psline[linestyle=dashed,dash=3pt 2pt](8,4)(9,6)%
\CArrow[OR]54 32{$\!\Ya\mu {\nn-1}$}\CArrow[WD]82 32{$\Ya\lambda {\nn}$, $\Lambda_{\nn,m}$}\CArrow[OL]84 82{$\Ya\nu {\nn-1}\!$}
\CArrow[OF]32 00{$\bu$}                                    \CArrow[OF]82 90{$\bu$}%
\LPut[L]54 32{$Q_{\nn-1,\nn}$} \LPut[R]82 84{$Q_{\nn-1,\nn}$}%
}
\rput(18,6){\scriptsize{
\CArrow[OL]06 14{$(1,-1)$}\CArrow[WD]54 14{$(-1,0)$~~~~~~}\CArrow[OR]86 54{$(N-1-m,1)$}
\CArrow[OL]14 12{$(0,-1)$}\CArrow[WD]32 12{$(-1,0)$~~~~}\CArrow[OR]54 32{$(N-2-m,1)$}
\CArrow[OL]12 00{$(-1,-1)$}                        \CArrow[OR]32 20{$(N-3-m,1)$}%
\rput(3,4.5){\tiny{$N-1-m$}}
\rput(2,2.5){\tiny{~~$N-3-m$}}%
}}
\rput(12,0){\scriptsize{
\psline[linestyle=dashed,dash=3pt 2pt](6,6)(5,4)%
\psline[linestyle=dashed,dash=3pt 2pt](7,4)(8,6)%
\CArrow[OL]54 32{$(2-N,-1)$}\CArrow[WU]72 32{~~~$(-1,0)$}\CArrow[OR]74 72{$(-m,1)$}
\CArrow[OL]32 00{$(1-N,-1)$}\CArrow[OR]72 80{$(-1-m,1)$}%
\rput(5,1.5){\tiny{$1-N-m$}}%
}}
\end{pspicture}
\end{center}
\caption{
$SU(\nn)$ partition function:
the framing indices for the horizontal lines are 
$N-1-m$, $N-3-m$,$\cdots$,$1-N-m$ from the top to the bottom. 
Those for the left and the right internal lines are $1$ and $-1$, respectively. 
}
\end{figure}
}

\newcommand\FigSUNiiN{%
\begin{figure}[h]
\psset{unit=\FigureSize cm}
\begin{center}
\begin{pspicture}(0,0)(18,17.5)
\psgrid
\rput(4,8){%
                       \CArrow[OB]4{10} 48{$\bu$}\CArrow[WD]68 48{$\Ya\lambda 1$}\CArrow[OB]6{10} 68{$\bu$}
                       \CArrow[OR]48 26{$\Ya\mu 1$}                          \CArrow[OL]68 86{$\Ya\nu 1$}
\CArrow[WF]26 06{$\bu$}\CArrow[OR]26 24{$\Ya\mu 2$} \CArrow[WD]84 24{$\Ya\lambda 3$}\CArrow[OL]86 84{$\Ya\nu 2$}\CArrow[WB]{10}6 86{$\bu$}
                       \CArrow[OR]24 02{$\Ya\mu 3$}                          \CArrow[OL]84 {10}2{$\Ya\nu 3$}
\CArrow[WF]02 {-2}2{$\bu$}\CArrow[OR]02 00{$\Ya\mu 4$}                       \CArrow[OL]{10}2 {10}0{$\Ya\nu 4$}\CArrow[WB]{12}2 {10}2{$\bu$}%
                       \LPut[U]68 48{$\Lambda$}
\LPut[L]48 26{$\QQ 12$}                        \LPut[R]86 68{$\QQp 12$}
\LPut[L]26 24{$\QQ 23$}\LPut[U]84 24{$\Lambda$}\LPut[R]84 86{$\QpQ 23$}
\LPut[L]24 02{$\QQ 34$}                        \LPut[R]{10}2 84{$\QQp 34$}
\LPut[L]02 00{$\QQ 45$}                        \LPut[R]{10}0 {10}2{$\QpQ 45$}%
}
\psline[linestyle=dashed,dash=3pt 2pt](4,6)(4,8)%
\psline[linestyle=dashed,dash=3pt 2pt](14,6)(14,8)%
                       \CArrow[OR]46 44{$\Ya\mu {2\nn-2}$}\CArrow[WD]{14}4 44{$\Ya\lambda {2\nn-1}$}\CArrow[OL]{14}6 {14}4{$\Ya\nu {2\nn-2}$}
                       \CArrow[OR]44 22{$\Ya\mu {2\nn-1}$}                                   \CArrow[OL]{14}4 {16}2{$\Ya\nu {2\nn-1}$}
\CArrow[WF]22 02{$\bu$}\CArrow[OF]22 20{$\bu$}                                            \CArrow[OF]{16}2 {16}0{$\bu$}\CArrow[WB]{18}2 {16}2{$\bu$}%
\LPut[L]46 44{$\QQ{2\nn-2}{2\nn-1}$}\LPut[U]{14}4 44{$\Lambda$}\LPut[R]{14}6 {14}4{$\QpQ{2\nn-2}{2\nn-1}$}
\LPut[L]44 22{$\QQ{2\nn-1}{2\nn  }$}                           \LPut[R]{14}4 {16}2{$\QQp{2\nn-1}{2\nn  }$}%
\end{pspicture}
\end{center}
\caption{
$SU(\nn)$ partition function with $N_f =2\nn$:
the framing index for the longitudinal internal lines are all $-1$.
}
\end{figure}
}

\begin{document}

%
\begin{titlepage}
\begin{flushright}
{April, 2008 \\
revised July, 2008} 
\end{flushright}
\vspace{0.5cm}
\begin{center}
{\Large \bf
Refined BPS state counting from Nekrasov's formula \\
and Macdonald functions}
\vskip1.0cm
{\large Hidetoshi Awata and Hiroaki Kanno}
\vskip 1.0em
{\it 
Graduate School of Mathematics \\
Nagoya University, Nagoya, 464-8602, Japan}
\end{center}
\vskip1.5cm

\begin{abstract}
It has been argued that Nekrasov's partition function
gives the generating function of refined BPS state counting 
in the compactification of $M$ theory on local Calabi-Yau spaces.
We show that a refined version of the topological vertex 
we previously proposed (hep-th/0502061) is a building block of
Nekrasov's partition function with two equivariant parameters.
Compared with another refined topological vertex
by Iqbal, Kozcaz and Vafa (hep-th/0701156), our refined vertex 
is expressed entirely in terms of the specialization of
the Macdonald symmetric functions which is related to the equivariant
character of the Hilbert scheme of points on ${\mathbb C}^2$. 
We provide diagrammatic rules for computing the partition function from
the web diagrams appearing in geometric engineering of
Yang-Mills theory with eight supercharges. 
Our refined vertex has a simple transformation law
under the flop operation of the diagram, which suggests that
homological invariants 
of the Hopf link are related to the Macdonald functions. 
\end{abstract}
\end{titlepage}


\renewcommand{\thefootnote}{\arabic{footnote}} \setcounter{footnote}{0}


\Section{Introduction}

The problem of instanton counting is one of the important aspects of nonperturbative dynamics 
in gauge and string theory. The result is encoded 
in the partition function of topological gauge and string theories, which is often computed exactly by
the duality and/or the localization principle. A celebrated example in gauge theory is
Nekrasov's partition function 
$Z_{Nek}(\epsilon_i, a_\ell, \Lambda)$, which reproduces 
the  Seiberg-Witten prepotential from the microscopic viewpoint of equivariant integration 
over the instanton moduli space \cite{Nek}. On the string theory side, 
the topological vertex 
$C_{\lambda_1 \lambda_2 \lambda_3}(q)$
is constructed based on the geometric transition, which is a duality of topological closed string 
to the Chern-Simons theory \cite{AMV,AKMV}. 
The topological vertex provides a building block of all genus topological string amplitudes
on local toric Calabi-Yau 3-fold. It is amusing that these two instanton counting problems are 
actually related in an appropriate  setup, which is expected from geometric engineering \cite{KKV, KMV}.

To compute the integration over the instanton moduli space of $SU(N)$ gauge theory,
Nekrasov considered the toric action on 
${\mathbb R}^4 \simeq {\mathbb C}^2 \ni (z_1, z_2) \to (e^{i \epsilon_1} z_1, e^{i \epsilon_2} z_2)$,
which induces the toric action on the moduli space of framed instantons. 
By the localization theorem, the integral becomes a sum
over the contributions from each fixed point of the toric action,
which is labeled by  the set of $N$ Young diagrams (the \lq\lq colored\rq\rq\  partitions)
$\{\lambda_{\ell}\}_{\ell=1}^{N},~
(\lambda_{\ell,1} \geq \lambda_{\ell,2} \geq \cdots \geq \lambda_{\ell,i} \geq \lambda_{\ell,i+1} \geq \cdots )$.
As a consequence we obtain Nekrasov's partition function \cite{rf:NakajimaYoshioka}:
\beq
Z_{Nek} (\epsilon_1, \epsilon_2, a_\ell , \Lambda) = \sum_{\{\lambda_{\ell}\}} \frac{\Lambda^{2 N |\lambda|}}
{\prod_{\alpha, \beta=1}^{N} n_{\alpha,\beta}^{\{\lambda_{\ell}\}} (\epsilon_1, \epsilon_2, a_\ell)}~,
\eeq
where $|\lambda|= \sum_{\ell=1}^{N} \sum_{1 \leq i} \lambda_{\ell,i}$ and 
\beqa
n_{\alpha,\beta}^{\{\lambda_{\ell}\}} 
&=& \prod_{s \in \lambda_\alpha}
\left(-\ell_{\lambda_\beta}(s)\epsilon_1 + (a_{\lambda_\alpha}(s) +1) \epsilon_2 + a_\beta - a_\alpha \right)  \CR
& &~~~\prod_{t \in \lambda_\beta}\left((\ell_{\lambda_\alpha}(t) +1)\epsilon_1
-  a_{\lambda_\beta}(t)\epsilon_2 + a_\beta - a_\alpha \right).
\eeqa
The parameter $\Lambda$ of instanton expansion is introduced as a dynamical scale in the renormalization. 
The vacuum expectation values of the scalar fields in the vector multiplets are $a_\alpha, 1 \leq \alpha \leq N$.
Mathematically they are equivariant parameters for the action of maximal torus on the gauge group. 
We denote the leg length and the arm length at $s=(i,j)$ with respect to  the Young diagram $\lambda$
by $\ell_\lambda(i,j)=\lambda_j^\vee - i,$ and $a_\lambda(i,j)=\lambda_i - j$, respectively.
The relation to the (topological) string theory becomes transparent, if we consider 
a five-dimensional (\lq\lq trigonometric\rq\rq, or $K$-theoretic) lift of
the partition function by promoting the factors 
$n_{\alpha,\beta}^{\{\lambda_{\ell}\}}(\epsilon_1, \epsilon_2, a_\ell)$
in the denominator to
\beq
N_{\alpha,\beta}^{\{\lambda_{\ell}\}} (\epsilon_1, \epsilon_2, a_\ell)
= \prod_{s \in \lambda_\alpha} \left( 1-t^{-\ell_{\lambda_\beta}(s)} q^{- a_{\lambda_\alpha}(s) -1} 
{\bf e}_\beta/{\bf e}_\alpha \right) 
\prod_{t \in \lambda_\beta}\left( 1- t^{\ell_{\lambda_\alpha}(t) +1} q^{a_{\lambda_\beta}(t)} 
{\bf e}_\beta/{\bf e}_\alpha \right),
\eeq
where $(q,t):=(e^{\epsilon_2},e^{-\epsilon_1})$ and ${\bf e}_\alpha := e^{-a_\alpha}$.
We can show that, when $q=t=e^{-g_s}$, Nekrasov's partition function is 
nothing but the topological string amplitude on an appropriate local toric Calabi-Yau manifold
\cite{IK-P1}--\cite{Zhou}.

All genus topological string amplitude on local Calabi-Yau 3-fold can be computed
by a diagrammatic rule, in terms of the topological vertex:
\beq
C_{\mu \lambda \nu} (q) = q^{\frac{\kappa(\nu)}{2}} s_{\lambda}(q^\rho) 
\sum_{\eta} s_{\mu/\eta}(q^{\lambda^\vee+\rho}) s_{\nu^\vee/\eta}(q^{\lambda+\rho})~,
\eeq
where $s_{\lambda/\mu}(x)$ is the (skew) Schur function and 
$q^{\lambda+\rho}$ means the substitution $x_i := q^{\lambda_i -i + \frac{1}{2}}$.
The partition $\lambda^\vee$ is defined by the transpose of the corresponding Young diagram. 
The definition of $\kappa(\lambda)$ is given in Appendix {\appNotation}\footnote{Our notations for partitions are 
summarized in Appendix  {\appNotation}.}. 
Then a natural question is: For generic parameters $(\epsilon_1, \epsilon_2)$, 
can we obtain $Z_{Nek}(\epsilon_i, {a}_\ell, \Lambda)$
in a similar manner by generalizing the topological vertex $C_{\mu \lambda \nu}(q)$?
This is the problem of constructing a refined topological vertex. An answer to this question has been given by 
Iqbal, Kozcaz and Vafa \cite{IKV}. The refined topological vertex they proposed is
\begin{align}
&C_{\mu\nu\lambda}^{(IKV)}(t,q)= \CR
& \left(\frac{q}{t}\right)^{\frac{||\nu||^2 + ||\lambda||^2}{2}} t^{\frac{\kappa(\nu)}{2}}
P_{\lambda^\vee}(t^{-\rho}; q,t)  
\sum_{\eta} \left(\frac{q}{t} \right)^{\frac{|\eta|+|\mu|-|\nu|}{2}}
s_{\mu^\vee/\eta}(q^{-\lambda}t^{-\rho}) 
s_{\nu/\eta}(t^{-\lambda^\vee} q^{-\rho})
~.
\end{align}
As before, $q^{\lambda} t^{\rho}$ etc. means 
the specialization $x_i := q^{\lambda_i} t^{\frac{1}{2} - i}$. 
On the other hand, before the proposal in \cite{IKV} we had introduced the following vertex in 
\cite{AK}\footnote{We have slightly changed the original definition in \cite{AK}
by improving the framing factor.}:
\begin{align}
& { \Ciio\mu\lambda\nu qt }
= \CR
& { \fla\nu qt }^{-1}
P_{\lambda}(t^{\rho}; q,t) 
\sum_{\eta} \left(\frac{q}{t}\right)^{\frac{|\eta|-|\nu|}{2}} 
~{\iota} P_{\mu^\vee/\eta^\vee}( -t^{\lambda^\vee}q^{\rho}; t,q)
P_{\nu/\eta}(q^{\lambda} t^{\rho}; q,t)~.  
\label{ourvertex}
\end{align}
It is convenient to introduce the conjugate vertex 
${ \Cooi\mu\lambda\nu qt } := { \Ciio{\mu^\vee}{\lambda^\vee}{\nu^\vee} tq } (-1)^{|\mu|+|\lambda|+|\nu|}$.
Note that the conjugation involves the exchange of $t$ and $q$. 
The refined vertex $C_{\mu\nu\lambda}^{(IKV)}(t,q)$ partly employs the Macdonald function 
$P_\lambda(x; q,t)$ \cite{Mac},
but there still remain the skew Schur functions\footnote{However,
this implies a nice interpretation in terms of plane partitions.}. 
Compared with it, our proposal eliminates the skew Schur functions completely and
the vertex is expressed in terms of the (skew) Macdonald function $P_{\lambda/\mu}(x;q,t)$.
The price for this elimination is that we have to introduce
the involution $\iota$ on the algebra of symmetric functions defined 
by $\iota(p_n) = -p_n$, where $p_n(x):= \sum_{i=1}^\infty x_i^n$ is the power sum function.
Since $\{p_n(x)\}_{n=1}^\infty$ forms a ``multiplicative''
basis of the algebra of  the symmetric functions,
the involution $\iota$ is uniquely defined by the above relation. 
Finally ${ \fla\lambda qt } := (-1)^{|\lambda|}q^{\frac{||\lambda||^2}{2}} t^{-\frac{||\lambda^\vee||^2}{2}}$ 
is the framing factor proposed recently by Taki \cite{Taki}.
The refined vertex $C_{\mu\nu\lambda}^{(IKV)}(t,q)$ has a nice interpretation as
the counting of \lq\lq unisotropic\rq\rq\  plane partitions, 
or by statistical mechanics of the melting crystal model \cite{ORV, IKV, IKS}. 
Although the relation of our vertex to such a statistical model is unclear, 
our vertex is more symmetric than $C_{\mu\nu\lambda}^{(IKV)}(t,q)$,
since we have replaced all the (skew) Schur functions in the topological vertex
by the (skew) Macdonald functions. 
However, it seems impossible to make $\Ciio\mu\lambda\nu qt$ 
completely symmetric under the cyclic permutation of partitions. 

In \cite{IKV} it is claimed (see also the arguments in \cite{Taki})
 that one can reproduce Nekrasov's partition function 
from the refined topological vertex $C_{\mu\nu\lambda}^{(IKV)}(t,q)$. 
%
As we mentioned already in \cite{AK},
our vertex \eqref{ourvertex} also reproduces the $SU(N)$ Nekrasov's partition function,
and in this article we will show it concretely.
Though $C_{\mu\nu\lambda}^{(IKV)}(t,q)$ and
$\Ciio\mu\lambda\nu qt$ are different, they give the same result 
as long as we put trivial representations to external edges, which is the case
when we compute Nekrasov's partition function by the method of topological vertex.
The Schur functions and the Macdonald functions are two different basis of
the space of symmetric functions and hence they satisfy the Cauchy formulas of the same type
(see section 5.2 and Appendix B). This is a technical reason why 
$C_{\mu\nu\lambda}^{(IKV)}(t,q)$ and $\Ciio\mu\lambda\nu qt$ 
give the same result after taking the summation over the partitions attached to internal edges. 
%
We should emphasize that, throughout this article except for appendix {\appDunkl},
$q$ and $t$ are treated as formal parameters
and we do not use 
any asymptotic computations such as $\lim_{N \to \infty} q^N =0$ otherwise stated.

In this paper we show that the refined topological vertex gives a building block of
the $K$-theoretic lift of Nekrasov's partition function. We would like to
point out the following possible application. 
It has been argued Nekrasov's partition function 
gives the generating function of refined BPS state counting 
in the compactification of $M$ theory on local Calabi-Yau spaces \cite{HIV, AK, IKV}.
 As far as we know (see appendix {\appSpin}, for examples), the refined BPS state counting always gives 
integers, which is a refined version of the conjecture of the integrality of
the Gopakumar-Vafa invariants \cite{GV1}
--\cite{KKV-M}. Recently the conjecture has been proven 
for local toric Calabi-Yau 3-folds \cite{Peng}
--\cite{Kon2}. The existence of the topological vertex is 
one of the important ingredients in the proofs. Hence one may expect that
the refined vertex is helpful in proving the integrality of the refined Gopakumar-Vafa
invariants for the local toric case. 
Macdonald functions are also related to
$q$-deformed Virasoro and $\cal W$ algebras
\cite{rf:AKOS}.
We hope these quantum groups play an important role 
in topological string theory and Yang-Mills theory.


The paper is organized as follows:
In section 2 we introduce the $K$-theoretic lift of 
Nekrasov's partition function following 
a mathematical formulation in \cite{NY2,GNY}.
The $K$-theoretic lift allows the Chern-Simons
coupling $m \in {\mathbb Z}$ and we find that
the framing factor $\fla\lambda qt$ 
of the refined topological vertex arises naturally
from the $m$ dependence of the partition function.
We also examine the symmetry of the partition function
under 
$r_L : (q,t) \to (q^{-1},t^{-1})$ and 
$r_R : (q,t) \to (t,q)$. 
This is a necessary condition for the $K$-theoretic lift
to be interpreted as a character of $Spin(4) = SU(2)_L \times SU(2)_R$.
In section 3 we review the idea of geometric engineering.
When we compute the partition function using
the refined topological vertex as a building block,
we will fix a preferred direction, for which we mainly choose 
the horizontal left arrow $(-1,0)$ in this paper. 
The (dual) toric diagram in geometric engineering has 
a feature in which we can arrange the diagram so that each vertex has 
a unique edge with the preferred direction. 
To emphasize the fact that we fix the preferred direction of the diagram,
we will call it web diagram in the following.
We define our refined topological vertex in section 4.
The gluing rules of the vertex for computing the partition function are
also provided. 
In section 5 we consider four-point functions obtained by
gluing two refined vertices. We show that the four-point function enjoys a 
rather simple transformation law under the flop operation of the web diagram.
Based on this transformation law we argue a possible relation of
our refined vertex to homological invariants 
of the Hopf link \cite{GSV, GIKV}.
In section 6 we discuss one-loop diagrams with some examples.
Finally, we present several examples of the computation of
the partition function in sections 7--9.
In appendix {\appPartition} 
we explain the equivalence between several expressions of the Nekrasov formula.
In appendix {\appMac} we give a definition of the Macdonald
symmetric functions and collect several useful formulas.
We present examples of the refined BPS state counting in appendix {\appSpin}.
In appendix {\appDunkl} we remark that  our refined topological vertex can be
expressed in terms of the $q$-Dunkl operator. Appendix {\appNotation} gives
a list of notations and some identities for partitions used in this paper.


The following notations are used through this article.
$q$ and  $t$ are formal parameters otherwise stated.
Let $\lambda$ be a Young diagram,
i.e. 
a partition $\lambda = (\lambda_1,\lambda_2,\cdots)$,
which is a sequence of nonnegative integers such that
$\lambda_{i} \geq \lambda_{i+1}$ and 
$|\lambda| = \sum_i \lambda_i < \infty$.
$\lambda^\vee $ is its conjugate (dual) diagram.
$\ell(\lambda) = \lambda^\vee_1$ is the length and 
$|\lambda| = \sum_i \lambda_i$ is the weight.
For each square $s=(i,j)$ in $\lambda$, 
\ba
a(s)&:=& \lambda_i-j,\qquad a'(s):=j-1,
\cr
\ell(s)&:=& \lambda^\vee_j-i,\qquad \ell'(s):=i-1,
\ea
are the arm length, arm colength, leg length and leg colength, 
respectively\footnote{In \cite{IKV} the definitions of the arm length 
and the leg length are exchanged.}.
Let 
$p_n(x) = \sum_{i=1}^\infty x_i^n$ 
be the power sum function in the set of variables $x=(x_1,x_2,\cdots)$.
If $|t^{-1}|<1$, 
the variable $q^\lambda t^\rho$ stands for 
$x_i = q^{\lambda_i} t^{\ha-i}$.
But for all $t\in\bC$, 
we define
\ba
p_n(c q^\lambda t^\rho) 
&:=&
c^n\sum_{i=1}^\infty (q^{n\lambda_i}-1)t^{n(\ha-i)}
+ { c^n \over t^{n\over 2} - t^{-{n\over 2}} }, 
\qquad\quad 
q, t, c\in\bC,
\cr
p_n(c q^\lambda t^\rho, c L t^{-\rho} ) 
&:=&
c^n\sum_{i=1}^\infty (q^{n\lambda_i}-1)t^{n(\ha-i)}
+ c^n {1 - L^n \over t^{n\over 2} - t^{-{n\over 2}} }, 
\qquad 
q, t, c, L\in\bC
\cr 
&=& c^n \sum_{i=1}^N q^{n\lambda_i} t^{n(\ha-i)},
\qquad 
q, t, c\in\bC,
\quad L=t^{-N}, N\in\bN.
\label{eq:powersum}
\ea
All symmetric functions in this article are treated as 
polynomials in the power sum symmetric functions 
$(p_1,p_2,\cdots)$.
Finally, we often use
$u:=(qt)^{\ha}$ and $v :=\left({q/t}\right)^{\ha}$.


\Section{Structure of Nekrasov's Partition Function}


\subsection{Partition function with Chern-Simons coupling}
\label{sec:PartitionFunctionCS}


The five-dimensional lift of Nekrasov's partition function of $SU(\nn)$ theory 
is given by the summation over the set of $\nn$ Young diagrams (or colored partitions)
$\{ \Ya\lambda\a \}_{\a=1}^{\nn}$, as follows:
\beq
Z^{\mathrm{inst}}(\epsilon_1, \epsilon_2; a_\a, \Lambda) \CR
= \sum_{ \{ \Ya\lambda\a \}} \frac{
\left(
e^{-\frac{\epsilon_1+ \epsilon_2}{2}}
\Lambda^2\right)^{\nn \cdot |\lambda|}}
{\prod_{\alpha, \beta=1}^{\nn}  
\NYNY }~.
\label{Klift-0}
\eeq
The product in the denominator is the equivariant Euler character of the tangent space to the instanton moduli 
space $M(\nn,k)$ at a fixed point of the toric action, which is labeled by $\{ \Ya\lambda\a \}_{\a=1}^{\nn}$ with $|\lambda|=k$ . 
Each factor is given by
\be
\NYNY
=
\prod_{s \in \lambda_\alpha}
\left( 1 - e^{
\ell_{\lambda_\beta}(s)\epsilon_1 - (a_{\lambda_\alpha}(s) + 1)\epsilon_2 
+ a_\alpha - a_\beta
}\right)
\prod_{t \in \lambda_\beta}
\left( 1 - e^{
-(\ell_{\lambda_\alpha}(t)+1)\epsilon_1 +a_{\lambda_\beta}(t)\epsilon_2 
+ a_\alpha - a_\beta
}\right).
\ee
The product 
$\prod_{\alpha, \beta=1}^{\nn} \NYNY$
consists of
$2\nn k$ factors, which agree to the complex dimensions of $M(\nn,k)$. 
In \cite{NY2, GNY} the five-dimensional lift is mathematically identified as 
the $K$-theoretic lift and it is computed as follows:
\beqa
& &Z_{m}^{\mathrm{inst}}(\epsilon_1, \epsilon_2; a_\a, \Lambda) \CR
&=& \sum_{k=0}^\infty 
\left(
e^{-\frac{1}{2}(\nn+m)(\epsilon_1+ \epsilon_2)}\Lambda^{2\nn} \right)^k
\sum_{i} (-1)^i {\mathrm{ch}} H^i(M(\nn,k), {\cal L}^{\otimes m}) \CR
&=& \sum_{\{\Ya\lambda\a\}} \frac{
\left(
e^{-\frac{1}{2}(\nn+m)(\epsilon_1+ \epsilon_2)}\Lambda^{2\nn} \right)^{|\lambda|}}
{\prod_{\alpha, \beta} \NYNY }
\cdot \exp \left(m \sum_\alpha \sum_{s \in \lambda_\alpha} (a_\alpha - \ell'(s)\epsilon_1 - a'(s) \epsilon_2) \right)~,
\label{Klift-m}
\eeqa
where $M(\nn,k)$ is the framed moduli space of rank $\nn$ torsion free sheaves $E$ on ${\mathbb P}^2$ with
$c_2(E)=k$. The line bundle ${\cal L}$ over $M(\nn,k)$ is defined by
\beq
{\cal L} := {\mathrm{det}} \left[ R^1 (p_{2})_* ({\cal E} \otimes (p_1)^*{\cal O}_{{\mathbb P}^2}(-\ell_\infty))\right]~,
\eeq
where ${\cal E}$ is the universal sheaf on ${\mathbb P}^2 \times M(\nn,k)$ and $p_{1,2}$ is 
the projection to the first or the second component. 
Physically $Z_{m}^{\mathrm{inst}}$ is the instanton part of the partition function of 
$SU(\nn)$ gauge theory on ${\mathbb R}^4 \times S^1$ with eight supercharges, 
and the power $m \in {\mathbb Z}$ of the line bundle ${\cal L}$ is identified 
as the coefficient of the five-dimensional Chern-Simons term \cite{Tac}.

Let $(q,t):=(e^{\epsilon_2},e^{-\epsilon_1})$,  
${\bf e}_\alpha := e^{-a_\alpha}$ and
$Q_{\alpha,\beta} := \QQ\alpha\beta$;%
\footnote{The different conventions $(q,t):=(e^{-\epsilon_2},e^{\epsilon_1})$ and 
$Q_{\beta,\alpha} := \QQ\alpha\beta$ 
are also used in the literature.}
then 
$Z_{m}^{\mathrm{inst}}(\epsilon_1, \epsilon_2; a_\a, \Lambda)$
is written as
\be
\ZQ{inst}{m}qt{{\bf e}_1,\cdots,{\bf e}_{\nn}}\Lambda
=
\sum_{\{\Ya\lambda\a\}}
{
\prod_{\a =1}^{\nn}
\left(v^{-\nn}\Lambda^{2\nn} 
\left(-{\bf e}_\a \right)^{-m}\right)^{|\Ya\lambda\a |}
\fla{\Ya\lambda\a }qt ^{-m}
\over
\prod_{\a ,\b  = 1}^{\nn}
\Nek{\Ya\lambda\a }{\Ya\lambda\b }{Q_{\a ,\b }}qt
}~,
\label{eq:NekrasovZinstM}
\ee
with $v:=(q/t)^\ha$ and 
\ba
\Nek{\lambda_\alpha}{\lambda_\beta}{Q_{\alpha,\beta}}qt 
&:=& 
N^{\{\lambda_{\alpha}\}}_{\beta,\alpha}
\left(\epsilon_1,\epsilon_2;a_\alpha\right)
\cr 
&=&
\!\prod_{s \in \lambda_\alpha}
\left( 1- q^{a_{\lambda_\alpha}(s)} t^{\ell_{\lambda_\beta}(s) +1} {Q_{\alpha,\beta}} \right)  
\!\prod_{t \in \lambda_\beta}
\left( 1- q^{- a_{\lambda_\beta}(t) - 1} t^{-\ell_{\lambda_\alpha}(t)} {Q_{\alpha,\beta}} \right),~~~
\ea
where
$|\lambda|$ is the number of boxes of $\lambda$ and
\be
\fla\lambda qt  
:= 
\prod_{s\in\lambda}(-1) q^{a'(s) + \ha} t^{ - \ell'(s) -\ha}
=
\prod_{(i,j)\in\lambda} 
(-1) q^{\lambda_i - j + \ha} t^{ -\lambda^\vee_j + i -\ha},
\ee
is the framing factor
\footnote{
In terms of 
$||\lambda||^2 
:= 
\sum_i \lambda_i^2 
= 
2\sum_{s\in\lambda}(a(s)+\ha)
$,
this is
$
\fla\lambda qt  
= 
(-1)^{|\lambda|} q^{\ha||\lambda||^2} t^{ -\ha||\lambda^\vee||^2}
$.
}
which has been proposed by Taki \cite{Taki}.
This is nothing but the $m$ dependent $(q,t)$ factor of 
the partition function.
Note that the framing factor satisfies the following symmetry:
\be
\fla\lambda qt 
=
\fla\lambda{q^{-1}}{t^{-1}} ^{-1}  
= 
\fla{\lambda^\vee}tq ^{-1} . \label{f-symmetry}
\ee
Let 
$u = (qt)^{\ha}$ and 
$v =\left({q/t}\right)^{\ha}$. 
We have the following six equivalent expressions of $\Nek{\lambda}{\mu}Qqt $:
%
\\
{\bf Proposition.}
%
\ba
\Nek{\lambda}{\mu}Qqt  
&=&
\prod_{(i,j)\in\mu } 
\left( 1 - Q\, q^{\lambda_i-j} t^{\mu^\vee_j-i+1} \right)
\prod_{(i,j)\in\lambda} 
\left( 1 - Q\, q^{-\mu_i+j-1} t^{-\lambda^\vee_j+i  } \right),
\label{eq:NekIp}
\\
\Nek{\lambda}{\mu}Qqt  
&=&
\prod_{(i,j)\in\lambda} 
\left( 1 - Q\, q^{\lambda_i-j} t^{\mu^\vee_j-i+1} \right)
\prod_{(i,j)\in\mu } 
\left( 1 - Q\, q^{-\mu_i+j-1} t^{-\lambda^\vee_j+i  } \right),
\label{eq:NekIm}
\ea
\ba
\Nek{\lambda}{\mu}Qqt  
&=&
\tPi\left(-v^{-1} Q\, q^{\lambda}t^{\rho},\  t^{\mu^\vee}q^{\rho}\right)
/
\tPi\left(-v^{-1} Q\, t^{\rho},\  q^{\rho}\right),
\label{eq:NekIIp}
\\
\Nek{\lambda}{\mu}Qqt  
&=&
\tPi\left(-v^{-1} Q\, t^{-\lambda^\vee}q^{-\rho},\  q^{-\mu}t^{-\rho}\right)
/
\tPi\left(-v^{-1} Q\, q^{-\rho},\  t^{-\rho}\right),
\ea
\ba
\Nek{\lambda}{\mu}Qqt  
&=&
\Pi\left(Q\, q^\lambda t^\rho,q^{-\mu}t^{-\rho};q,t\right)
/
\Pi\left(Q\, t^\rho,t^{-\rho};q,t\right),
\label{eq:NekIIIp}
\\
\Nek{\lambda}{\mu}Qqt  
&=&
\Pi\left(Q\, t^{\mu^\vee} q^\rho,t^{-\lambda^\vee}q^{-\rho};t^{-1},q^{-1}\right)
/
\Pi\left(Q\, q^\rho,q^{-\rho};t^{-1},q^{-1}\right).
\ea
where
\ba
\tPi(-x,y)
& := &
\Exp{
-\sum_{n>0}
{1
\over n} p_n(x) p_n(y)
}
=
\prod_{i,j}(1- x_i y_j),
\label{eq:Pizero}
\\
\Pi(v x,y;q,t)
& := &
\Exp{
\sum_{n>0}{1\over n}
{
t^{{n\over 2}}-t^{-{n\over 2}}
\over 
q^{{n\over 2}}-q^{-{n\over 2}}
}
p_n(x) p_n(y)
}
=
\left\{
\begin{array}{ll}
\ds{
\prod_{i,j}
{(u x_i y_j ; q)_\infty
\over 
 (v x_i y_j ; q)_\infty},
}
\quad &|q|<1
\cr
\ds{
\prod_{i,j}
{(u^{-1} x_i y_j ; q^{-1})_\infty
\over 
(v^{-1} x_i y_j ; q^{-1})_\infty},
}
\quad &|q^{-1}|<1.
\end{array}
\right.
\ea
Here 
$(x;q)_\infty$ is the $q$-shifted factorial 
$
(x;q)_\infty := \prod_{k\in\bZ_{\geq 0}} (1-q^k x)
$.
%
Note that when $|q^{-1}|, |t^{-1}| < 1$,
(\ref{eq:NekIIp}) is written as
\be
\prod_{i,j=1}^\infty
\left(1-Q\, q^{\lambda_i-j} t^{\mu^\vee_j-i+1}\right)
/
\left(1-Q\, t^{1-i} q^{-j}\right).
\ee
and also when $|q|<1$,
(\ref{eq:NekIIIp}) is
\be
\prod_{i,j=1}^\infty
{
\left(Q\, q^{\lambda_i-\mu_j} t^{j-i+1} ; q\right)_\infty
\over
\left(Q\, q^{\lambda_i-\mu_j} t^{j-i  } ; q\right)_\infty 
}
{
\left(Q\, t^{j-i  } ; q\right)_\infty
\over
\left(Q\, t^{j-i+1} ; q\right)_\infty  
}.
\ee
%
The equivalence of these six expressions are proved in app.\ {\appPartition}.
The first formula,
(\ref{eq:NekIp}),
is given by
\cite{rf:NakajimaYoshioka},
and 
(\ref{eq:NekIIp})--(\ref{eq:NekIIIp})
by \cite{rf:NekrasovOkounkov}. 
In this article, we mainly use 
(\ref{eq:NekIm}) and (\ref{eq:NekIIp}).


\subsection{Another form of the partition function}


Let us transform Nekrasov's partition function $Z_{m}^{\mathrm{inst}}$ so that it becomes 
transparent, and compare it with the amplitude constructed by the method of topological vertex.
Using (\ref{eq:appsetFormulaII}), i.e.
\be
\prod_{(i,j)\in\lambda}q^{\mu_i-j}
\prod_{(i,j)\in\mu}q^{-\lambda_i+j-1}
=
\prod_{(i,j)\in\mu}q^{\mu_i-j}
\prod_{(i,j)\in\lambda}q^{-\lambda_i+j-1},
\ee
we can show, 
from (\ref{eq:NekIp}), that
\ba
\Nek{\mu}{\lambda}{Q^{-1}}qt 
&=&
\Nek\lambda\mu{v^2Q}qt 
Q^{-|\lambda|-|\mu|} 
\fla\mu qt 
/
\fla\lambda qt 
\cr
&=&
\Nek{\mu^\vee}{\lambda^\vee}{Q}tq 
(vQ)^{-|\lambda|-|\mu|} 
\fla\mu qt 
/
\fla\lambda qt~. 
\label{inversion}
\ea
Here we use (\ref{eq:NDual}).
Hence we have
\ba
\prod_{\a <\b }^{\nn}
\Nek{\Ya\lambda\b }{\Ya\lambda\a }{Q_{\b ,\a }}qt 
&=&
\prod_{\a <\b }^{\nn}
\Nek{{\Yav\lambda\b }}{{\Yav\lambda \a }}{Q_{\a ,\b }}tq 
\cr
&\times&
\prod_{\a =1}^{\nn} 
\left(
v^{\nn-1}\prod_{\b =1}^{\a -1} Q_{\b ,\b +1}^\b  \prod_{\b =\a }^{\nn-1}Q_{\b ,\b +1}^{\nn-\b }
\right)^{-|\Ya\lambda \a |}
\hskip -10pt
\fla{\Ya\lambda \a }qt ^{-\nn+2\a -1},~~~
\ea
and thus Nekrasov's formula (\ref{eq:NekrasovZinstM}) 
is rewritten as
\be
\Zm{inst}{m}
=
\sum_{\Ya\lambda {1},\cdots,\Ya\lambda {\nn}}
{
\prod_{\a =1}^{\nn}
\Lambda_{\a ,m}{}^{|\Ya\lambda \a |}
\fla{\Ya\lambda \a }qt ^{\nn-m-2\a +1}
\over
\prod_{\a <\b }^{\nn}
\Nek{\Ya\lambda \a }{\Ya\lambda \b }{Q_{\a ,\b }}qt 
\Nek{{\Yav\lambda \b }}{{\Yav\lambda \a }}{Q_{\a ,\b }}tq 
\prod_{\a =1}^{\nn}
\Nek{\Ya\lambda \a }{\Ya\lambda \a }{1}qt 
},
\ee
where
\be
\Lambda_{\a ,m} 
:=
v^{-1}\Lambda^{2\nn} \left(-{\bf e}_\a \right)^{-m}
\prod_{\b =1}^{\a -1} Q_{\b ,\b +1}^\b  \prod_{\b =\a }^{\nn-1}Q_{\b ,\b +1}^{\nn-\b }.
\label{eq:Lambdai}
\ee
For example, for $SU(2)$ theory we have
\beqa
\Zm{inst}{m}
&=& 
\sum_{\lambda_1, \lambda_2}
\frac{
\left(v^{-1}\Lambda^4 Q_{H} \right)^{|\lambda|}
(-\Q_1)^{-m|\lambda_1|} (-\Q_2)^{-m|\lambda_2|} 
\fla{\lambda_1}qt^{ 1-m} 
\fla{\lambda_2}qt^{-1-m} }
{
\Nek{\Ya\lambda 1 }{\Ya\lambda 2 }{Q_{H}}qt 
\Nek{{\Yav\lambda 2 }}{{\Yav\lambda 1 }}{Q_{H}}tq 
\Nek{\Ya\lambda 1 }{\Ya\lambda 1}{1}qt
\Nek{\Ya\lambda 2 }{\Ya\lambda 2}{1}qt 
} 
\CR
&=& 
\sum_{\lambda_1, \lambda_2}
\frac{
\left(v^{-1}\Lambda^4 Q_{H} \right)^{|\lambda|}
(-\Q_1)^{-m|\lambda_1|} (-\Q_2)^{-m|\lambda_2|} 
\fla{\lambda_1}qt^{ 1-m} 
\fla{\lambda_2}qt^{-1-m} }
{\Nek{\Ya\lambda 1 }{\Ya\lambda 1}{1}qt
\Nek{\Ya\lambda 2 }{\Ya\lambda 2}{1}qt } 
\cr
& &\times 
{
\tPi\left(-v^{-1}Q_H t^\rho,\ q^\rho \right)
\tPi\left(-v     Q_H q^\rho,\ t^\rho \right)
\over
\tPi\left(-v^{-1}Q_H q^{\Ya \lambda 1}t^\rho,\ t^{\Yav\lambda 2}q^\rho \right)
\tPi\left(-v     Q_H t^{\Yav\lambda 2}q^\rho,\ q^{\Ya \lambda 1}t^\rho \right)
}~,
 \label{SU2}
\eeqa
where 
\be
\Nek{\lambda}{\lambda}{1}qt
=
\prod_{s \in \lambda} 
\left( 1-q^{a(s)} t^{\ell(s)+1}  \right)
\left( 1- q^{- a(s)-1} t^{-\ell(s)}\right),
\ee
and $Q_{H} = Q_{12}$. It is this form of Nekrasov's partition function that is
obtained from the refined topological vertex with formulas of the Macdonald 
functions.


\subsection{Symmetry as a character of $Spin(4)$ }

If Nekrasov's partition function gives the generating function of refined BPS state counting 
in the compactification of $M$ theory on local Calabi-Yau spaces, it has to be a character of 
$Spin(4) \simeq SU(2)_L \times SU(2)_R$, since the spin of massive BPS particle in
five dimensions is a representation of $Spin(4)$. 
In general, if a function $f(u, v)$ in two variables $(u,v)$ is invariant under 
both $u \to u^{-1}$ and $v \to v^{-1}$, it is a linear combination of $Spin(4)$ characters
\beq
f(t,q) = \sum_{(s_L, s_R)} a_{(s_L, s_R)} \chi_{s_L}(u) \chi_{s_R}(v)~, \label{Spin-decomp}
\eeq
where
\beq
\chi_n (z) := z^n + z^{n-2} + \cdots + z^{-n+2} + z^{-n} 
= \frac{z^{n+1} - z^{-n-1}}{z - z^{-1}}~,
\eeq
is the character of the irreducible representation of $SU(2)$ with spin $n/2$.
Hence, if the $k$-instanton part 
$Z^{(k)}(q,t)$ 
of the partition function is invariant 
under the transformations 
$r_L : (q,t) \to (q^{-1},t^{-1})$ and $r_R : (q,t) \to (t,q)$. 
$Z^{(k)}(q,t)$ is expanded as
\beq
Z^{(k)}(q,t) = \sum_{(s_L, s_R)} a_{(s_L, s_R)}^{(k)} \chi_{s_L}(u) \chi_{s_R}(v)~, \label{character} 
\eeq
with rational coefficients $a_{(s_L, s_R)}^{(k)}$. Recall that $u= \sqrt{qt}$ and $v=\sqrt{q/t}$. 
Actually we will find an appropriate scaling of $Z^{(k)}(q,t)$ depending on the instanton number $k$ is 
necessary for the genuine invariance under the above transformations. 
If the partition function takes the form \eqref{character},
then the $k$-instanton part $F^{(k)}(q,t)$ 
of the free energy is also a linear combination of
$Spin(4)$ characters. Furthermore, if the pole structure of the free energy is appropriate, we can factor out
the character of the half-hypermultiplet and subtract the multicovering contributions
to obtain the expansion of the total free energy in the Gopakumar-Vafa form:
\beqa
F &=& \log Z = \sum_{k=0}^\infty  F^{(k)} (Q_\beta;q,t) \CR
&=& \sum_{\beta \in H_2(X, {\mathbb Z})} 
\sum_{(j_L, j_R)} \sum_{n=1}^\infty
\frac{N_\beta^{(j_L,j_R)} u^n v^n}{n(u^n v^n -1)(u^n- v^n)} \chi_{n\cdot j_L}(u) \chi_{n\cdot j_R}(v)  
Q_\beta^n~. \label{GVrefined} 
\eeqa
The coefficients $N_\beta^{(j_L,j_R)}$ of the expansion \eqref{GVrefined}  are conjectured to be
nonnegative integers, since from the viewpoint of the Calabi-Yau compactification of $M$ theory
they are interpreted as multiplicities of the five-dimensional BPS particles arising from 
$M2$ branes wrapping on a two-cycle $\beta \in H_2(X, {\mathbb Z})$ in the Calabi-Yau 3-fold $X$.
We have checked the integrality of the refined BPS state counting from the $SU(2)$ and $SU(3)$ 
partition functions up to instanton number $2$. The result is presented in appendix {\appSpin}. 

Since the transformation 
$(q,t) \to (t^{-1},q^{-1})$ 
is compensated by the transpose of colored partitions, we have
\beq
\prod_{\alpha, \beta=1}^{\nn} 
\NY{\lambda_\alpha}{ \lambda_\beta}  {q^{-1}}{t^{-1}}{Q_{\alpha,\beta}} 
= \prod_{\alpha, \beta=1}^{\nn} 
\NY{\lambda_\alpha^\vee}{\lambda_\beta^\vee}tq{Q_{\alpha,\beta}} ~.
\eeq
By \eqref{inversion}, we also find that
\beq
\prod_{\alpha, \beta=1}^{\nn} 
\NY{\lambda_\alpha}{ \lambda_\beta} {t^{-1}}{q^{-1}}{Q_{\alpha,\beta}^{-1}}
= \left( \frac{q}{t} \right)^{\nn |\lambda|}  \prod_{\alpha, \beta=1}^{\nn}  
\NY{\lambda_\alpha}{ \lambda_\beta}tq{Q_{\alpha,\beta}} ~.
\eeq
Therefore, we obtain
\beqa
 \sum_{\{\lambda_\a\}, |\lambda| =k}
  \frac{1}{ \prod_{\alpha, \beta=1}^{\nn} \NY{\lambda_\alpha}{ \lambda_\beta} {q^{-1}}{t^{-1}}{Q_{\alpha,\beta}} }
 &=&
 \sum_{\{\lambda_\a\}, |\lambda| =k}
 \frac{1}{ \prod_{\alpha, \beta=1}^{\nn}  \NY{\lambda_\alpha}{ \lambda_\beta}tq{Q_{\alpha,\beta}} }~, \CR
\sum_{\{\lambda_\a\}, |\lambda| =k}
  \frac{1}{ \prod_{\alpha, \beta=1}^{\nn} \NY{\lambda_\alpha}{ \lambda_\beta}{t^{-1}}{q^{-1}}{Q_{\alpha,\beta}^{-1}} }
 &=&  \left( \frac{t}{q} \right)^{\nn |\lambda|} 
 \sum_{\{\lambda_\a\}, |\lambda| =k}
 \frac{1}{ \prod_{\alpha, \beta=1}^{\nn} \NY{\lambda_\alpha}{ \lambda_\beta} tq{Q_{\alpha,\beta}} }~. \CR
\eeqa
Thus if we can prove
\beq
 \sum_{\{\lambda_\a\}, |\lambda| =k}
  \frac{1}{ \prod_{\alpha, \beta=1}^{\nn} \NY{\lambda_\alpha}{ \lambda_\beta} tq{Q_{\alpha,\beta}^{-1}} }
 =
 \sum_{\{\lambda_\a\}, |\lambda| =k}
 \frac{1}{ \prod_{\alpha, \beta=1}^{\nn} \NY{\lambda_\alpha}{ \lambda_\beta}tq{Q_{\alpha,\beta}} }~, \label{Qinv}
\eeq
the partition function 
$Z^{\mathrm{inst}}_m$
is invariant under both of the reflections $r_L$ and $r_R$. 
It is easy to see that the property \eqref{Qinv} is valid for $\nn =2$, 
since the exchange of two partitions effectively induces ${\bf e}_\alpha \to {\bf e}_\alpha^{-1}$.
For $\nn > 2$ the validity of \eqref{Qinv} seems nontrivial. The overall reflection of the roots 
${\bf e}_\alpha \to {\bf e}_\alpha^{-1}$ cannot be induced by any permutation of colored partitions.
However, we have checked by explicit computations that \eqref{Qinv}  is true for $\nn = 3$ and $k=1,2$. 
This is consistent with the computation in appendix {\appSpin}, where we obtain the results of 
refined BPS state counting from Nekrasov's partition function of $SU(3)$ gauge theory.




The symmetry of Nekrasov's partition function with the Chern-Simons coupling 
can also be derived easily. 
Because of  (\ref{eq:nDual}), the factor $\Nek{\lambda}{\mu}Qqt $ 
enjoys the following duality relations:
\be
\Nek{\lambda}{\mu}{vQ}qt  
= \Nek{\mu}{\lambda}{v^{-1} Q}{q^{-1}}{t^{-1}}
= \Nek{\mu^\vee}{\lambda^\vee}{v^{-1} Q}tq .
\label{eq:NDual}
\ee
From the expression 
(\ref{eq:NekIm}),
we also have
\be
\Nek{\lambda}{\mu}{vQ}qt 
=
\Nek{\mu}{\lambda}{v Q^{-1}}qt 
Q^{|\lambda|+|\mu|} 
\fla\lambda qt 
/
\fla\mu qt .
\label{eq:NDualQ}
\ee
Since 
\be
\Nek{\lambda}{\mu}{v^2 Q}qt  
\Nek{\mu}{\lambda}{v^2 Q^{-1}}qt  
=
\Nek{\lambda}{\mu}Qqt  
\Nek{\mu}{\lambda}{Q^{-1}}qt  
v^{2(|\lambda|+|\mu|)},
\ee
we find that
\ba
\Nek{\lambda}{\mu}Qqt  
\Nek{\mu}{\lambda}{Q^{-1}}qt  
v^{|\lambda|+|\mu|}
&=&
\Nek{\lambda}{\mu}{Q^{-1}}{q^{-1}}{t^{-1}} 
\Nek{\mu}{\lambda}{Q     }{q^{-1}}{t^{-1}} 
v^{-|\lambda|-|\mu|}
\cr
&=&
\Nek{\lambda^\vee}{\mu^\vee}{Q^{-1}}tq  
\Nek{\mu^\vee}{\lambda^\vee }{Q     }tq  
v^{-|\lambda|-|\mu|}.
\ea
Thus, Nekrasov's partition function $\Zm{inst}{m}$ 
has the following symmetries:
\ba
\ZQ{inst}{m}qt{ \Q_1,\cdots,\Q_{\nn} }{\Lambda}
&=&
\ZQ{inst}{-m}{q^{-1}}{t^{-1}}{ \Q_1^{-1},\cdots,\Q_{\nn}^{-1} }{\Lambda}
\cr
&=&
\ZQ{inst}{-m}tq{ \Q_1^{-1},\cdots,\Q_{\nn}^{-1} }{\Lambda} .
\ea



\Section{Geometric Engineering and Toric Geometry}


In this section following \cite{KKV, KMV, IK-P2, HIV}, we review the toric geometry that is
necessary for geometric engineering. Geometric engineering tells how
to obtain ${\cal N}=2$ $SU(N_c)$ super Yang-Mills theory with $N_f$ fundamental matters
from type II(A)  string theory on local Calabi-Yau manifold $K_S$,
the canonical bundle of a 4-cycle $S$. The (toric) geometry of the 4-cycle $S$ 
can be described by the (dual) toric diagram.
The prescription of the geometric engineering implies that the toric diagram of $S$
has $N_c$ horizontal internal edges (\lq\lq color\rq\rq\ $D5$ branes) and 
$N_f$ horizontal external edges (\lq\lq flavor\rq\rq\ $D5$ branes).
For example, the vertical distance of \lq\lq color\rq\rq\ $D5$ branes represents 
vacuum expectation values of the Higgs fields or the mass of $W$ bosons.
The matter fermions are given by 
fundamental strings connecting a \lq\lq color\rq\rq\ $D5$ brane
and a \lq\lq flavor\rq\rq\ $D5$ brane.
The vertical distance of a \lq\lq color\rq\rq\ $D5$ brane  
and a \lq\lq flavor\rq\rq\ $D5$ brane  represents the mass of the corresponding matter fermion.

One of the properties of toric diagrams that arise from geometric engineering
is that each vertex has a unique horizontal edge. In the following
we will consider toric diagrams in which we specify the horizontal edges as distinguished. 
In the computation by the method of topological vertex, we cut the internal horizontal edges.
Then the contribution of each component is given by an amplitude of 
\lq\lq the vertex on a strip\rq\rq \cite{IK-P3}. By gluing these amplitudes
we obtain the partition function for the local toric Calabi-Yau manifold $K_S$. 


In the compactification of type IIA string theory on local Calabi-Yau manifold,
${\cal N}=2$ supersymmetric $SU(N)$ gauge theory is geometrically engineered by ALE fibration 
of $A_{N-1}$ type over the rational curve ${\bf P}^1$.
The fiber consists of a chain of $N-1$ rational curves whose intersection form is given by the minus of
the Cartan matrix of $A_{N\!-\!1}$. The holomorphic 2-cycles in the fiber are in one-to-one correspondence
with the positive roots of $A_{N-1}$.
The (dual) toric diagram takes the form of \lq\lq ladder\rq\rq\ diagram with
$N$ parallel horizontal edges. In the toric diagram the faces correspond to compact $4$-cycles (divisors).

\FigLadder

In the ladder diagram of  ALE fibration over ${\bf P}^1$,
we find $N-1$ divisors, all of which are ${\bf P}^1$
fibration over ${\bf P}^1$, namely the Hirzebruch surfaces. The degree of the Hirzebruch
surface can be determined by the (relative) slopes of the vertical edges of the face. 
We will denote the Hirzebruch surface of degree $n$ by ${\bf F}_n$. 
It is known that for each $N$ there are $N+1$ types of such geometry, which we label by $m=0,1, \cdots N$  \cite{IK-P2, HIV}.
The integer $m$ is related to the coupling constant of five-dimensional Chern-Simons coupling \cite{Tac}.
Let us call such geometry toric $SU(N)_m$ geometry.
We can characterize the toric $SU(N)_m$ geometry by saying that its compact $4$-cycles are
$\{ {\bf F}_{N-2+m}, {\bf F}_{N-4+m}, \cdots  {\bf F}_{-N+2+m} \}$.

\FigDivisor

The K\"ahler parameters of $SU(N)_m$ geometry are $T_B$ of the base space ${\bf P}^1$ 
and $T_{F_i}~(i=1, \cdots, N-1)$ of the fiber which is a chain of $(N-1)$ ${\bf P}^1$'s. 
In the subdiagram of Figure 2,  the rational curves of both side edges correspond to the 
fiber of ${\bf F}_{N-2k+2}$, and their K\"ahler parameters are $T_{F_k}$. On the 
other hand, if we denote the K\"ahler parameters of the upper and the lower edges by
$T_{B_k}$ and $T_{B_{k+1}}$, respectively. The difference is related to the degree of 
the Hirzebruch surface as follows:
\beq
T_{B_k} - T_{B_{k+1}} = (N-2k+m) T_{F_k}~. \label{recursion}
\eeq
From the recursion relation \eqref{recursion} we find, if $N+m = 2r+1$ is odd, that
\beqa
T_{B_r} &:=& T_B~, \CR
T_{B_i} &=& T_B + \sum_{j=i}^{r} (N+m-2j) T_{F_j}~, \quad (1 \leq i \leq r-1)~, \CR
T_{B_i} &=& T_B + \sum_{j=r+1}^{i-1} (2j -N -m) T_{F_j}~, \quad (r+1 \leq i \leq N)~, \label{odd}
\eeqa
and if $N+m = 2r$ is even,
\beqa
T_{B_r} &=& T_{B_{r+1}} := T_B~, \CR
T_{B_i} &=& T_B + \sum_{j=i}^{r-1} (N+m -2j) T_{F_{j}}~, \quad (1 \leq i \leq r-1)~, \CR
T_{B_i} &=& T_B + \sum_{j=r+1}^{i-1} (2j -N -m) T_{F_{j}}~, \quad (r+2 \leq i \leq N)~. \label{even}
\eeqa
 In \eqref{odd} and \eqref{even} we take the first relations as initial conditions
in solving \eqref{recursion}.

From the slope of each edge in Figure 1, we can also compute the framing index
by the rule to be explained in section 4.2. Let us denote the index of left, right, upper and lower edges by
$n_{L,k}, n_{R,k}, n_{B,k}$ and $n_{B,k+1}$, respectively.
Then we compute
\beqa
n_{B, k} &=& (m-k+1,1) \wedge (-N+k, 1) = N+m-2k+1, \CR
n_{L,k} &=& (-1, 0) \wedge (N-k-2,-1) = 1, \CR
n_{R,k} &=& (m-k,1) \wedge (-1, 0) = 1.
\eeqa
Note that $n_{L,k}$ and $n_{R,k}$ are independent of $k$. By definition
the framing index changes the sign, if we reverse the orientation of the edge,
or replace the representation associated to the edge by its transpose.
We will use these framing indices in the computation of the partition function
by gluing the refined topological vertices.


\Section{Refined Topological Vertex}


In \cite{AK},
we defined the refined topological vertex
which is written not by the Schur functions 
but by the Macdonald functions.
Here we slightly modify it by improving the framing factor.
%


\subsection{Refined topological vertex} 
\label{sec:RefinedTV}


Let $P_{\lambda/\mu} (x;q,t)$ and $\langle P_\lambda|P_\lambda\rangle_{q,t}$ 
be the Macdonald function in the infinite number of variables
$x=(x_1,x_2,\cdots)$
and its scalar product, respectively, 
defined in Appendix {\appMac}.
%
%
We introduce an involution $\iota$ acting on the power sum function
$p_n(x)$ by
$\iota(p_n) = -p_n$.
For example,
\be
\iota p_n(q^\lambda t^\rho) 
= 
- \sum_{i=1}^\infty (q^{n\lambda_i}-1)t^{n(\ha-i)}
- { 1\over t^{n\over 2} - t^{-{n\over 2}} }. 
\label{eq:involution}
\ee
Note that
$
\iota p_n(t^\rho) 
= 
-p_n(t^{\rho}) 
= 
p_n(t^{-\rho}). 
$


We define a vertex $V_{\mu \lambda}{}^\nu$ as follows:%
%
\footnote{
Although we will show that the Nekrasov formula is represented by our vertex $V_{\mu \lambda }{}^\nu$,
one can also produce it through the following vertex without the involution $\iota$
$$
U_{\mu\lambda}{}^\nu 
=
\tP_\lambda  (t^\rho;q,t)
\sum_\sigma\tP_{\mu/\sigma}(q^{-\lambda  } t^{-\rho};q^{-1},t) \ 
\tP_{\nu/\sigma}(q^{\lambda } t^{\rho};q^{-1},t)
\langle \tP_\sigma|\tP_\sigma\rangle_{q,t}.
$$
}
\be
V_{\mu \lambda}{}^\nu 
:= 
\tP_\lambda(t^\rho;q,t)
\sum_\sigma
\iota \tP_{\mu ^\vee /\sigma^\vee}(-t^{\lambda^\vee}q^{\rho};t,q) \ 
\tP_{\nu /\sigma}(q^{\lambda} t^{\rho};q,t)
v^{|\sigma|},
\ee
where 
\be
P_\lambda(t^{\rho};q,t)
=
\prod_{s\in\lambda} 
{
(-1) t^{\ha} q^{a(s)}
\over 
1-q^{a(s)} t^{\ell(s)+1} 
},
\qquad
P_{\lambda^\vee}(-q^{\rho};t,q)
=
\prod_{s\in\lambda} 
{
(-1)q^{-\ha} q^{-a(s)}
\over 
1-q^{-a(s)-1} t^{-\ell(s)} 
},
\label{eq:LargeNPrincipalSpecialization}
\ee
which follows by substituting $Q=0$ into (\ref{eq:Specialization}).
From (\ref{eq:ACMac}), 
$V_{\mu \lambda}{}^\nu $
is rewritten as
\be
V_{\mu \lambda}{}^\nu 
= 
\tP_\lambda(t^\rho;q,t)
\sum_\sigma
\iota \tP_{\mu/\sigma}(q^{-\lambda}t^{-\rho};q,t) \ 
\tP_{\nu /\sigma}(q^{\lambda} t^{\rho};q,t)
\langle P_\sigma|P_\sigma\rangle_{q,t}\ 
g_\mu(q,t),
\label{eq:RTV}
\ee
with
\be
g_\lambda(q,t)  
:=
{
v^{|\lambda|}
\over 
\langle P_\lambda|P_\lambda\rangle_{q,t}
}
=
\prod_{s\in\lambda}
\left({q\over t}\right)^\ha 
{
1-q^{a (s)  } t^{  \ell (s)+1}
\over 
1-q^{a (s)+1} t^{  \ell(s)   }
},
\ee
which satisfies
\be
g_\lambda(q,t) 
= g_\lambda(q^{-1},t^{-1})
= g_{\lambda^\vee}(t,q)^{-1}.
\ee
%
%
From \eqn{eq:RTV}, \eqn{eq:ACprincipalMac} and \eqn{eq:Wsymm}, 
one can show the symmetry%
%
\footnote{
If we replace Macdonald functions 
$P_{\lambda/\mu} (x;q,t)$'s
in 
$V_{\mu \lambda}{}^\nu $ 
by ``normalized'' Macdonald functions 
$
\widetilde P_{\lambda/\mu} (x;q,t) := 
P_{\lambda/\mu} (x;q,t)
\sqrt{g_\lambda(q,t)/g_\mu(q,t)}
$,
then the $g$ factors in 
(\ref{eq:SymmVi})--(\ref{eq:SymmViii}) 
disappear. Because
$
\widetilde P_{\lambda} (x; q,t) 
\widetilde P_{\lambda^\vee} (y; t,q) 
=
P_{\lambda} (x; q,t) 
P_{\lambda^\vee} (y; t,q) 
$,
all results in this article remain the same even if we use the normalized Macdonald functions  
$\widetilde P_{\lambda/\mu} (x;q,t)$.
}
%
\ba
g_\lambda(q,t)^{-1}  
V_{\lambda\bullet}{}^\bullet 
&=& V_{\bullet\lambda}{}^\bullet 
= V_{\bullet \bullet}{}^\lambda ,
\label{eq:SymmVi}
\\
g_\mu(q,t)^{-1}  
V_{\mu \bullet}{}^\nu 
&=&
g_\nu(q,t)^{-1}  
V_{\nu \bullet}{}^\mu ,
\label{eq:SymmVii}
\\
V_{\bullet \lambda}{}^\nu 
&=&
V_{\bullet \nu}{}^\lambda.
\label{eq:SymmViii}
\ea


Incorporating the framing factor, 
we define our refined topological vertices $\Ciio\mu \lambda \nu{q}{t}$ and $\Cooi\mu \lambda\nu{q}{t}$
as follows:
\ba
\Ciio\mu \lambda  \nu{q}{t}
&:=&
V_{\mu\lambda  }{}^\nu  v^{-|\nu|} \fla\nu qt ^{-1}
\\
&=&
\tP_\lambda  (t^\rho;q,t)
\sum_\sigma\iota \tP_{\mu^\vee /\sigma^\vee}(-t^{\lambda^\vee}
q^{\rho};t,q) \ 
\tP_{\nu /\sigma}(q^{\lambda } t^{\rho};q,t)
v^{|\sigma|-|\nu|}
\fla\nu qt ^{-1},
\cr
\Cooi\mu \lambda\nu{q}{t}
&:=&
\Ciio{\mu^\vee}{\lambda^\vee}{\nu^\vee}{t}{q}
(-1)^{|\lambda|+|\mu|+|\nu|}
\\
&=&
\tP_{\lambda^\vee}(-q^{\rho};t,q)
\sum_\sigma
\tP_{\nu^\vee/\sigma^\vee}(-t^{\lambda^\vee} q^{\rho};t,q) \
\iota \tP_{\mu  /\sigma}(q^{\lambda  } t^{\rho};q,t)
v^{-|\sigma|+|\nu|}
\fla\nu qt . \nonumber
\ea
The lower and the upper indices correspond to 
the incoming and the outgoing representations,
respectively,
and the edges of the topological vertex are ordered clockwise.
Although only the refined vertices of the above types are mainly used in this article,
the following vertices may also be useful:
\ba
\Coii\mu \lambda  \nu{q}{t}
&:=&
\Ciio\mu\lambda\nu{q}{t}
v^{|\mu|+|\nu|}
\fla\mu qt \fla\nu qt 
\\
&=&
\tP_\lambda  (t^\rho;q,t)
\sum_\sigma\iota \tP_{\mu^\vee /\sigma^\vee}(-t^{\lambda^\vee}
q^{\rho};t,q) \ 
\tP_{\nu /\sigma}(q^{\lambda  } t^{\rho};q,t)
v^{|\sigma|+|\mu|}
\fla\mu qt ,
\cr
\Cioo\mu\lambda \nu{q}{t}
&:=&
\Cooi\mu \lambda \nu{q}{t}
v^{-|\mu|-|\nu|}
\fla\mu qt ^{-1}\fla\nu qt ^{-1}
=
\Coii{\mu^\vee}{\lambda^\vee}{\nu^\vee}{t}{q}
(-1)^{|\lambda|+|\mu|+|\nu|}
\\
&=&
\tP_{\lambda^\vee}(-q^{\rho};t,q)
\sum_\sigma
\tP_{\nu^\vee/\sigma^\vee}(-t^{\lambda^\vee} q^{\rho};t,q) \
\iota \tP_{\mu/\sigma}(q^{\lambda  } t^{\rho};q,t)
v^{-|\sigma|-|\mu|}
\fla\mu  qt ^{-1}. \nonumber
\ea
\FigTV

Note that, when $q=t$,  
the topological vertex in \cite{AKMV} is 
\be
C_{\mu\lambda \nu} (q)
=
s_\lambda  (q^\rho)
\sum_\sigma s_{\mu/\sigma}(q^{\lambda^\vee + \rho}) \ s_{\nu^\vee
/\sigma}(q^{\lambda   + \rho})
\prod_{s\in\nu} q^{a(s)-\ell(s)}.
\ee
Since
$
s_{\mu/\sigma}(q^{\lambda^\vee + \rho }) =
\iota s_{\mu/\sigma}(q^{-\lambda  - \rho})
$, 
which follows from \eqn{eq:Maya},
our refined topological vertex 
$
\lim_{t\rightarrow q}\Ciio\mu \lambda \nu{q}{t}
$
coincides with the topological vertex 
$
C_{\mu\lambda \nu^\vee} (q)
$.
It is well-known that in the operator formalism the Schur functions are realized in terms of free fermions. 
Although we have no fermionic realization of the Macdonald functions, they are described by bosons
as shown in \cite{rf:AOS}. Our refined topological vertex has a bosonic realization 
by using that for the Macdonald functions. 


\subsection{Gluing rules} 


Here we show our gluing rules for constructing the partition function from a web diagram. 
Let us consider a graph with trivalent vertices and edges.
Each edge is associated with
an integer vector $\vv = (v_{1},v_{2})\in\bZ^2$.
Hence the trivalent vertex with edges indexed by $(i,j,k)$ in the counterclockwise ordering
is associated with a triplet of integer vectors $(\vv_i,\vv_j,\vv_k)$.
If we choose these vectors to be outgoing,  they should satisfy the following conditions
\be
\vv_i + \vv_j + \vv_k = \vo,
\qquad
\vv_i\wedge\vv_j = 1,
\qquad
( 
\vv_j\wedge\vv_k = 
\vv_k\wedge\vv_i =1), \label{normalize}
\ee
with
$\vv_i \wedge \vv_j := v_{i,1} v_{j,2} - v_{i,2} v_{j,1} $.
These correspond to the Calabi-Yau condition and the smoothness condition. 
Since the refined topological vertex has no cyclic symmetry, we should specify a preferred direction. 
Therefore one of these three vectors should be the preferred one and we denote it by white arrow.
Note that if we choose the middle edge as the preferred direction; $\vv_j = (-1, 0)$, then
the condition \eqref{normalize} implies that $\vv_i = (a,1), \vv_k=(b,-1)$ with $a+b=1$.

\FigFeynman

Let
$(\vv_i,\vv_j,\vv_k)$ and
$(\vv_k,\vv_{i'},\vv_{j'})$ be the vectors associated with 
the vertices at the origin and at the end of the vector $\vv_k$ of the $k$th edge, respectively. 
If we choose so that
$\vv_i$ and $\vv_j$ are incoming and 
$\vv_k$ and $\vv_{i'}$ are outgoing, 
then the framing index $n_k$ of the $k$th edge is defined by
\be
n_k := 
\vv_i\wedge\vv_{i'}= 
\vv_j\wedge\vv_{j'}.
\ee
Each edge is associated also with 
a Young diagram $\lambda$ and  
a K\"ahler parameter $Q\in\bC$
so that the propagator for the $k$th edge is defined as
\be
Q_k^{|\Ya\lambda k|} {\fla{\Ya\lambda k}qt}^{n_k},
\ee
and we glue the amplitudes 
by summing over the representation $\lambda$ on each edge.


\Section{Four-Point Functions} 


Here we show how to calculate the partition functions.
The building blocks for them are the following four-point functions.


\subsection{Building blocks} 


Assume that each vertex has a horizontal edge, which we take as the preferred direction.
Fix the orientation of the preferred direction, say $(-1,0)$;
then we have four possibilities of the configuration of two horizontal edges [Fig. 5].
Although the slopes and directions of ``vertical," i.e. nonhorizontal, edges can be arbitrary,
we show in Figure 5 the simplest one whose internal edge is orthogonal to the preferred direction and
we tentatively take the orientation of ``vertical" edges from the top to the bottom.
The framing index is $1$, $0$, $0$ and $-1$, respectively.
They are independent of the slope of ``vertical" edges, 
but change the sign according to the orientations.

\FigFourpoint

We order the three edges at each vertex in the clockwise direction such that 
the preferred direction is in the middle position.
This fixes the ordering of three edges uniquely.
The lower and upper indices of the refined vertex correspond to 
the incoming and the outgoing representation.
Then the following four-point functions are building blocks for the partition function: 
\ba
\Ziiio\mu{\Ya\lambda 1}{\Ya\lambda 2}\nu Qqt
&:=&
\sum_\eta 
\Ciio\mu{\Ya\lambda 1}\eta qt
\Ciio\eta{\Ya\lambda 2}\nu qt
Q^{|\eta|} \fla\eta qt ,
\cr
\Ziioo{\mu}{\Ya\lambda 1}{\Ya\lambda 2}{\nu}Qqt
&:=&  
\sum_{\eta}
\Ciio{\mu}{\Ya\lambda 1}{\eta}{q}{t}
\Cooi{\nu}{\Ya\lambda 2}{\eta}{q}{t}
Q^{|\eta|},
\cr
\Zioio{\mu}{\Ya\lambda 1}{\Ya\lambda 2}{\nu}Qqt
&:=&  
\sum_{\eta}
\Cooi{\eta}{\Ya\lambda 1}{\mu}{q}{t}
\Ciio{\eta}{\Ya\lambda 2}{\nu}{q}{t}
Q^{|\eta|},
\cr
\Ziooo\mu{\Ya\lambda 1}{\Ya\lambda 2}\nu Qqt
&:=&
\sum_\eta 
\Cooi\eta{\Ya\lambda 1}\mu qt
\Cooi\nu{\Ya\lambda 2}\eta qt
Q^{|\eta|} \fla\eta qt ^{-1}.
\label{eq:FourPointFunctions}
\ea
Note that
\ba
\Ziooo\mu{\Ya\lambda 1}{\Ya\lambda 2}\nu Qqt
&=&
\sum_{\eta^\vee} 
\Ciio{\eta^\vee}{\Yav\lambda 1}{\mu^\vee} tq
\Ciio{\nu^\vee}{\Yav\lambda 2}{\eta^\vee} tq
(-1)^{|\mu|+|\Ya\lambda 1|+|\Ya\lambda 2|+|\nu|}
Q^{|\eta|} \fla\eta tq 
\cr
&=&
\Ziiio{\nu^\vee}{\Yav\lambda 2}{\Yav\lambda 1}{\mu^\vee} Qtq 
(-1)^{|\mu|+|\Ya\lambda 1|+|\Ya\lambda 2|+|\nu|}.
\ea
We will show that
$\Zioio{\mu}{\Ya\lambda 1}{\Ya\lambda 2}{\nu}Qqt$
is related with
$\Ziioo{\mu}{\Ya\lambda 1}{\Ya\lambda 2}{\nu}Qqt$
by the flop.
%
%
If we take the orientation of ``vertical" edges from the bottom to the top,
the sign of the framing index changes.
The corresponding four-point functions are written by 
$\Coii\mu\lambda\nu qt$'s and 
$\Cioo\mu\lambda\nu qt$'s and
they are the same as 
those in (\ref{eq:FourPointFunctions}) up to 
the framing factors
$\fla\mu qt ^{\pm 1} \fla\nu qt ^{\pm 1}$
for the outer ``vertical" edges.


Although we have fixed a preferred direction in this article,
we can change it in some special cases.
Let
\be
\Zoiio\bullet\nu\bullet\mu Qqt
:=
\sum_\eta
\Ciio\bullet\eta\mu qt
\Cooi\bullet\eta\nu qt
Q^{|\eta|},
\ee
then from (\ref{eq:SymmViii}),
we have the following symmetry 
\be
\Ziioo\bullet\mu\nu\bullet Qqt
=
\Zoiio\bullet\nu\bullet\mu Qqt
v^{|\mu|-|\nu|}
\fla\mu qt
/
\fla\nu qt,
\label{eq:Slice}
\ee
which changes the preferred direction.
\FigSlice


\subsection{OPE formula} 


Next, we turn to showing some formulas for calculating the partition functions.
Let us denote a symmetric function $f$ in the set of variables
$(x^1_1,x^1_2,\cdots,x^2_1,x^2_2,\cdots,x^N_1,x^N_2,\cdots)$
by
$
f\left(x^1,x^2,\cdots,x^N\right)
$
or 
$
f\left(\{x^i\}_{i=1}^N\right) 
$.
To calculate the partition functions,
the essential part is the following Cauchy formula for the Macdonald function,
\be
\sum_\lambda
\tP_\lambda(x;q,t) \tP_{\lambda^\vee} (y;t,q)
=
\Pi_0(x,y),
\label{eq:conjugateCauchy}
\ee
or, more generally,
\be
\sum_\lambda
\tP_{\lambda/\mu} (x;q,t) \tP_{\lambda^\vee/\nu^\vee} (y;t,q)
=
\Pi_0(x,y)
\sum_\lambda
\tP_{\mu^\vee/\lambda^\vee} (y;t,q) \tP_{\nu/\lambda} (x;q,t),
\label{eq:skewCauchy}
\ee
with $\Pi_0(x,y)$ in (\ref{eq:Pizero}) and the adding formula
\be
\sum_\mu
\tP_{\lambda/\mu} (x;q,t) 
\tP_{\mu/\nu} (y;q,t) 
= 
\tP_{\lambda/\nu} (x,y;q,t).
\label{eq:addSkewMacdonald}
\ee
Note that for $c\in\bC$, 
$\tPi(cx,y) = \tPi(x,cy)$,
and for our involution $\iota$ in (\ref{eq:involution}),
$\tPi(\iota x,y) = \tPi(x,\iota y) = \tPi(x,y)^{-1}$.
Using these we have the following lemma.
\\
{\bf Lemma.}
Let $x$, $y$, $z$ and $w$ be sets of variables
and $\alpha$, $\beta$ and $\gamma\in\bC$. Then 
\ba
&&  \hskip -4pt
\sum_{\Ya\sigma 1,\eta,\Ya\sigma 2}
\tP_{{\mu^\vee}/{\Yav\sigma 1} } \left(x;t,q\right) 
\tP_{{\eta}/{\Ya\sigma 1}} \left(y;q,t\right) 
\tP_{{\eta^\vee}/{\Yav\sigma 2} } \left(z;t,q\right) 
\tP_{{\nu}/{\Ya\sigma 2}} \left(w;q,t\right) 
\alpha^{|\Ya\sigma 1|}
\beta^{|\eta|}
\gamma^{|\Ya\sigma 2|}
\cr
&=&
\sum_{\eta}
\tP_{{\mu^\vee}/{\eta^\vee} } \left(x,\alpha\beta z;t,q\right) 
\tP_{{\nu}/{\eta}} \left(\beta\gamma y,w;q,t\right) 
\left(\alpha\beta\gamma\right)^{|\eta|}
\tPi\left(y, \beta z \right).
\label{eq:FourPointFormula}
\ea
%
\proof
Let $\alpha=a/b$, $\beta=b/c$, $\gamma=c/d$; 
then, from 
(\ref{eq:scaleTrans}),
(\ref{eq:skewCauchy}) and 
(\ref{eq:addSkewMacdonald}),
the left-hand side of the above equation is 
\ba
&&  
\sum_{\Ya\sigma 1,\eta,\Ya\sigma 2}
\tP_{{\mu^\vee}/{\Yav\sigma 1} } \left({x\over a};t,q\right) 
\tP_{{\eta}/{\Ya\sigma 1}} \left(b y;q,t\right) 
\tP_{{\eta^\vee}/{\Yav\sigma 2} } \left({z\over c};t,q\right) 
\tP_{{\nu}/{\Ya\sigma 2}} \left(d w;q,t\right) 
a^{|\mu|} d^{-|\nu|}
\cr
&=&
\sum_{\Ya\sigma 1,\eta,\Ya\sigma 2}
\tP_{{\mu^\vee}/{\Yav\sigma 1} } \left({x\over a};t,q\right) 
\tP_{{\Yav\sigma 1}/{\eta^\vee} } \left({z\over c};t,q\right) 
\tP_{{\Ya\sigma 2}/{\eta}} \left(b y;q,t\right) 
\tP_{{\nu}/{\Ya\sigma 2}} \left(d w;q,t\right) 
a^{|\mu|} d^{-|\nu|}
\tPi\left(b y, {z\over c} \right)
\cr
&=&
\sum_{\eta}
\tP_{{\mu^\vee}/{\eta^\vee} } \left({x\over a},{z\over c};t,q\right) 
\tP_{{\nu}/{\eta}} \left(b y,d w;q,t\right) 
a^{|\mu|} d^{-|\nu|}
\tPi\left(b y, {z\over c} \right)
\cr
&=&
\sum_{\eta}
\tP_{{\mu^\vee}/{\eta^\vee} } \left(x,{a\over c}z;t,q\right) 
\tP_{{\nu}/{\eta}} \left({b\over d}y,w;q,t\right) 
\left({a\over d}\right)^{|\eta|}
\tPi\left(y, {b\over c}z \right),
\ea
and the lemma is proven.
\qed

%
%
Successively using this lemma, we obtain the following OPE formula,
which is useful for calculating more general diagrams.
\\
{\bf Proposition.}
Let $x^i$'s  be sets of variables,
$c_{i,i+1}\in\bC$ and
$c_{i,j}:=\prod_{k=i}^{j-1} c_{i,i+1}$. Then
\ba
\sum_{\{\Ya\lambda 1,\Ya\lambda 2,\cdots,\Ya\lambda{2N-1}\}}
\prod_{i=1}^N
\tP_{{\Yav\lambda{2i-2}}/{\Yav\lambda{2i-1}} } \left(x^{2i-1};t,q\right) 
\tP_{{\Ya \lambda{2i  }}/{\Ya \lambda{2i-1}} } \left(x^{2i  };q,t\right) 
\prod_{i=1}^{2N-1}
c_{i,i+1}^{|\Ya\lambda i|}
&&
\cr
=
\sum_{\eta}
\tP_{{\Yav\lambda{0}}/{\eta^\vee} } 
\left(\{c_{1,2i-1}x^{2i-1}\}_{i=1}^N;t,q\right) 
\tP_{{\Ya \lambda{2N  }}/{\eta} } 
\left(\{x^{2i} c_{2i,2N}\}_{i=1}^N;q,t\right)
c_{1,2N}^{|\eta|} 
&& 
\cr
\times
\prod_{1\leq i<j\leq N}\tPi\left( x_{2i}\ , \ c_{2i,2j-1} x_{2j-1}\right),
&&
\label{eq:OPEFormula}
\ea
for any integer $N\geq 2$. 

Therefore the number of Young diagrams to perform summation reduces from $2N-1$ to one.
If ${\Ya\lambda 0}$ or ${\Ya\lambda {2N}}$ is the trivial representation, 
then since
$\tP_{\bullet/\lambda}(x;q,t)=\delta_{\bullet,\lambda}$,
the number of Young diagrams for perform summation becomes zero.
The trace over ${\Ya\lambda 0}={\Ya\lambda {2N}}$ is also calculated 
by the trace formula explained in the next section.
If we realize the Macdonald polynomials by bosons as in \cite{rf:AOS},
these OPE formulas come from the operator product expansion of vertex operators.


\subsection{Computations of four-point functions}


Here we apply the OPE formula (\ref{eq:FourPointFormula}) to the above building blocks.
Let $x^\a $ and $y^\a $ be the set of variables as
$x^\a = q^{\Ya\lambda\a } t^\rho $ and 
$y^\a = t^{{\Yav\lambda\a }} q^{\rho}$, respectively.
Then
\ba
&&
\Ziiio\mu{\Ya\lambda 1}{\Ya\lambda 2}\nu Qqt
=
\tP_{\Ya\lambda 1} \left(t^{\rho};q,t\right)
\tP_{{\Ya\lambda 2} } \left(t^{\rho};q,t\right)
\fla\nu qt^{-1} 
v^{-|\nu|}
\cr
&&
\times\hskip-7pt
\sum_{\Ya\sigma 1,\eta,\Ya\sigma 2}
\hskip-4pt
\tP_{{\mu}^\vee /{\Yav\sigma 1} } \left(-\iota y^1 ;t,q\right) 
\tP_{\eta/\Ya\sigma 1} \left(x^1 ;q,t\right) 
\tP_{{\eta}^\vee/{\Yav\sigma 2} } \left(-\iota y^2;t,q\right) 
\tP_{\nu/\Ya\sigma 2} \left(x^2 ;q,t\right) 
v^{|\Ya\sigma 1|+|\Ya\sigma 2|}
\left(v^{-1} Q \right)^{|\eta|},~
\cr
&&
\Ziioo{\mu}{\Ya\lambda 1}{\Ya\lambda 2}{\nu}Qqt
=
\tP_{\Ya\lambda 1} \left(t^{\rho};q,t\right)
\tP_{{\Yav\lambda 2} } \left(-q^{\rho};t,q\right)
\cr
&&
\times\hskip-7pt
\sum_{\Ya\sigma 1,\eta,\Ya\sigma 2}
\hskip-4pt
\tP_{{\mu}^\vee /{\Yav\sigma 1} } \left(-\iota y^1 ;t,q\right) 
\tP_{\eta/\Ya\sigma 1} \left(x^1 ;q,t\right) 
\tP_{{\eta}^\vee/{\Yav\sigma 2} } \left(-y^2;t,q\right) 
\tP_{\nu/\Ya\sigma 2} \left(\iota x^2 ;q,t\right) 
v^{|\Ya\sigma 1|-|\Ya\sigma 2|}
Q^{|\eta|},
\cr
&&
\Zioio{\mu}{\Ya\lambda 1}{\Ya\lambda 2}{\nu}Qqt
=
\tP_{{\Yav\lambda 1} } \left(-q^{\rho};t,q\right)
\tP_{\Ya\lambda 2} \left(t^{\rho};q,t\right)
\fla\mu qt 
\fla\nu qt^{-1} 
v^{|\mu|-|\nu|}
\cr
&&
\times\hskip-7pt
\sum_{\Ya\sigma 1,\eta,\Ya\sigma 2}
\hskip-4pt
\tP_{{\mu}^\vee /{\Yav\sigma 1} } \left(-y^1 ;t,q\right) 
\tP_{\eta/\Ya\sigma 1} \left(\iota x^1 ;q,t\right) 
\tP_{{\eta}^\vee/{\Yav\sigma 2} } \left(-\iota y^2;t,q\right) 
\tP_{\nu/\Ya\sigma 2} \left(x^2 ;q,t\right) 
v^{-|\Ya\sigma 1|+|\Ya\sigma 2|}
Q^{|\eta|}.
\ea
%
From (\ref{eq:FourPointFormula}), they reduce to
\ba
\Ziiio\mu{\Ya\lambda 1}{\Ya\lambda 2}\nu Qqt
&=&  
\tP_{\Ya\lambda 1} \left(t^{\rho};q,t\right)
\tP_{{\Ya\lambda 2} } \left(t^{\rho};q,t\right)
\fla\nu qt^{-1} 
v^{-|\nu|}
\cr 
&& \hskip -65pt
\times
\sum_{\eta}
\iota \tP_{{\mu}^\vee /\eta^\vee } \left(-y^1,-Q y^2;t,q\right) 
\tP_{\nu /\eta} \left(Q x^1 ,x^2 ;q,t\right) 
(vQ) ^{|\eta|}
\tPi\left(-v^{-1}Q  x^1, y^2 \right)^{-1},
\cr
\Ziioo{\mu}{\Ya\lambda 1}{\Ya\lambda 2}{\nu}Qqt
&=&
\tP_{\Ya\lambda 1} \left(t^{\rho};q,t\right)
\tP_{{\Yav\lambda 2} } \left(-q^{\rho};t,q\right)
\cr
&& \hskip -65pt
\times
\sum_{\eta}
\tP_{{\mu}^\vee/{\eta}^\vee} 
\left(- \iota y^1 , -v Q y^2;t,q\right)
\tP_{\nu/\eta} \left(v^{-1} Qx^1, \iota x^2 ;q,t\right) 
Q^{|\eta|}
\tPi\left(-Q x^1,y^2\right),
\cr
\Zioio{\mu}{\Ya\lambda 1}{\Ya\lambda 2}{\nu}Qqt
&=&
\tP_{{\Yav\lambda 1} } \left(-q^{\rho};t,q\right)
\tP_{\Ya\lambda 2} \left(t^{\rho};q,t\right)
\fla\mu qt 
\fla\nu qt^{-1} 
v^{|\mu|-|\nu|}
\cr
&& \hskip -65pt
\times
\sum_{\eta}
\tP_{{\mu}^\vee/{\eta}^\vee} 
\left(-y^1 , -v^{-1} Q\iota y^2;t,q\right)
\tP_{\nu/\eta} \left(v Q\iota x^1,x^2 ;q,t\right) 
Q^{|\eta|}
\tPi\left(-Q x^1,y^2\right).
\ea
Since
$
\tPi\left(-Q x^1,y^2\right)
/
\tPi\left(-Q t^\rho,q^\rho\right)
=
\Nek{\Ya\lambda 1}{\Ya\lambda 2}{vQ}qt 
$,
the instanton part, 
such as \break
$
\Ziiio\mu{\Ya\lambda 1}{\Ya\lambda 2}\nu Qqt
/
\Ziiio\bullet\bullet\bullet\bullet Qqt
$
is written not by 
$\tPi\left(-v^{-1}Q x^1,y^2\right)$'s
but by
$\Nek{\Ya\lambda 1}{\Ya\lambda 2}{Q}qt $'s.


Note that
\be
\Ziioo{\mu}{\Ya\lambda 1}{\Ya\lambda 2}{\nu}Qqt
=
\Ziioo{\nu}{\Ya\lambda 2}{\Ya\lambda 1}{\mu}Q{q^{-1}}{t^{-1}}
=
\Ziioo{{\nu}^\vee}{{\Yav\lambda 2} }{{\Yav\lambda 1} }{{\mu}^\vee}Qtq
(-1)^{|\Ya\lambda 1|+|\Ya\lambda 2|+|\mu|+|\nu|}.
\ee


\subsection{Flop operation} 


The flop invariance of the topological vertex is shown in \cite{IK-P3, KM}.
We can show the flop invariance of the refined topological vertex as follows.
First,
\be
\Zioio{\mu}{\Ya\lambda 1}{\Ya\lambda 2}{\nu}Qqt
=
\Ziioo{\mu}{\Ya\lambda 2}{\Ya\lambda 1}{\nu}{Q^{-1}}qt
Q^{|\mu|+|\nu|}
{
\tPi\left(-Q x^1,y^2\right)
\over 
\tPi\left(-Q ^{-1}x^2,y^1\right)
}
{
\fla\mu qt
\over
\fla\nu qt
}.
\ee
%
Next, from
(\ref{eq:NDual}) and
(\ref{eq:NDualQ}), we have
\ba
{
\tPi\left(-Q x^1,y^2\right)
/
\tPi\left(-Q t^\rho,q^\rho\right)
\over 
\tPi\left(-Q^{-1}x^2, y^1\right)
/
\tPi\left(-Q^{-1} t^\rho, q^\rho\right)
}
&=&
{
\Nek{\Ya\lambda 1}{\Ya\lambda 2}{vQ}qt 
\over
\Nek{\Ya\lambda 2}{\Ya\lambda 1}{vQ^{-1}}qt 
}
\cr
= 
{
\Nek{\Ya\lambda 1}{\Ya\lambda 2}{vQ}qt 
\over 
\Nek{{\Yav\lambda 1}}{{\Yav\lambda 2} }{v^{-1} Q^{-1}}tq 
}
&=& 
Q^{|\Ya\lambda 1|+|\Ya\lambda 2|}
{\fla{\Ya\lambda 1}qt \over \fla{\Ya\lambda 2}qt }.~~
\ea
Thus we obtain the following flop invariance%
\footnote{
The flop invariance of $C_{\mu\nu\lambda}^{(IKV)}(t,q)$ 
has recently been discussed in \cite{Taki2}.
}
\ba
{
\Zioio{\mu}{\Ya\lambda 1}{\Ya\lambda 2}{\nu}Qqt
\over 
\Zioio{\mu'}{\bullet}{\bullet}{\nu'}Qqt
}
=
{
\Ziioo{\mu}{\Ya\lambda 2}{\Ya\lambda 1}{\nu}{Q^{-1}}qt
\over 
\Ziioo{\mu'}{\bullet}{\bullet}{\nu'}{Q^{-1}}qt
}
Q^{|\Ya\lambda 1|+|\Ya\lambda 2|+|\mu|+|\nu|}
{\fla\mu qt \over \fla\nu qt}
&&\hskip-4pt
{\fla{\Ya\lambda 1}qt \over \fla{\Ya\lambda 2}qt }
\cr
\times
Q^{-|\mu'|-|\nu'|}
{ \fla{\nu'} qt \over \fla{\mu'} qt}
&&\hskip-4pt .
\label{eq:Flop}
\ea
The denominator corresponds to the perturbative part.

\FigFlop
%


Combining (\ref{eq:Flop}) with (\ref{eq:Slice}), we have
\be
{
\Zioio{\bu}{\nu}{\mu}{\bu}Qqt
\over 
\Zioio{\bu}{\bullet}{\bullet}{\bu}Qqt
}
Q^{-|\mu|-|\nu|}
=
{
\Ziioo{\bu}{\mu}{\nu}{\bu}{Q^{-1}}qt
\over 
\Ziioo{\bu}{\bullet}{\bullet}{\bu}{Q^{-1}}qt
}
{\fla\nu qt \over \fla\mu qt}
=
{
\Zoiio{\bu}{\nu}\bu{\mu}{Q^{-1}}qt
\over 
\Zoiio{\bu}{\bullet}{\bullet}{\bu}{Q^{-1}}qt
}
v^{|\mu|-|\nu|},
\ee
which changes the preferred direction also.

\FigFlopSlice


\subsection{Finite $N$ Macdonald polynomial and homological invariants} 


When 
$\Ya\lambda 1$ or $\Ya\lambda 2$ is the trivial representation,
the amplitudes of the above diagrams are written by the Macdonald polynomials
with a finite number of variables.
%
%
Note that
$\tP_{\lambda} \left(a x^1, b\iota x^2 ;q,t\right)$
with 
$x^\a = q^{\Ya\lambda\a } t^\rho $
and
$a,b\in\bC$
is the Macdonald function in the power sum functions
\be
p_n\left(a x^1\right)+\iota p_n\left(b x^2\right)
=
\sum_{i=1}^\infty
\left\{ 
a^n \left(q^{n\Yai\lambda 1i} - 1\right)
-
b^n \left(q^{n\Yai\lambda 2i} - 1\right)
\right\}
t^{n(\ha-i)}
+ 
{ a^n - b^n \over t^{n\over 2} - t^{-{n\over 2}} }.
\ee
For $N\in\bN$ and $N \geq \ell(\lambda)$,
\be
p_n\left(q^\lambda t^\rho\right)+\iota p_n\left(t^{-N+\rho}\right)
=
\sum_{i=1}^\infty
\left(q^{n\lambda_i} - 1\right)
t^{n(\ha-i)}
+ 
{1 - t^{-n N} \over t^{n\over 2} - t^{-{n\over 2}} }
=
\sum_{i=1}^{N}
\left(
q^{\lambda_i} t^{\ha-i}
\right)^n,
\ee
which are the power sum symmetric polynomials in $N$ variables.
Therefore
$
\tP_{\lambda} 
\left(q^\lambda t^\rho , t^{-N-\rho} ; q,t \right)
$
is the Macdonald polynomial in $N$ variables
$\{q^{\lambda_i} t^{\ha-i}\}_{1\leq i\leq N}$.
%
%
On the other hand, from 
(\ref{eq:appsetFormulaI}) and 
(\ref{eq:Specialization})
\ba
\Nek{\lambda}{\bullet}{v Q}qt
&=&
\prod_{(i,j)\in\lambda}
\left( 1 - vQ q^{\lambda_i-j} t^{1-i} \right)
=
\prod_{(i,j)\in\lambda}
\left( 1 - vQ q^{j-1}  t^{1-i} \right)
\cr
&=&
{
\tP_{\lambda^\vee}(q^\rho,vQq^{-\rho};t,q)
\over 
\tP_{\lambda^\vee}(q^\rho;t,q)
}
=
{
\tP_\lambda(t^\rho,v^{-1}Q^{-1}t^{-\rho};q,t)
\over 
\tP_\lambda(t^\rho;q,t)
}
Q^{|\lambda|} 
{\fla\lambda qt},
\cr
\Nek{\bullet}{\lambda}{v Q}qt
&=&
\prod_{(i,j)\in\lambda}
\left( 1 - vQ q^{-\lambda_i+j-1} t^{i} \right)
=
\prod_{(i,j)\in\lambda}
\left( 1 - v^{-1} Q q^{1-j} t^{i-1} \right)
\cr
&=&
{
\tP_\lambda(t^\rho,v^{-1}Qt^{-\rho};q,t)
\over 
\tP_\lambda(t^\rho;q,t)
}
=
{
\tP_{\lambda^\vee}(q^\rho,vQ^{-1}q^{-\rho};t,q)
\over 
\tP_{\lambda^\vee}(q^\rho;t,q)
}
Q^{|\lambda|} 
{\fla\lambda qt}^{-1}.
\ea
Therefore, some factors in 
$
\Znoniioo{\mu}{\Ya\lambda 1}{\Ya\lambda 2}{\nu}
/
\Znoniioo{\bullet}{\bullet}{\bullet}{\bullet}
$
and 
$
\Znonioio{\mu}{\Ya\lambda 1}{\Ya\lambda 2}{\nu}
/
\Znonioio{\bullet}{\bullet}{\bullet}{\bullet}
$
might be written by the Macdonald polynomials 
in a finite number of variables.
%
%
For example, if $\mu$ and one of the 
$\Ya\lambda \alpha$ ($\alpha = 1$ or $2$) are the trivial representation,
\ba
{
\Ziioo{\bullet}{\lambda}{\bullet}{\nu}{Q^{-1}}qt
\over 
\Ziioo{\bullet}{\bullet}{\bullet}{\bullet}{Q^{-1}}qt
}
&=&
\tP_\nu \left(q^\lambda t^\rho, v Qt^{-\rho} ;q,t\right) 
\tP_{\lambda} \left(t^{\rho}, v^{-1}Qt^{-\rho};q,t\right)
v^{-|\nu|}
Q^{-|\lambda|-|\nu|}
\fla\lambda qt
,
\cr
{
\Zioio{\bullet}{\bullet}{\lambda}{\nu}Qqt
\over 
\Ziioo{\bullet}{\bullet}{\bullet}{\bullet}Qqt
}
&=&
\tP_\nu \left(q^\lambda t^\rho, v Qt^{-\rho} ;q,t\right) 
\tP_{\lambda} \left(t^{\rho}, v^{-1}Qt^{-\rho};q,t\right)
v^{-|\nu|} f_\nu(q,t)^{-1}.
\ea
When $v^{\pm1} Q=t^{-N}$ with $N\in\bN$,
they are written by the Macdonald polynomials in $N$ variables.
These are candidates for the $SU(N)$ homological invariants.

Note that 
\be
{\cal W}_{\lambda,\nu} (q,t)
:= 
\Ciio\bullet\lambda\nu qt 
v^{|\nu|} \fla\nu qt
=
\tP_\lambda\left(      t^\rho;q,t\right)
\tP_\nu \left(q^\lambda t^\rho;q,t\right),
\ee
has a nice symmetry \cite{Mac}(Ch.\ VI.6): 
\be
{\cal W}_{\lambda,\nu} (q,t)
= {\cal W}_{\nu,\lambda} (q,t).
\ee
When $t=q$, 
${\cal W}_{\lambda,\nu} (q,q)$
gives a large $N$ limit of the Hopf link invariants.


\Section{One-Loop Diagrams} 


Some one-loop diagrams which correspond to the 
trace of the vertex operators
can be calculated by the following trace formula.


\subsection{Trace formula} 


First, we have:
\\
{\bf Lemma.}
Let $x$ and $y$ be sets of variables and $a$, $b$ and  $c:=ab\in\bC$.
If $|c|<1$, then
\ba
\sum_{\lambda,\mu}
\tP_{\lambda^\vee/ \mu^\vee}\left(x;t,q\right) 
\tP_{\lambda/ \mu}\left(y;q,t\right) 
a^{|\lambda|}b^{|\mu|}
&=&
\prod_{k\geq 0}
{
\tPi\left(ac^k x,y\right)
\over
1-c^{k+1}
}
\cr
&=&
\exp\left\{-\sum_{n>0}{1\over n}{p_n(ax)p_n(-y)-c^n\over 1-c^n} \right\}.
\label{eq:TwoTraceFormula}
\ea
%
\proof
As in \cite{Mac}(Ch.\ I.5), 
let $F(x,y)$
denote the left-hand side of the above equation.
Then it follows from the Cauchy formula (\ref{eq:skewCauchy}) that
\ba
F(x,y)
&=&
\sum_{\lambda,\mu}
\tP_{\lambda^\vee/ \mu^\vee}\left(ax;t,q\right) 
\tP_{\lambda/ \mu}\left(y;q,t\right) 
(ab)^{|\mu|}
\cr
&=&
\sum_{\lambda,\mu}
\tP_{\mu^\vee/\lambda^\vee}\left(ax;t,q\right) 
\tP_{\mu/\lambda}\left(y;q,t\right) 
(ab)^{|\mu|}\tPi\left(ax,y\right)
\cr
&=&
\sum_{\lambda,\mu}
\tP_{\mu^\vee/\lambda^\vee}\left(abx;t,q\right) 
\tP_{\mu/\lambda}\left(y;q,t\right) 
a^{|\mu|}b^{|\lambda|}
\tPi\left(ax,y\right).
\ea
Therefore
\be
F(x,y)=
F(cx,y)
\tPi\left(ax,y\right)
=
F(0,y)
\prod_{k\geq 0} \tPi\left(ac^k x,y\right),
\qquad |c|<1.
\ee
But
\be
F(0,y)
=
\sum_\lambda
\tP_{\lambda/\lambda} (y;q,t) c^{|\lambda|}
=
\sum_{\lambda}
c^{|\lambda|}
=
\prod_{n>0}
(1-c^n)^{-1},
\qquad |c|<1,
\ee
and the lemma is proven.
\qed

From the above lemma and 
(\ref{eq:OPEFormula}) 
we obtain the following trace formula.
\\
{\bf Proposition.}
For $N\in\bN$, 
let $x^i=x^{2N+i}$'s  be sets of variables,
$\Ya\lambda 0 = \Ya\lambda{2N}$,
$c_{i,i+1}=c_{2N+i,2N+i+1}\in\bC$, 
$c_{i,j}:=\prod_{k=i}^{j-1} c_{i,i+1}$ and
$c:=c_{1,2N+1}=\prod_{i=1}^{2N}c_{i,i+1}$.
If $|c|<1$, then
\ba
&&\hskip -30pt
\sum_{\{\Ya\lambda 1,\Ya\lambda 2,\cdots,\Ya\lambda{2N}\}}
\prod_{i=1}^N
\tP_{{\Yav\lambda{2i-2}}/{\Yav\lambda{2i-1}} } \left(x^{2i-1};t,q\right) 
\tP_{{\Ya \lambda{2i  }}/{\Ya \lambda{2i-1}} } \left(x^{2i  };q,t\right) 
\cdot
\prod_{i=1}^{2N}
c_{i,i+1}^{|\Ya\lambda i|}
\cr
&=&
\prod_{k\geq 0} {1\over 1-c^{k+1}}
\prod_{i=1}^N \prod_{j=i+1}^{i+N}
\tPi\left( x^{2i}, \ c_{2i,2j-1} c^k  x^{2j-1}\right)
\cr
&=&
\exp\left\{-\sum_{n>0}{1\over n}
{1\over 1-c^n} 
\left\{
\sum_{i=1}^N \sum_{j=i+1}^{i+N}
c_{2i,2j-1}^n p_n\left(x^{2i}\right) p_n\left(-x^{2j-1}\right)
-c^n
\right\}\right\}.
\label{eq:TraceFormula}
\ea
\proof
From
(\ref{eq:OPEFormula}) and 
(\ref{eq:TwoTraceFormula}),
the left-hand side of the above equation is 
\ba
&&
\sum_{\lambda,\eta}
\tP_{{\lambda^\vee}/{\eta^\vee} } 
\left(\{c_{1,2i-1}x^{2i-1}\}_{i=1}^N;t,q\right)
\tP_{{\lambda}/{\eta} } 
\left(\{x^{2i}c_{2i,2N}\}_{i=1}^N;q,t\right) 
c_{1,2N}^{|\eta|} c_{2N,2N+1}^{|\lambda|}
\cr
&&\hskip 6truecm
\times
\prod_{1\leq i<j\leq N} \tPi\left( x^{2i}, \ c_{2i,2j-1} x^{2j-1}\right)
\cr
&=&
\prod_{1\leq i<j\leq N} \tPi\left( x^{2i}, \ c_{2i,2j-1} x^{2j-1}\right)
\cdot
\prod_{k\geq 0}
{
\tPi\left(   
\{x^{2i}c_{2i,2N+1}\}_{i=1}^N  ,
\{c_{1,2i-1}c^k x^{2i-1}\}_{i=1}^N 
\right)
\over 
1-c^{k+1}
},
\ea
here
$
\tPi\left(\{x^i\}_{i=1}^N ,\{y^j\}_{j=1}^M \right)
=
\prod_{i=1}^N \prod_{j=1}^M
\tPi(x^i,y^j)
$.
Then the left-hand side of (\ref{eq:TraceFormula}) reduces to 
\ba
&&\hskip-10pt
\prod_{1\leq i<j\leq N} 
\tPi\left( x^{2i}, \ c_{2i,2j-1} x^{2j-1}\right)
\cdot
\prod_{k\geq 0} 
{1\over 1-c^{k+1}}
\prod_{i,j=1}^N 
\tPi\left(x^{2i},\ c_{2i,2N+2j-1} c^k x^{2N+2j-1}\right)
\cr
&=&
\prod_{k\geq 0}
{1\over 1-c^{k+1}}
\!\!\prod_{1\leq i<j\leq N} 
\tPi\left( x^{2i}, \ c_{2i,2j-1} c^k x^{2j-1} \right)
\!\prod_{1\leq j\leq i\leq N} 
\tPi\left( x^{2i}, \ c_{2i,2N+2j-1} c^k x^{2j-1}\right), \cr
& &
\ea
which equals to the second line of (\ref{eq:TraceFormula}).
\qed

From this trace formula, we can calculate one-loop diagram 
if the loop does not contain preferred directions and 
also the framing factors cancel out.


\subsection{Examples for $N=2$ and $4$} 



For an example of the trace formula for $N=2$, let
\ba
Z_2 
&:=&
\sum_{\mu,\nu}
\Coii\nu\bullet\lambda qt
\Cioo\nu\bullet\lambda qt
\Lambda^{|\lambda|}
Q^{|\nu|}
\cr
&=&
\sum_{\mu,\nu, \sigma_1, \sigma_2}
\tP_{\nu^\vee/\Yav\sigma 1}\left(-\iota q^\rho;t,q\right)
\tP_{\lambda     /\Ya \sigma 1}\left(       t^\rho;t,q\right)
\tP_{\lambda^\vee/\Yav\sigma 2}\left(-      q^\rho;t,q\right)
\tP_{\nu     /\Ya \sigma 2}\left( \iota t^\rho;t,q\right)
v^{|\Ya\sigma 1|-|\Ya\sigma 2|}
\Lambda^{|\lambda|}
Q^{|\nu|}.
\cr &&
\ea
Then from (\ref{eq:TraceFormula}) with
$(c_{1,2},c_{2,3},c_{3,4},c_{4,5})=(v,\Lambda,v^{-1},Q)$ and
$(x^1,x^2,x^3,x^4)=(-\iota q^\rho,t^\rho,-q^\rho,\iota t^\rho)$,
it follows that $c=Q\Lambda$ and 
\be
\begin{pmatrix}
  c_{2,3}  &  c_{2,5}  \cr
\ c_{4,5}\ &\ c_{4,7}\ 
\end{pmatrix}
=
\begin{pmatrix}
  \Lambda &   c/v       \cr
\ Q     \ &\ cv\ 
\end{pmatrix},
\ee
and thus
\be
Z_2
=
\prod_{k\geq 0} 
{
\tPi\left(t^\rho,-\Lambda c^k q^\rho\right)
\tPi\left(t^\rho,-Q      c^k q^\rho\right)
\over
\tPi\left(t^\rho,-v c^{k+1}q^\rho\right)
\tPi\left(t^\rho,-v^{-1}c^{k+1}q^\rho\right)
}
{1\over 1-c^{k+1}}.
\ee
From (\ref{eq:powersum}), we obtain
\be
Z_2
=
\exp\left\{
-\sum_{n>0} {1\over n} {1\over 1-c^n}
\left\{
{
(\Lambda^n + Q^n) - (v^n + v^{-n})c^n  
\over 
(t^{n\over 2} - t^{-{n\over 2}})
(q^{n\over 2} - q^{-{n\over 2}})
}
-c^n
\right\}\right\}.
\ee
If we separate out the part
$
Z_2^{\rm pert}
:=
Z_2(\Lambda = 0)
=
\exp\left\{
-\sum_{n>0}
Q^n
/
(n
(t^{n\over 2} - t^{-{n\over 2}})
(q^{n\over 2} - q^{-{n\over 2}})
)
\right\},
$
then 
$
Z_2^{\rm inst}
:=
Z_2/Z_2^{\rm pert}
$
is
\be
Z_2^{\rm inst}
=
\exp\left\{
-\sum_{n>0} {1\over n} {\Lambda^n\over 1-c^n}
{
(Q^n - u^n)(Q^n - u^{-n})
\over 
(t^{n\over 2} - t^{-{n\over 2}})
(q^{n\over 2} - q^{-{n\over 2}})
}
\right\}.
\label{eq:LoopChiy}
\ee
As we will see in section 7.2, this gives the equivariant $\chi_y$ genus of 
the Hilbert scheme of points on ${\mathbb C}^2$. 

\FigTrace


For an example for $N=4$, let
\ba
Z_4 
&:=&
\sum_{\{\Ya\mu\alpha\}}
\prod_{\alpha =1}^4
\Ciio{\Ya\mu\alpha}\bullet{\Ya\mu{\alpha+1}} qt
\fla{\Ya\mu{\alpha+1}}qt 
\cr
&=&
\sum_{\{\Ya\mu\alpha\}}
\prod_{\alpha =1}^4
\tP_{\Yav\mu\alpha/\Yav\sigma\alpha}(-\iota q^\rho)
\tP_{\Yav\mu{\alpha+1}/\Yav\sigma\alpha}(t^\rho)
v^{|\Ya\sigma\alpha|-|\Ya\mu\alpha|}Q^{|\Ya\mu\alpha|},
\ea
with $\Ya\mu 5 = \Ya\mu 1$.
Then from (\ref{eq:TraceFormula}) with
$(c_{2\a -1,2\a }, c_{2\a ,2\a +1})=(v, v^{-1} Q_\a )$ and
$(x^{2\a },x^{2\a -1})=(t^\rho,-\iota q^\rho)$,
it follows that
\be
Z_4=
\prod_{k\geq 0} {1\over 1-c^{k+1}}
\prod_{i=1}^4 \prod_{j=i+1}^{i+4}
\tPi\left( t^\rho, \ -c^k c_{2i,2j-1} q^\rho\right)^{-1},
\ee
where $c=Q_1 Q_2 Q_3 Q_4$ and
\be
\begin{pmatrix}
  c_{2,3}  &  c_{2, 5}  &  c_{2, 7}  &  c_{2, 9} \cr
  c_{4,5}  &  c_{4, 7}  &  c_{4, 9}  &  c_{4,11} \cr
  c_{6,7}  &  c_{6, 9}  &  c_{6,11}  &  c_{6,13} \cr
\ c_{8,9}\ &\ c_{8,11}\ &\ c_{8,13}\ &\ c_{8,15}\ 
\end{pmatrix}
=
v^{-1}
\begin{pmatrix}
 \ Q_1\ &\ Q_1 Q_2\ &\ Q_1 Q_2 Q_3\ & c\  \cr
   Q_2  &  Q_2 Q_3  &  Q_2 Q_3 Q_4  & c\cr
   Q_3  &  Q_3 Q_4  &  Q_3 Q_4 Q_1  & c\cr
   Q_4  &  Q_4 Q_1  &  Q_4 Q_1 Q_2  & c
\end{pmatrix}.
\ee
Thus
\be
Z_4 
=
\exp\left\{
\sum_{n>0} {1\over n} {1\over 1-c^n}
\left\{
{
v^{-n}
\sum_{\a =1}^4 
\left(
Q_{\a }^n + 
Q_{\a }^n Q_{\a +1}^n +
Q_{\a }^n Q_{\a +1}^n Q_{\a +2}^n +
c^n
\right)
\over 
(t^{n\over 2} - t^{-{n\over 2}})
(q^{n\over 2} - q^{-{n\over 2}})
}
-c^n
\right\}\right\},
\ee 
where $Q_{i+4} = Q_i$.


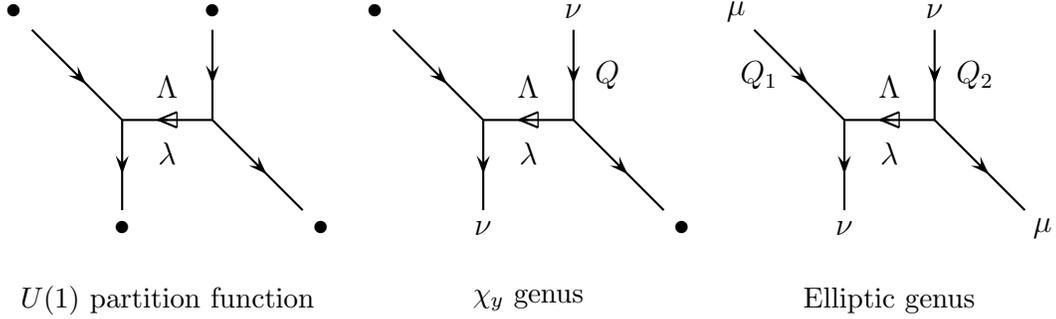
\Section{$U(1)$ Partition Function, $\chi_y$ Genus and Elliptic Genus}


Nekrasov's $U(1)$ partition function, 
the $\chi_y$ genus and the elliptic genus 
are realized  by our refined topological vertex,
as shown in \cite{AK}.
Since the diagrams for $U(1)$ theory have trivial framing, 
the vertex in \cite{AK} and the improved vertex in the present paper 
give the same answer. 


\subsection{$U(1)$ partition function}


First, the $U(1)$ partition function is written as follows.
Let
\be
Z 
:= 
\sum_\lambda\Lambda^{|\lambda|}
\Ciio\bullet\lambda\bullet{q}{t}
\Cooi\bullet\lambda\bullet{q}{t}.
\ee
Then
\ba
Z
&=&
\sum_\lambda\Lambda^{|\lambda|}
\tP_\lambda(t^\rho ;q,t) \ \tP_{\lambda^\vee}(-q^{\rho} ;t,q)
\cr
&=&
\sum_\lambda
\prod_{s\in\lambda}
v^{-1}\Lambda
{1\over 
(1-q^{ a(s)  } t^{ \ell(s)+1})
(1-q^{-a(s)-1} t^{-\ell(s)  })},
\ea
from
(\ref{eq:LargeNPrincipalSpecialization}).
This agrees with the $U(1)$ Nekrasov's formula
$\ZQ{inst}{0}qt{{\bf e}_1}{\Lambda^{1\over 2}}$
in (\ref{eq:NekrasovZinstM}).
%
Using the Cauchy-formula (\ref{eq:conjugateCauchy})
we have
\ba
Z
&=&
\Exp{-\sum_{n>0}{1\over n}
{\Lambda^n\over (t^{n\over 2} - t^{-{n\over 2}}) (q^{n\over 2} - q^{-{n\over 2}}) }
}
\cr
&=&
\Exp{\mp\sum_{n>0}{1\over n}\sum_{i,j}\left(\Lambda\,t^{\ha -i}q^{\pm(\ha -j)}\right)^n},
\qquad   |q^{\mp1}|, |t^{-1}| < 1
\cr
&=& 
\prod_{i,j\geq 1}
(1-\Lambda\, t^{\ha -i} q^{\pm(\ha -j)})^{\pm1},
\qquad   |q^{\mp1}|, |t^{-1}| < 1.
\ea

\FigUi


\subsection{$\chi_y$ genus}


Next, the $\chi_y$ genus is realized as follows.
Let
\be
\tZ :=
\sum_{\lambda,\nu } 
Q^{|\nu |} \Lambda^{|\lambda|} 
\Ciio\bullet\lambda\nu{q}{t} 
\Cooi\bullet\lambda\nu{q}{t}.
\ee
Then
\be
\tZ
=
\sum_{\lambda, \nu } 
Q^{|\nu |}
\Lambda^{|\lambda|} 
\tP_\lambda\left(t^{\rho};q,t\right)
\tP_{\lambda^\vee} \left(-q^{\rho};t,q\right)
\tP_\nu \left(q^{\lambda}t^{\rho};q,t\right)
\tP_{\nu^\vee} \left(-t^{\lambda^\vee}q^{\rho};t,q\right).
\ee
From 
(\ref{eq:LargeNPrincipalSpecialization}) and
(\ref{eq:conjugateCauchy}) we have
\be
\tZ
=
\sum_{\lambda} 
\tPi(-Q q^{\lambda}t^{\rho},\, t^{\lambda^\vee}q^{\rho})
\prod_{s\in\lambda}
v^{-1}\Lambda
{1\over 
(1-q^{ a(s)  } t^{ \ell(s)+1})
(1-q^{-a(s)-1} t^{-\ell(s)  })}.
\ee
If we separate out the part 
$
\tZm{pert}{}
:=
\sum_{\nu } 
Q^{|\nu |} 
\Ciio\bullet\bullet\nu{q}{t} 
\Cooi\bullet\bullet\nu{q}{t}
=
\tPi(-Q t^{\rho},\, q^{\rho})
$,
which is independent of $\Lambda$, 
then
$
\Zm{inst}{}
:= \tZ/\tZm{pert}{}
$
is from (\ref{eq:NekIIp})
\ba
\Zm{inst}{}
&=&
\sum_{\lambda} 
\left(v^{-1}\Lambda\right)^{|\lambda|}
{
\Nek{\lambda}{\lambda}{vQ}qt 
\over
\Nek{\lambda}{\lambda}1qt 
}
\cr
&=&
\sum_{\lambda} 
\prod_{s\in\lambda}
v^{-1}\Lambda
{
1-vQ q^{ a(s)  } t^{ \ell(s)+1}\over 
1-  q^{ a(s)  } t^{ \ell(s)+1}
}
{
1-vQ q^{-a(s)-1} t^{-\ell(s)  }\over 
1-         q^{-a(s)-1} t^{-\ell(s)  }
}.
\label{eq:chiy}
\ea
This agrees with the $\chi_y$ genus (20) of \cite{rf:LiLiuZhou}
with $vQ = y$, $v^{-1}\Lambda = Q^{\rm LLZ}$ 
and $(q,t)=(1/t_1,t_2)$ or $(1/t_2,t_1)$.
%


If our refined topological vertex had cyclic symmetry, 
then this $\chi_y$ genus $\Zm{inst}{}$ would agree with $Z_2^{\rm inst}$ in section 6.2,
and hence the following identity should hold
\ba
&&\hskip-60pt
\sum_{\lambda} 
\Lambda^{|\lambda|}
\prod_{s\in\lambda}
{
1-Q q^{ a(s)  } t^{ \ell(s)+1}\over 
1-  q^{ a(s)  } t^{ \ell(s)+1}
}
{
1-Q q^{-a(s)-1} t^{-\ell(s)  }\over 
1-         q^{-a(s)-1} t^{-\ell(s)  }
}
\cr
&=&
\exp\left\{
\sum_{n>0} {1\over n} {\Lambda^n\over 1-\Lambda^n Q^n}
{
(1-t^n Q^n)(1-q^{-n}Q^n)
\over 
(1-t^n)(1-q^{-n})
}
\right\}.
\label{eq:conjecture}
\ea
From (\ref{eq:Specialization}),
this 
is close to the Cauchy formula for the Macdonald functions in power sums 
$p_n = {(1-t^n Q^n)/(1-t^n)}$ and
$(-\Lambda)^n{(1-q^{-n}Q^n)/(1-q^{-n})}$, i.e.
\ba
&&\hskip-60pt
\sum_{\lambda} 
(-\Lambda)^{|\lambda|}
\prod_{s\in\lambda}
{
1-Q q^{ a'(s)  } t^{1-\ell'(s)}\over 
1-  q^{ a(s)  } t^{ \ell(s)+1}
}
{
1-Q q^{a'(s)-1} t^{-\ell'(s)}\over 
1-         q^{-a(s)-1} t^{-\ell(s)  }
}
q^{-a'(s)} t^{\ell'(s)}
\cr
&=&
\exp\left\{
-\sum_{n>0} {1\over n} {\Lambda^n}
{
(1-t^n Q^n)(1-q^{-n}Q^n)
\over 
(1-t^n)(1-q^{-n})
}
\right\}.
\ea
Although we have no proof for (\ref{eq:conjecture}), 
computer calculations support that
$\Zm{inst}{}=Z_2^{\rm inst}$,
which strongly suggests a kind of symmetry of web diagrams.
See also the discussions in the recent papers \cite{IKS, PS}. 



\subsection{Elliptic genus}


Finally, the elliptic genus is written as follows.
Let
\be
\tZ :=
\sum_{\lambda,\mu,\nu} 
Q_1^{|\mu |}\Lambda^{|\lambda|} Q_2^{|\nu |} 
\Ciio\mu\lambda\nu{q}{t}
\Cooi\mu\lambda\nu{q}{t}.
\ee
Then
\ba
\tZ
&=&
\sum_{\lambda,\mu,\nu} 
\tP_\lambda(t^\rho;q,t)
\sum_\sigma
\iota \tP_{\mu^\vee /\sigma^\vee}( -t^{\lambda^\vee} q^{\rho};t,q) \ 
   \tP_{\nu /\sigma}(q^{\lambda} t^{\rho};q,t)
Q_1^{|\mu |} \Lambda^{|\lambda|} Q_2^{|\nu |} 
\cr
&&\hskip8pt
\times
\tP_{\lambda^\vee}(-q^{\rho};t,q)
\sum_\eta 
\iota \tP_{\mu /\eta}(q^{\lambda} t^{\rho};q,t) \
   \tP_{\nu^\vee /\eta^\vee}(-t^{\lambda^\vee} q^{\rho};t,q) 
v^{ |\sigma|-|\eta| }.
\ea
From 
(\ref{eq:LargeNPrincipalSpecialization}) and 
the trace formula (\ref{eq:TraceFormula}) with
$(c_{1,2},c_{2,3},c_{3,4},c_{4,5})=(v,Q_2,v^{-1},Q_1)$ and
$(x^1,x^2,x^3,x^4)=
(-\iota t^{\lambda^\vee}q^\rho, q^\lambda t^\rho, -t^{\lambda^\vee}q^\rho, \iota q^\lambda t^\rho)$,
it follows that 
\ba
\tZ
&=&
\sum_{\lambda} 
\prod_{s\in\lambda}
v^{-1}\Lambda
{1\over 
(1-q^{ a(s)  } t^{ \ell(s)+1})
(1-q^{-a(s)-1} t^{-\ell(s)  })}
\cr
&&\hskip8pt\times
\prod_{k\geq 0}
{
\tPi(-Q_1c^k q^{\lambda}t^{\rho},\, t^{\lambda^\vee}q^{\rho})\
\tPi(-Q_2 c^k q^{\lambda}t^{\rho},\, t^{\lambda^\vee}q^{\rho})
\over 
\tPi(-v^{-1} c^{k+1} q^{\lambda}t^{\rho},\, t^{\lambda^\vee}q^{\rho})
\tPi(-v c^{k+1} q^{\lambda}t^{\rho},\, t^{\lambda^\vee} q^{\rho})
}
{1\over 1-c^{k+1} },
\ea
with $c=Q_1 Q_2$ and $|c|<1$.
If we factor out the $\Lambda$-independent part
\ba
\tZm{pert}{}
&:=&
\sum_{\mu,\nu} 
Q_1^{|\mu |} Q_2^{|\nu |} 
\Ciio\mu\bullet\nu{q}{t}
\Cooi\mu\bullet\nu{q}{t}
\cr
&=&
\prod_{k\geq 0}
{
\tPi(-Q_1c^k t^{\rho},\, q^{\rho})\
\tPi(-Q_2 c^k t^{\rho},\, q^{\rho})
\over 
\tPi(-v^{-1} c^{k+1} t^{\rho},\, q^{\rho})
\tPi(-v c^{k+1} t^{\rho},\,  q^{\rho})
}
{1\over 1-c^{k+1} },
\ea
then 
$
\Zm{inst}{}
:= \tZ/\tZm{pert}{}
$
is from (\ref{eq:NekIIp}),
\ba
\Zm{inst}{}
&=&
\sum_{\lambda} 
\left(v^{-1}\Lambda\right)^{|\lambda|}
\prod_{k\geq 0}
{
\Nek{\lambda}{\lambda}{ v Q_1c^k }qt 
\Nek{\lambda}{\lambda}{ v Q_2c^k }qt 
\over
\Nek{\lambda}{\lambda}{        c^{k  } }qt 
\Nek{\lambda}{\lambda}{ v^2 c^{k+1} }qt 
}
\cr
&=&
\sum_{\lambda} 
\prod_{k\geq 0}
\prod_{s\in\lambda}
v^{-1}\Lambda
{
\left(1-    vQ_1^{k+1} Q_2^{k    }q^{ a(s)  } t^{ \ell(s)+1}\right)
\left(1-    vQ_1^{k    } Q_2^{k+1}q^{ a(s)  } t^{
\ell(s)+1}\right)\over 
\left(1-      Q_1^{k    } Q_2^{k    }q^{ a(s)  } t^{
\ell(s)+1}\right)
\left(1-v^2 Q_1^{k+1} Q_2^{k+1}q^{ a(s)  } t^{ \ell(s)+1}\right)
}
\cr
&&\hskip15mm\times
{
\left(1-   vQ_1^{k+1} Q_2^{k    }q^{-a(s)-1} t^{-\ell(s)}\right)
\left(1-   vQ_1^{k    } Q_2^{k+1}q^{-a(s)-1} t^{-\ell(s)}\right)
\over 
\left(1-      Q_1^{k    } Q_2^{k    }q^{-a(s)-1}
t^{-\ell(s)}\right)
\left(1-v^2 Q_1^{k+1} Q_2^{k+1}q^{-a(s)-1} t^{-\ell(s)}\right)
}.
\ea
This agrees with the elliptic genus (24) of \cite{rf:LiLiuZhou}
with $Q_1 Q_2 = p$, $v Q_1 = y$, $v^{-1}\Lambda = y^{-1} Q^{\rm LLZ}$
and $(q,t)=(t_1,1/t_2)$ or $(t_2,1/t_1)$.


\Section{$SU(\nn)$ Partition Function}


Nekrasov's $SU(\nn)$ partition function
is also realized  by our refined topological vertex,
as mentioned in \cite{AK}.


\subsection{Pure $SU(2)$ partition function}


The pure $SU(2)$ partition function without Chern-Simons couplings
is written as follows.
Let
\ba
\ZlaQ{ \Ya\lambda 1,\Ya\lambda 2 }{ \Q_1,\Q_2 }qt 
&:=&
\sum_\mu
\Ciio\bullet{\Ya\lambda 1}\mu{q}{t}
\Ciio\mu{\Ya\lambda 2}\bullet{q}{t}
Q_{1,2}^{|\mu |} \fla\mu qt 
\cr
&=&
\sum_\mu
\tP_{\Ya\lambda 1} \left(t^{\rho};q,t\right)
\tP_\mu \left(q^{\Ya\lambda 1} t^\rho ;q,t\right) 
\iota \tP_{\mu^\vee} \left(-t^{\lambda_2^\vee} q^{\rho };t,q\right) 
\tP_{\Ya\lambda 2} \left(t^{\rho};q,t\right)
\left(v^{-1} Q_{1,2}\right)^{|\mu |}
\cr
&=&
\tPi\left(-v^{-1} Q_{1,2}\, q^{\Ya\lambda 1} t^{\rho},\ t^{{\Yav\lambda 2}}
q^{\rho}\right)^{-1}
\tP_{\Ya\lambda 1} \left(t^{\rho};q,t\right)
\tP_{\Ya\lambda 2} \left(t^{\rho};q,t\right),
\ea
from (\ref{eq:conjugateCauchy}),
where $Q_{\alpha,\beta} := \QQ\alpha\beta$.
The dual part is
\ba
\ZlaQ{ {\Yav\lambda 2},{\Yav\lambda 1} }{ \Q_2^{-1},\Q_1^{-1} }tq 
&=&
\sum_\nu
\Ciio\bullet{{\Yav\lambda 2}}{\nu^\vee}{t}{q}
\Ciio{\nu^\vee}{{\Yav\lambda 1}}\bullet{t}{q}
Q_{1,2}^{|\nu |} \fla{\nu^\vee}tq 
\cr
&=&
\sum_\nu
\Cooi\bullet{{\Ya\lambda 2}}{\nu}{q}{t}
\Cooi{\nu}{{\Ya\lambda 1}}\bullet{q}{t}
Q_{1,2}^{|\nu |} \fla{\nu}qt ^{-1}
(-1)^{|\Ya\lambda 1|+|\Ya\lambda 2|}.
\ea
Then, from
(\ref{eq:LargeNPrincipalSpecialization}) and  
(\ref{eq:conjugateCauchy}),
it follows that
\ba
\tZ 
&:=&
\sum_{\Ya\lambda 1,\Ya\lambda 2}
\ZlaQ{  \Ya\lambda 1,\Ya\lambda 2 }{ \Q_1,\Q_2 }qt  
\ZlaQ{ {\Yav\lambda 2},{\Yav\lambda 1} }{ \Q_2^{-1},\Q_1^{-1} }tq 
(\Lambda Q_{1,2})^{|\Ya\lambda 1|+|\Ya\lambda 2|}
\fla{\Ya\lambda 1}qt  / \fla{\Ya\lambda 2}qt 
\cr
&=&
\sum_{\Ya\lambda 1,\Ya\lambda 2}
\tPi\left(-v^{-1} Q_{1,2}\, q^{\Ya\lambda 1} t^{\rho},\ t^{{\Yav\lambda 2}}
q^{\rho}\right)^{-1}
\tPi\left(-v^{-1} Q_{2,1}\, t^{{\Yav\lambda 1}} q^{\rho},\ q^{\Ya\lambda 2}
t^{\rho}\right)^{-1}
\cr
&\times&
\prod_{s\in\Ya\lambda 2 }
v^{-1}\Lambda
{1\over \left(1-q^{a(s)}t^{\ell(s)+1}\right)\left(1-q^{-a(s)-1}t^{-\ell(s)}\right)}
\cr
&\times&
\prod_{s\in\Ya\lambda 1 }
v^{-1}\Lambda
{1\over \left(1-q^{a(s)}t^{\ell(s)+1}\right)\left(1-q^{-a(s)-1}t^{-\ell(s)}\right)}.
\label{eq:suiiZ}
\ea
If we factor out the $\Lambda$-independent part
$
\tZm{pert}{}
:=
\ZlaQ{\bullet,\bullet}{\Q_1,\Q_2}qt 
\ZlaQ{\bullet,\bullet}{\Q_2^{-1},\Q_1^{-1}}tq 
$,
then
$
\Zm{inst}{}
:= \tZ/\tZm{pert}{}
$ 
agrees with the $SU(2)$ Nekrasov's formula 
$\ZQ{inst}{0}qt{{\bf e}_1,{\bf e}_2}{\Lambda^{1\over 4}}$
in (\ref{eq:NekrasovZinstM}).

\FigSUii


\subsection{Pure $SU(\nn)$ partition function}


The pure $SU(\nn)$ partition function with Chern-Simons terms is written as follows.
Let
\ba
&&
\hskip-20pt
\ZlaQ{\Ya\lambda {1},\cdots,\Ya\lambda {\nn}}{\Q_1,\cdots,\Q_{\nn}}qt 
\cr
&:=&  
\sum_{ \{\Ya\mu \a \} }
\prod_{\a =1}^{\nn}
\Ciio{\Ya\mu {\a -1}}{\Ya\lambda \a }{\Ya\mu \a }{q}{t}
\prod_{\a =1}^{\nn-1} 
Q_{\a ,\a +1}^{|\Ya\mu \a  |} \fla{\Ya\mu \a }qt 
\cr
&=& 
\sum_{ \{\Ya\mu \a \} }
\prod_{\a =1}^{\nn}\sum_{\Ya\sigma  \a }
\iota \tP_{{\Yav\mu {\a -1}}/{\Yav\sigma  \a }} \left(-t^{{\Yav\lambda \a }}
q^{\rho} ;t,q\right) 
\tP_{\Ya\lambda \a } \left(t^{\rho};q,t\right)
\tP_{\Ya\mu \a /\Ya\sigma  \a } \left(q^{\Ya\lambda \a } t^\rho ;q,t\right) 
\prod_{\a =1}^{\nn-1}
v^{|\Ya\sigma  \a |-|\Ya\mu \a |}
Q_{\a ,\a +1}^{|\Ya\mu \a  |},
\cr
&&
\ea
with 
$Q_{\a ,\b } =  \QQ\a\b  $
and
$\Ya\mu 0 = \Ya\mu {\nn} = 0$.
Note that $\Ya\sigma  1=\Ya\sigma  {\nn}=0$.
From the OPE formula (\ref{eq:OPEFormula}), we have
\be
\ZlaQ{\Ya\lambda {1},\cdots,\Ya\lambda {\nn}}{\Q_1,\cdots,\Q_{\nn}}qt 
=
\prod_{\a <\b }
\tPi\left(-v^{-1} Q_{\a ,\b }\, q^{\Ya\lambda \a } t^{\rho},\ t^{{\Yav\lambda \b }}
q^{\rho}\right)^{-1}
\prod_{\a =1}^{\nn}
\tP_{\Ya\lambda \a } \left(t^{\rho};q,t\right).
\ee
%

The dual part is
\be
\ZlaQ{{\Yav\lambda \nn},\cdots,{\Yav\lambda 1}}{\Q_\nn^{-1},\cdots,\Q_1^{-1}}tq
=
\sum_{ \{\Ya\mu \a \} }
\prod_{\a =1}^{\nn}
\Cooi{\Ya\mu \a }{\Ya\lambda \a }{\Ya\mu {\a -1}}{q}{t}
\prod_{\a =1}^{\nn-1} 
Q_{\a ,\a +1}^{|\Ya\mu \a  |} \fla{\Ya\mu \a }qt ^{-1}
(-1)^{|\Ya\lambda \a |},
\ee
with 
$Q_{\alpha,\beta} := \QQ\alpha\beta$.
Then, using $\Lambda_{\a ,m}$ in
(\ref{eq:Lambdai}),
\ba
\tZm{}{m} 
&:=& 
\sum_{\Ya\lambda {1},\cdots,\Ya\lambda {\nn}}
\ZlaQ{\Ya\lambda 1,\cdots,\Ya\lambda \nn}{\Q_1,\cdots,\Q_{\nn}}qt 
\ZlaQ{{\Yav\lambda \nn},\cdots,{\Yav\lambda 1}}{\Q_\nn^{-1},\cdots,\Q_1^{-1}}tq 
\prod_{\a =1}^{\nn} 
\Lambda_{\a ,m}{}^{|\Ya\lambda \a |}
\fla{\Ya\lambda \a }qt ^{\nn-m-2\a +1}
\cr
&=&
\sum_{\Ya\lambda {1},\cdots,\Ya\lambda {\nn}}
\prod_{\a <\b }
\tPi\left(-v^{-1} Q_{\a ,\b }\, q^{\Ya\lambda \a } t^{\rho},\ t^{{\Yav\lambda \b }}
q^{\rho}\right)^{-1}
\tPi\left(-v^{-1} Q_{\b ,\a }\, t^{{\Yav\lambda \a }} q^{\rho},\ q^{\Ya\lambda \b }
t^{\rho}\right)^{-1}
\cr
&& \quad\times\quad
\prod_{\a =1}^\nn 
\fla{\Ya\lambda \a }qt ^{-m}
\prod_{s\in\Ya\lambda \a }
{v^{-1}\Lambda^{2\nn}\left(-Q_\a \right)^{-m}
\over 
\left(1-q^{a(s)}t^{\ell(s)+1}\right)\left(1-q^{-a(s)-1}t^{-\ell(s)}\right)},
\label{eq:suNZ}
\ea
with $\Ya\mu 0 = \Ya\mu {\nn} = \Ya\nu 0 = \Ya\nu \nn = 0$.
If we factor out the $\Lambda$-independent part
$
\tZm{pert}{}
:=
\ZlaQ{\bullet,\cdots,\bullet}{\Q_1,\cdots,\Q_{\nn}}qt 
\ZlaQ{\bullet,\cdots,\bullet}{\Q_\nn^{-1},\cdots,\Q_1^{-1}}tq 
$,
then
$
\Zm{inst}{m}
:= { \tZm{}{m}/\tZm{pert}{} }
$ 
agrees with the $SU(\nn)$ Nekrasov's formula
$\ZQ{inst}{m}qt{{\bf e}_1,\cdots,{\bf e}_{\nn}}{\Lambda}$
in (\ref{eq:NekrasovZinstM}).


\FigSUN


\Section{$SU(\nn)$ with $N_f = 2 \nn $}


The partition functions with fundamental matters are also realized by 
the refined topological vertex as follows. 
Let
\ba
&& \hskip -20pt
\ZlaQ{\Ya\lambda {1},\cdots,\Ya\lambda {2\nn-1}}{\Q_1,\cdots,\Q_{2\nn-1}}qt 
\cr 
&:=&  
\sum_{ \{\Ya\mu \a \} }
\prod_{\a =1}^{\nn}
\Ciio{\Ya\mu {2\a -2}}{\Ya\lambda {2\a -1}}{\Ya\mu {2\a -1}}{q}{t}
\Cooi{\Ya\mu {2\a   }}{\Ya\lambda {2\a   }}{\Ya\mu {2\a -1}}{q}{t}
\prod_{\a =1}^{2\nn-1} 
Q_{\a ,\a +1}^{|\Ya\mu \a  |}
\cr
&=& 
\sum_{ \{\Ya\mu \a \} }
\prod_{\a =1}^{\nn}
\sum_{\Ya\sigma  {2\a -1}}
\iota \tP_{{\Yav\mu {2\a -2}}/{\Yav\sigma  {2\a -1}}}
\left(-t^{{\Yav\lambda {2\a -1}}} q^{\rho} ;t,q\right) 
\tP_{\Ya\lambda {2\a -1}} \left(t^{\rho};q,t\right)
\tP_{\Ya\mu {2\a -1}/\Ya\sigma  {2\a -1}} \left(q^{\Ya\lambda {2\a -1}} t^\rho ;q,t\right) 
\cr
&&\times
\sum_{\Ya\sigma  {2\a }}
\tP_{{\Yav\mu {2\a -1}}/{\Yav\sigma  {2\a }}}
\left(-t^{{\Yav\lambda {2\a }}} q^{\rho} ;t,q\right) 
\tP_{{\Yav\lambda {2\a   }}} \left(-q^{\rho};t,q\right)
\iota \tP_{\Ya\mu {2\a }/\Ya\sigma  {2\a }} \left(q^{\Ya\lambda {2\a }} t^\rho ;q,t\right) 
\cr
&&\times
\prod_{\a =1}^{\nn}
v^{|\Ya\sigma  {2\a -1}|-|\Ya\sigma  {2\a }|}
\prod_{\a =1}^{2\nn-1} 
Q_{\a ,\a +1}^{|\Ya\mu \a  |},
\ea
with 
$\Ya\mu 0 = \Ya\mu {2\nn} = \Ya\sigma  0 = \Ya\sigma  {2\nn} = 0$.
As in the pure $SU({\nn})$ case,
from 
(\ref{eq:OPEFormula}) we have
\ba
&& \hskip -20pt
\ZlaQ{\Ya\lambda {1},\cdots,\Ya\lambda {2\nn-1}}{\Q_1,\cdots,\Q_{2\nn-1}}qt 
\cr
&=&
\prod_{\a <\b }
\tPi\left(-
v^{(-1)^\a  + (-1)^\b \over 2}
Q_{\a ,\b }\, q^{\Ya\lambda \a } t^{\rho},\ t^{{\Yav\lambda \b }}q^{\rho}\right)^{(-1)^{\a +\b +1}}
\prod_{\a =1}^{\nn}
\tP_{ \Ya\lambda {2\a -1}} \left(t^{\rho};q,t\right)
\tP_{{\Yav\lambda {2\a   }}} \left(-q^{\rho};t,q\right)
\cr
&=&
\prod_{\a <\b }
{
\tPi\left(-Q_{2\a ,2\b -1}\, q^{\Ya\lambda {2\a }} t^{\rho},\
t^{{\Yav\lambda {2\b -1}}} q^{\rho}\right)
\tPi\left(-Q_{2\a -1,2\b }\, q^{\Ya\lambda {2\a -1}} t^{\rho},\
t^{{\Yav\lambda {2\b }}} q^{\rho}\right)
\over 
\tPi\left(-vQ_{2\a ,2\b }\, q^{\Ya\lambda {2\a }} t^{\rho},\
t^{{\Yav\lambda {2\b }}} q^{\rho}\right)
\tPi\left(-v^{-1} Q_{2\a -1,2\b -1}\, q^{\Ya\lambda {2\a -1}} t^{\rho},\
t^{{\Yav\lambda {2\b -1}}}q^{\rho}\right)
}
\cr
&&\times
\prod_{\a =1}^{\nn}
\tP_{ \Ya\lambda {2\a -1}} \left(t^{\rho};q,t\right)
\tP_{{\Yav\lambda {2\a   }}} \left(-q^{\rho};t,q\right).
\ea
%
The dual part is
\be
\ZlaQ{{\Yav\lambda {2\nn-1}},\cdots,{\Yav\lambda 1}}{{\Q'}_{2\nn-1}^{-1},\cdots,{\Q'}_1^{-1}}tq 
=
\sum_{ \{\Ya\nu \a \} }
\prod_{\a =1}^{\nn}
\Cooi{\Ya\nu {2\a -1}}{\Ya\lambda {2\a -1}}{\Ya\nu {2\a -2}}{q}{t}
\Ciio{\Ya\nu {2\a -1}}{\Ya\lambda {2\a   }}{\Ya\nu {2\a    }}{q}{t}
\prod_{\a =1}^{2\nn-1} 
{Q'}_{\a ,\a +1}^{|\Ya\nu \a  |}
(-1)^{|\Ya\lambda {2\a -1}|},
\ee
with $Q'_{\a ,\b } = \QpQp\a\b $ and $\Q'_{2\a -1} = \Q_{2\a -1}$.


When 
$\Ya\lambda {2\a }$ for even integers $2\a $ is a trivial representation, 
let 
\ba
\tZ
&:=&
\sum_{\Ya\lambda 1,\Ya\lambda 3,\cdots,\Ya\lambda {2\nn-1}}
\ZlaQ{\Ya\lambda 1,\bullet,\Ya\lambda 3,\cdots,\bullet,\Ya\lambda {2\nn-1}}{\Q_1,\cdots,\Q_{2\nn-1}}qt 
\ZlaQ{{\Yav\lambda {2\nn-1}},\bullet,\cdots,{\Yav\lambda 3},\bullet,{\Yav\lambda 1}}{{\Q'}_{2\nn-1}^{-1},\cdots,{\Q'}_1^{-1}}tq 
\prod_{\a =1}^{\nn} 
\Lambda_\a ^{|\Ya\lambda {2\a -1}|} 
\fla{\Ya\lambda {2\a -1}}qt ^{-1},
\cr
\Lambda_\a 
&:=&
v^{-1}\Lambda^{2\nn}
\prod_{\b =1}^{\a -1} { \Q _{2\b -1}\over \Q'_{2\b   } }
\prod_{\b =\a }^{\nn} { \Q'_{2\b   }\over \Q _{2\b -1} }.
\ea
In addition, let
$Z^{\rm inst} := \tZ/\tZ^{\rm pert}$ 
with 
$
\tZ^{\rm pert} := 
\ZlaQ{\bullet,\cdots,\bullet}{\Q_1,\cdots,\Q_{2\nn-1}}qt 
\ZlaQ{\bullet,\cdots,\bullet}{{\Q'}_{2\nn-1}^{-1},\cdots,{\Q'}_1^{-1}}tq 
$.
Then
\ba
Z^{\rm inst} 
&=&
\sum_{ \{ \Ya\lambda {2\a -1} \} }
{
\prod_{\a =1}^{\nn} 
\Lambda_\a ^{|\Ya\lambda {2\a -1}|}
\fla{\Ya\lambda {2\a -1}}qt ^{-1}
\over
\prod_{\a <\b } 
\left(
\Nek{\Ya\lambda \a }{\Ya\lambda \b }{Q_{\a ,\b }}qt 
\Nek{{\Yav\lambda \b }}{{\Yav\lambda \a }}{Q'_{\a ,\b }}tq 
\right)^{(-1)^{\a +\b }}
\prod_{\a =1}^{\nn} 
\Nek{\Ya\lambda {2\a -1}}{\Ya\lambda {2\a -1}}1qt 
}
\cr
&=&
\sum_{ \{ \Ya\lambda {2\a -1} \} }
{
\prod_{\a =1}^{\nn} 
\Lambda^{2\nn |\Ya\lambda {2\a -1}|}
\over
\prod_{\a <\b } 
\left(
\Nek{\Ya\lambda \a }{\Ya\lambda \b }{Q_{\a ,\b }}qt 
\Nek{{\Ya\lambda \b }}{{\Ya\lambda \a }}{Q'_{\b ,\a }}qt 
\right)^{(-1)^{\a +\b }}
\prod_{\a =1}^{\nn} 
\Nek{\Ya\lambda {2\a -1}}{\Ya\lambda {2\a -1}}1qt 
}, \cr
&&
\ea
gives the $SU(\nn )$ partition function with $N_f=2\nn $.
 
\FigSUNiiN


\section*{Acknowledgments}

We would like thank T. Eguchi, A. Iqbal, Y. Konishi, H. Konno, S. Minabe,
S. Moriyama, H. Nakajima, N. Nekrasov, H. Ochiai, N. Reshetikhin, J. Shiraishi, 
M. Taki, K. Yoshioka and C. Vafa
for discussions and helpful correspondence.
In particular, we are grateful to M. Taki for sharing his result \cite{Taki2}
before he submitted the paper to arXiv. 
Part of the results in this paper was presented at the following workshops: 
``Infinite analysis 
2005'' (27--30 September, 2005) at
Tambara Institute of Mathematical Sciences, University of Tokyo;
``Strings 2006'' (19--24 June, 2006) at 
Beijin friendship hotel; and
``Progress of String Theory and Quantum Field Theory''
(7-10 December, 2007) at Osaka City University.
We 
would like to thank the organizers for the invitation to the workshops and for the hospitality.
The work of H.K.  is supported in part by a Grant-in-Aid for Scientific Research
[\#19654007] from the Japan Ministry of Education, Culture, Sports, Science and Technology.


\Section*{Appendix A : Proof of the Proposition in Sect. \ref{sec:PartitionFunctionCS} }
\renewcommand{\theequation}{A.\arabic{equation}}\setcounter{equation}{0}
\renewcommand{\thesubsection}{A.\arabic{subsection}}\setcounter{subsection}{0}


\subsection{Combinatorial identities} 



We have the following formula for the Young diagrams,
which translates the summation in squares into that in lows:
\\
{\bf Lemma.}
For all integers 
$\NNi\lambda\geq\ell(\lambda)$ and
$\NNi\mu\geq\ell(\mu)$,
\be
(1-q)\sum_{(i,j)\in\lambda} q^{j-1} t^{-i+1} 
=
\sum_{i=1}^{\NNi\lambda} \left(1-q^{\lambda_i}\right) t^{-i+1},
\label{eq:partitionFormulaI}
\ee
\be
(1-q)\sum_{(i,j)\in\mu } q^{\lambda_i-j} t^{\mu^\vee_j-i} 
=
\left( 
\sum_{i=1}^{\NNi\mu}
\sum_{j=i}^{\NNi\mu}
-t^{-1}
\sum_{i=1}^{\NNi\mu}
\sum_{j=i+1}^{\NNi\mu + 1}
\right)
q^{\lambda_i-\mu_j} t^{j-i}.
\label{eq:partitionFormulaII}
\ee

\proof
(\ref{eq:partitionFormulaI}) follows from
$\sum_{j=1}^{\lambda} q^{j-1} = {(1-q^{\lambda})/(1-q)}$.

The left-hand side of (\ref{eq:partitionFormulaII}) reduces to 
\ba
(1-q)
\sum_{i=1}^{\ell(\mu)} 
\sum_{k=0}^{\ell(\mu)-i}
t^k q^{\lambda_i-\mu_{i+k}} 
\sum_{\ell=0}^{\mu_{i+k}-\mu_{i+k+1}-1} q^\ell 
&=&
\sum_{i=1}^{\ell(\mu)} 
\sum_{k=0}^{\ell(\mu)-i}
t^k \left(q^{\lambda_i-\mu_{i+k}} - q^{\lambda_i-\mu_{i+k+1}} \right)
\cr
&=&
\sum_{i=1}^{\NNi\mu} 
\sum_{j=i}^{\NNi\mu}
t^{j-i} \left(q^{\lambda_i-\mu_j} - q^{\lambda_i-\mu_{j+1}} \right),
\ea
which equals the right-hand side of (\ref{eq:partitionFormulaII}). 
\qed

From
\be
\sum_{1\leq i < j \leq N+1}
q^{\lambda_i-\mu_j} t^{j-i} 
=
\sum_{1\leq i < j \leq N} 
q^{\lambda_i-\mu_j} t^{j-i} 
+ 
\sum_{1\leq i \leq N} 
q^{\lambda_i-\mu_{N+1}} t^{N+1-i}, 
\ee
(\ref{eq:partitionFormulaII}) is rewritten as
\be
(1-q)\sum_{(i,j)\in\mu } q^{\lambda_i-j} t^{\mu^\vee_j-i+1} 
=
(t-1)\sum_{1\leq i<j\leq {\NNi\mu}} q^{\lambda_i-\mu_j} t^{j-i} +
t\sum_{i=1}^{\NNi\mu} q^{\lambda_i} \left(q^{-\mu_i}- t^{N-i}\right).
\label{eq:partitionFormulaIII}
\ee
%
Note that if $t=q$ and $\lambda= \mu$, 
 (\ref{eq:partitionFormulaIII}) reduces to
the formula of the Maya diagram: 
the length from a black box to a white one or black one is
$(\lambda_i-i) + (\lambda^\vee_j-j)+1$ (the hook length) or
$(\lambda_i-i) - (\lambda_j-j)$, respectively:
\be
\sum_{(i,j)\in\lambda} 
q^{(\lambda_i-i) + (\lambda^\vee_j-j)+1} 
+
\sum_{1\leq i<j\leq {\NNi\lambda} } 
q^{(\lambda_i-i) - (\lambda_j-j)}
=
\sum_{1\leq i\leq {\NNi\lambda} } 
\sum_{i< j\leq \lambda_i + {\NNi\lambda} }
q^{j-i}.
\ee

%
By using  (\ref{eq:partitionFormulaI}), we have: 
\\
{\bf Lemma.}
For all integers 
$\NNi\lambda \geq\ell(\lambda)$ and 
$\NNii\lambda\geq\ell(\lambda^\vee)$,
\be
\left( t^{\ha}-t^{-\ha}\right)
\sum_{i=1}^{\NNi\lambda} \left(q^{\lambda_i} -1\right)t^{\ha-i}
+
\left(q^{\ha}-q^{-\ha}\right)
\sum_{i=1}^{\NNii\lambda} \left(t^{-\lambda^\vee_i}-1\right)q^{i-\ha }
= 0.
\label{eq:ACsum}
\ee

\proof
Similar to  (\ref{eq:partitionFormulaI}),
for all integers 
$\NNii\lambda\geq\ell(\lambda^\vee)$,
\be
\sum_{i=1}^{\NNii\lambda} \left(1-t^{-\lambda^\vee_i}\right) q^{i-1}
=
(1-t^{-1})\sum_{(i,j)\in\lambda^\vee } t^{1-j} q^{i-1}
=
(1-t^{-1})\sum_{(i,j)\in\lambda} t^{1-i} q^{j-1}. 
\ee
Therefore, with  (\ref{eq:partitionFormulaI}),
\be
(1-q)
\sum_{i=1}^{\NNii\lambda} \left(1-t^{-\lambda^\vee_i}\right) q^{i-1}
=
(1-t^{-1})
\sum_{i=1}^{\NNi\lambda}\left(1-q^{\lambda_i}\right) t^{1-i}.
\ee
\qed

In the power sum function (\ref{eq:powersum}), 
(\ref{eq:ACsum}) is written as 
\be
\left( t^{{n\over 2}}-t^{-{n\over 2}}\right)
p_n \left(q^{\lambda}t^{\rho}, L t^{-\rho}\right)
+
\left( q^{{n\over 2}}-q^{-{n\over 2}}\right)
p_n \left(t^{-\lambda^\vee}q^{-\rho}, L q^{\rho}\right)
%
= 0,
\qquad
L\in\bC.
\label{eq:ACsumP}
\ee
Note that if $t=q$, 
 (\ref{eq:ACsum}) reduces to the formula of the Maya diagram: 
the black boxes and the white ones are at 
$\lambda_i -i +\ha$ and $-(\lambda^\vee_i -i +\ha)$
of the Maya diagram, respectively: 
\be
\sum_{i=1}^{\NNi\lambda} q^{\lambda_i -i +\ha }
+
\sum_{i=1}^{\NNii\mu} q^{-\lambda^\vee_i +i -\ha }
=
\sum_{i=1-{\NNi\lambda}}^{\NNii\mu} q^{i-\ha }.
\label{eq:Maya}
\ee
Hence
$ 
\sum_{i\geq 1} q^{\lambda_i -i +\ha }
+
\sum_{i\geq 1} q^{-\lambda^\vee_i +i -\ha }
= 
\sum_{i\in \bZ} q^{i-\ha }
= 
q^{-\ha } \delta(q).
$ 


\subsection{Factors in Nekrasov's formula}


We have the following formula for the Young diagrams, 
which implies the equivalence among several expressions of Nekrasov's formula.
\\
{\bf Proposition.}
The following 
$\fnek{\Li}{\Lii}i\pm{\lambda}{\mu}qt $'s ($i=1,2,3$) 
are all the same.
\ba
v\fnek{\Li}{\Lii}1+{\lambda}{\mu}qt 
&:=&
\sum_{(i,j)\in\mu } \left(q^{\lambda_i}-\Li\right) q^{\ha-j} t^{\mu^\vee_j-i+\ha}
+
\sum_{(i,j)\in\lambda} \left(q^{-\mu_i}-\Lii\right) q^{j-\ha} t^{-\lambda^\vee_j+i-\ha},
\cr
v\fnek{\Li}{\Lii}1-{\lambda}{\mu}qt 
&:=&
\sum_{(i,j)\in\lambda} q^{\lambda_i-j+\ha} t^{\ha-i} \left(t^{\mu^\vee_j}-\Lii\right) 
+
\sum_{(i,j)\in\mu } q^{-\mu_i+j-\ha} t^{i-\ha} \left(t^{-\lambda^\vee_j}-\Li\right),
~~~~~~~~
\ea
%
\ba
v\fnek{\Li}{\Lii}2+{\lambda}{\mu}qt 
&:=&
p_1\left(q^{ \lambda     } t^{ \rho}, \Li  t^{-\rho}\right)
p_1\left(t^{ \mu^\vee    } q^{ \rho}, \Lii q^{-\rho}\right)
-
p_1\left(                  t^{ \rho}, \Li  t^{-\rho}\right)
p_1\left(                  q^{ \rho}, \Lii q^{-\rho}\right),
\cr
v\fnek{\Li}{\Lii}2-{\lambda}{\mu}qt 
&:=&
p_1\left(t^{-\lambda^\vee} q^{-\rho}, \Li  q^{\rho}\right)
p_1\left(q^{-\mu         } t^{-\rho}, \Lii t^{ \rho}\right)
-
[\ \lambda = \mu = 0\ ],
\ea
%
\ba
v\fnek{\Li}{\Lii}3+{\lambda}{\mu}qt 
&:=&\left\{
p_1\left(q^{ \lambda     } t^{ \rho},\ \Li  t^{-\rho}\right)
p_1\left(q^{-\mu         } t^{-\rho},\ \Lii t^{ \rho}\right)
-
[\ \lambda = \mu = 0\ ]
\right\}{t^{-\ha} - t^{\ha}\over q^\ha - q^{-\ha}},
\cr
v\fnek{\Li}{\Lii}3-{\lambda}{\mu}qt 
&:=&\left\{
p_1\left(t^{-\lambda^\vee} q^{-\rho},\ \Li  q^{ \rho}\right)
p_1\left(t^{ \mu^\vee    } q^{ \rho},\ \Lii q^{-\rho}\right)
-
[\ \lambda = \mu = 0\ ]
\right\}
{q^\ha - q^{-\ha}\over t^{-\ha} - t^{\ha}},
\cr
&&
\ea
with $v := (q/t)^{\ha}$
and $L_1$, $L_2\in\bC$.
Here 
$p_1$ is the power sum function in (\ref{eq:powersum}) and
$
[\ \lambda = \mu = 0\ ]
$'s 
stand for terms substituting $\lambda = \mu = 0$ into the foregoing ones.

\proof
It is clear that
\ba
\fnek{\Li}{\Lii}1\pm{\lambda}{\mu}qt 
v 
&=& 
\fnek{\Lii}{\Li}1\pm{\mu}{\lambda}{q^{-1}}{t^{-1}} 
/v
= 
\fnek{\Lii}{\Li}1\mp{\mu^\vee}{\lambda^\vee}tq 
/v,
\cr
\fnek{\Li}{\Lii}2\pm{\lambda}{\mu}qt 
v 
&=& 
\fnek{\Lii}{\Li}2\mp{\mu}{\lambda}{q^{-1}}{t^{-1}} 
/v
= 
\fnek{\Lii}{\Li}2\pm{\mu^\vee}{\lambda^\vee}tq 
/v,
\cr
\fnek{\Li}{\Lii}3\pm{\lambda}{\mu}qt 
v 
&=& 
\fnek{\Lii}{\Li}3\pm{\mu}{\lambda}{q^{-1}}{t^{-1}} 
/v
= 
\fnek{\Lii}{\Li}3\mp{\mu^\vee}{\lambda^\vee}tq 
/v.
\ea
Therefore, it suffices to show
$
\neknon 1+{\lambda}{\mu}qt 
=\neknon 2+{\lambda}{\mu}qt 
=\neknon 3+{\lambda}{\mu}qt 
$.
First, applying (\ref{eq:ACsumP}) yields
$
\neknon 2+{\lambda}{\mu}qt 
=\neknon 3+{\lambda}{\mu}qt 
$.
%
Next, we prove that 
$
\neknon 1+{\lambda}{\mu}qt 
=\neknon 3+{\lambda}{\mu}qt 
$.
From (\ref{eq:partitionFormulaII})$\times t$, we have,
for all integers 
$\NNi{\lambda\mu}\geq\ell(\lambda)$, $\ell(\mu)$,
\be
\sum_{(i,j)\in\mu } q^{\lambda_i-j} t^{\mu^\vee_j-i+1} 
= 
{1\over 1-q}
\left[\
t
\sum_{i=1}^{\NNi{\lambda\mu}}
\sum_{j=i}^{\NNi{\lambda\mu}}
-
\sum_{i=1}^{\NNi{\lambda\mu}}
\sum_{j=i+1}^{\NNi{\lambda\mu} + 1}
\ \right]
q^{\lambda_i-\mu_j} t^{j-i}.
\ee
By replacing 
$q$, $t$ and $\lambda$ in
(\ref{eq:partitionFormulaII}) with
$1/q$, $1/t$ and $\mu $, respectively,
\be
\sum_{(i,j)\in\lambda } q^{-\mu_i+j-1} t^{-\lambda^\vee_j+i} 
=
{1\over 1-q}
\left[\
t
\sum_{j=1}^{\NNi{\lambda\mu}}
\sum_{i=j+1}^{\NNi{\lambda\mu} + 1}
-
\sum_{j=1}^{\NNi{\lambda\mu}}
\sum_{i=j}^{\NNi{\lambda\mu}}
\ \right]
q^{\lambda_i-\mu_j} t^{j-i}.
\ee
Adding these two equations, we have 
\ba
\fnek{\Li}{\Lii}1+{\lambda}{\mu}qt 
&+&
\Li\sum_{(i,j)\in\mu } q^{-j} t^{i}
+
\Lii\sum_{(i,j)\in\lambda} q^{j-1} t^{1-i}
\cr
&=&
\sum_{(i,j)\in\mu } q^{\lambda_i-j} t^{\mu^\vee_j-i+1} 
+\sum_{(i,j)\in\lambda } q^{-\mu_i+j-1} t^{-\lambda^\vee_j+i} 
\cr
&=&
{1\over 1-q}
\left[\ 
t
\sum_{i=1}^{\NNi{\lambda\mu}+1}
\sum_{j=1}^{\NNi{\lambda\mu}}
-
\sum_{i=1}^{\NNi{\lambda\mu}}
\sum_{j=1}^{\NNi{\lambda\mu}+1}
\ \right]
q^{\lambda_i-\mu_j} t^{j-i}
\cr
&=&
{1\over 1-q}
\left[\ 
t
\sum_{i=1}^{\NNi{\lambda\mu}+1}
\sum_{j=1}^{\NNi{\lambda\mu}}
-
\sum_{i=1}^{\NNi{\lambda\mu}}
\sum_{j=1}^{\NNi{\lambda\mu}+1}
\ \right]
\left(q^{\lambda_i-\mu_j} - 1\right)t^{j-i}.
\label{eq:nekiiiNlambdamu}
\ea
Thus, the following lemma with 
$\NNi{\lambda}=\NNi{\mu}=\NNi{\lambda\mu}$
shows that
$
\neknon 1+{\lambda}{\mu}qt 
=\neknon 3+{\lambda}{\mu}qt 
$.
\qed

{\bf Lemma.}
For any integers
$\NNi\lambda \geq \ell(\lambda)$,
$\NNi\mu \geq \ell(\mu)$ and
$\NNii\mu \geq \ell(\mu^\vee)$,
%
\ba
\fnek{\Li}{\Lii}2+{\lambda}{\mu}qt 
&=&
\sum_{i=1}^{\NNi\lambda} \sum_{j=1}^{\NNii\mu}
\left(q^{\lambda_i} t^{\mu^\vee_j} - 1\right)
 t^{1-i} q^{-j}
\cr
&+&
\sum_{(i,j)\in\lambda} 
q^{\lambda_i-j} t^{1-i} 
\left(q^{-\NNii\mu}-\Lii\right)
+
\sum_{(i,j)\in\mu } 
t^{\mu^\vee_j-i+1} q^{-j}
\left(t^{-\NNi\lambda}-\Li\right),
\cr
%
\fnek{\Li}{\Lii}3+{\lambda}{\mu}qt 
&=&
{1\over 1-q}
\left[\ t 
\sum_{i=1}^{\NNi\lambda + 1} \sum_{j=1}^{\NNi\mu    } - 
\sum_{i=1}^{\NNi\lambda    } \sum_{j=1}^{\NNi\mu + 1}  
\ \right]
\left(q^{\lambda_i-\mu_j} - 1\right) t^{j-i  }
\cr
&-&
\Li\sum_{(i,j)\in\mu } q^{-j} t^{i}
-
\Lii\sum_{(i,j)\in\lambda} q^{j-1} t^{1-i}.
\ea

\proof
\ba
&&\hskip-10pt
t^{-1}\fnek{\Li}{\Lii}2+{\lambda}{\mu}qt 
\cr
&=&
\left(
\sum_{i=1}^{\NNi\lambda} 
\left(q^{\lambda_i} -1\right) t^{-i} - {1-\Li \over 1-t}
\right)\left(
\sum_{j=1}^{\NNii\mu}
\left(t^{\mu_j^\vee} -1\right) q^{-j} - {1-\Lii \over 1-q}
\right)
-
{1-\Li\over 1-t}
{1-\Lii\over 1-q}
\cr
&=&
\left(
\sum_{i=1}^{\NNi\lambda} 
q^{\lambda_i} t^{-i} - {t^{-\NNi\lambda}-\Li \over 1-t}
\right)\left(
\sum_{j=1}^{\NNii\mu}
t^{\mu_j^\vee} q^{-j} - {q^{-\NNii\mu}-\Lii \over 1-q}
\right)
-
[\ \lambda = \mu = 0\ ]
\cr
&=&
\sum_{i=1}^{\NNi\lambda} \sum_{j=1}^{\NNii\mu}
\left(q^{\lambda_i} t^{\mu^\vee_j} - 1\right)
 t^{1-i} q^{-j}
+
\sum_{i=1}^{\NNi\lambda} {1-q^{\lambda_i}\over 1-q} t^{1-i} 
\left(q^{-\NNii\mu}-\Lii\right)
+
\sum_{j=1}^{\NNii\mu} {t^{\mu_j^\vee-1}\over 1-t^{-1}} q^{-j}
\left(t^{-\NNi\lambda}-\Li\right).
\nonumber
\ea
\ba
&&\hskip-12pt
{1-q \over t-1} 
\fnek{\Li}{\Lii}3+{\lambda}{\mu}qt 
\cr
&=&
\left(
\sum_{i=1}^{\NNi\lambda} 
\left(q^{\lambda_i} -1\right) t^{-i} - {1-\Li \over 1-t}
\right)\left(
\sum_{j=1}^{\NNi\mu}
\left(q^{-\mu_j} -1\right) t^{j} - {1-\Lii \over 1-t^{-1}}
\right)
-
{1-\Li\over 1-t}
{1-\Lii\over 1-t^{-1}}
\cr
%
&=&
\left(
\sum_{i=1}^{\NNi\lambda} 
q^{\lambda_i} t^{-i} - {t^{-\NNi\lambda}-\Li\over 1-t}
\right)\left(
\sum_{j=1}^{\NNi\mu}
q^{-\mu_j} t^{j} - {t^{\NNi\mu}-\Lii \over 1-t^{-1}}
\right)
-
[\ \lambda = \mu = 0\ ]
\cr
&=&
\sum_{i=1}^{\NNi\lambda} 
\sum_{j=1}^{\NNi\mu}
\left( q^{\lambda_i-\mu_j} - 1 \right) t^{j-i} 
-
\sum_{j=1}^{\NNi\mu}
\left( q^{-\mu_j} -1 \right) t^j {t^{-\NNi\lambda}-\Li \over 1-t}
-
\sum_{i=1}^{\NNi\lambda} 
\left( q^{\lambda_i} -1 \right)t^{-i} {t^{\NNi\mu}-\Lii \over 1-t^{-1}}.
\nonumber
\ea
\qed
%

%
Note that 
$\neknon 2\pm{\lambda}{\mu}qt $ and
$\neknon 3\pm{\lambda}{\mu}qt $
are independent of $\NNi\lambda$'s, 
if they are sufficiently large.
%
%
%
Let $\fnekrasov{\Li}{\Lii}\lambda\mu qt := \fnek{\Li}{\Lii}i\pm{\lambda}{\mu}qt $;
then it satisfies
\be
\fnekrasov{\Li}{\Lii}{\lambda}{\mu}qt 
v 
= 
\fnekrasov{\Lii}{\Li}{\mu}{\lambda}{q^{-1}}{t^{-1}} 
/v
= 
\fnekrasov{\Lii}{\Li}{\mu^\vee}{\lambda^\vee}tq 
/v.
\label{eq:nDual}
\ee


Let 
\be
\fNek{\Li}{\Lii}i{\pm}{\lambda}{\mu}{vQ}qt 
:= 
\Exp{-
\sum_{n>0}{Q^n\over n}v^n
\fnek{\Li^n}{\Lii^n}i{\pm}{\lambda}{\mu}{q^n}{t^n}
},
\ee
then we have:
\\
{\bf Corollary.}
The following 
$\fNek{\Li}{\Lii}i\pm{\lambda}{\mu}{Q}qt $'s ($i=1,2,3$) 
are all the same.
\ba
\fNek{\Li}{\Lii}1+{\lambda}{\mu}{vQ}qt  
&=&
\prod_{(i,j)\in\mu } 
{
1 - Q\, q^{\lambda_i-j+\ha} t^{\mu^\vee_j-i+\ha}
\over
1 - Q\Li\, q^{\ha-j} t^{i-\ha}
}
\prod_{(i,j)\in\lambda} 
{
1 - Q\, q^{-\mu_i+j-\ha} t^{-\lambda^\vee_j+i-\ha}
\over
1 - Q\Lii\, q^{j-\ha} t^{\ha-i}
},
\cr
\fNek{\Li}{\Lii}1-{\lambda}{\mu}{vQ}qt  
&=&
\prod_{(i,j)\in\lambda} 
{
1 - Q\, q^{\lambda_i-j+\ha} t^{\mu^\vee_j-i+\ha}
\over
1 - Q\Lii\, q^{j-\ha} t^{\ha-i}
}
\prod_{(i,j)\in\mu } 
{
1 - Q\, q^{-\mu_i+j-\ha} t^{-\lambda^\vee_j+i-\ha}
\over
1 - Q\Li\, q^{\ha-j} t^{i-\ha}
},
~~~~~~~
\ea
\ba
\fNek{\Li}{\Lii}2+{\lambda}{\mu}{vQ}qt  
&=&
{
\tPi\left(
- Q\, 
\left\{q^{\lambda}t^{\rho},\Li t^{-\rho}\right\},
\ 
\left\{t^{\mu^\vee}q^{\rho}, \Lii q^{-\rho}\right\}
\right)
\over
\tPi\left(
- Q\, 
\left\{t^{\rho},\Li t^{-\rho}\right\},
\ 
\left\{q^{\rho}, \Lii q^{-\rho}\right\}
\right)
},
\cr
\fNek{\Li}{\Lii}2-{\lambda}{\mu}{vQ}qt  
&=&
{
\tPi\left(
- Q\, 
\left\{t^{-\lambda^\vee}q^{-\rho},\Li q^{\rho}\right\},
\ 
\left\{q^{-\mu}t^{-\rho}, \Lii t^{\rho}\right\}
\right)
\over
\tPi\left(
- Q\, 
\left\{q^{-\rho},\Li q^{\rho}\right\},
\ 
\left\{t^{-\rho}, \Lii t^{\rho}\right\}
\right)
},
\ea
\ba
\fNek{\Li}{\Lii}3+{\lambda}{\mu}{vQ}qt  
&=&
{
\Pi\left(
vQ\, 
\left\{q^{\lambda}t^{\rho},\Li t^{-\rho}\right\},
\ 
\left\{q^{-\mu}t^{-\rho}, \Lii t^{\rho}\right\}
;q,t\right)
\over
\Pi\left(
vQ\, 
\left\{t^{\rho},\Li t^{-\rho}\right\},
\ 
\left\{t^{-\rho}, \Lii t^{\rho}\right\}
;q,t\right)
},
\cr
\fNek{\Li}{\Lii}3-{\lambda}{\mu}{vQ}qt  
&=&
{
\Pi\left(
vQ\, 
\left\{t^{-\lambda^\vee}q^{-\rho},\Li q^{\rho}\right\},
\ 
\left\{t^{\mu^\vee}q^{\rho}, \Lii q^{-\rho}\right\}
;t^{-1},q^{-1}\right)
\over
\Pi\left(
vQ\, 
\left\{q^{-\rho},\Li q^{\rho}\right\},
\ 
\left\{q^{\rho}, \Lii q^{-\rho}\right\}
;t^{-1},q^{-1}\right)
},
\ea
By letting $\Li$ and $\Lii=0$,
we obtain six expressions of $\Nek{\lambda}{\mu}Qqt $ in Nekrasov's formula.
This completes the proof of the proposition 
in section \ref{sec:PartitionFunctionCS}.


\Section*{Appendix B : Formula for the Macdonald Symmetric Function}


\renewcommand{\theequation}{B.\arabic{equation}}\setcounter{equation}{0}
\renewcommand{\thesubsection}{B.\arabic{subsection}}\setcounter{subsection}{0}


Here we recapitulate basic properties of 
the Macdonald symmetric function \cite{Mac}.


\subsection{Definition for the Macdonald symmetric function}


Bases of the ring of symmetric functions in an infinite number of variables 
$x=(x_1,x_2,\cdots)$
are indexed by the Young diagram,
i.e. the partition
$\lambda =(\lambda_1,\lambda_2,\cdots)$,
which is a sequence of nonnegative integers such that
$\lambda_{i} \geq \lambda_{i+1}$ and 
$|\lambda| = \sum_i \lambda_i < \infty$.
%
%
For example, the monomial symmetric function 
is defined by
$
m_{\lambda}(x)=\sum_{\sigma}
x_1^{\lambda_{\sigma(1)}}
x_2^{\lambda_{\sigma(2)}}
\cdots ,
$
where the summation is over all distinct permutations of 
$(\lambda_1,\lambda_2,\cdots )$.
%
%
The power sum symmetric function $p_{\lambda}(x)$ is
defined by
\be
p_{\lambda}(x)=
p_{\lambda_1}(x)
p_{\lambda_2}(x)\cdots ,
\qquad 
p_n(x)=\sum_{i=1}^{\infty}x_i^n.
\ee
%
%
We introduce an inner product on the ring of symmetric functions
in the following manner:
for any symmetric functions $f$ and $g$, in power sums $p_\lambda$'s,
\be
\langle f(p) | g(p)\rangle_{q,t} 
:= f(p^*)\, g(p)\,\vert_{{\rm constant\, part}},\qquad
p_n^* := n {1-q^n \over 1-t^n} {\partial \over \partial
p_n}.
\ee


The Macdonald symmetric function
$P_{\lambda}=P_{\lambda}(x;q,t)$
is uniquely specified by the following orthogonality and normalization:
\ba
  &&
  \langle P_{\lambda} | P_{\mu}\rangle_{q,t} =0\qquad { \rm if } \;
  \lambda\neq \mu,\\
  &&
  P_{\lambda}(x;q,t)
  =
  m_{\lambda}(x) + \sum_{\mu<\lambda} u_{\lambda\mu}m_{\mu}(x),
  \quad
  u_{\lambda\mu}\in {\mathbb Q}(q,t).
\ea
Here we used the dominance partial ordering on the Young diagrams defined as
$\lambda\geq\mu \Leftrightarrow |\lambda|=|\mu|$ and
$\lambda_1+\cdots+\lambda_i\geq\mu_1+\cdots+\mu_i$ for all
$i$.


The scalar product is given by 
\be
\langle P_\lambda|P_\lambda\rangle_{q,t}
=
\prod_{s\in\lambda}
{
1-q^{a (s)+1} t^{  \ell(s)   }
\over 
1-q^{a (s)  } t^{  \ell (s)+1}
},
\ee
which satisfies 
\be
\langle P_\lambda|P_\lambda\rangle_{q,t}
=
\left({q\over t}\right)^{|\lambda|}
\langle P_\lambda|P_\lambda\rangle_{q^{-1},t^{-1}}
=
\langle P_{\lambda^\vee} |P_{\lambda^\vee} \rangle_{t,q}^{-1}.
\ee
If we define
\be
g_\lambda(q,t)
:=
{
v^{|\lambda|}
\over 
\langle P_\lambda|P_\lambda\rangle_{q,t}
},
\ee
with $v = (q/t)^\ha$,  
then
\be
g_\lambda(q,t)=
g_\lambda(q^{-1},t^{-1})=
g_{\lambda^\vee}(t,q)^{-1}.
\ee
%

The skew Macdonald symmetric function 
$\tP_{\lambda/\mu}(x;q,t)$ is defined by
\be
\tP_{\lambda/\mu}(x;q,t) 
:=
g_\mu(q,t)
\tP_{\mu}^*\left(v^{-1} x;q,t\right) \, \tP_\lambda(x;q,t), 
\ee
where $*$ acts on the power sum as
$p_n^* := n {1-q^n \over 1-t^n} {\partial \over \partial p_n}$.
Finally let $\iota\tP_{\lambda/\mu}(x;q,t)$ be 
the skew Macdonald function with the involution $\iota$ 
acting on the power sum $p_n$ as 
$\iota(p_n) = -p_n$.

Let
$x=(x_1,x_2,\cdots)$ and
$y=(y_1,y_2,\cdots)$
be two sets of variables. Then we have
\be
\sum_\mu
\tP_{\lambda/\mu} (x;q,t) 
\tP_{\mu/\nu} (y;q,t) 
= 
\tP_{\lambda/\nu} (x,y;q,t),
\label{eq:AppaddSkewMacdonald}
\ee
where
$P_{\lambda/\nu} (x,y;q,t)$
denotes the skew Macdonald function in the set of variables 
\break
$(x_1,x_2,\cdots,y_1,y_2,\cdots)$.


\subsection{Symmetries and Cauchy formulas}
 

Next, we turn to showing the basic properties of 
the (skew) Macdonald symmetric function.
%
The Macdonald function enjoys the symmetries
\be
\tP_{\lambda/\mu} (cx; q,t) = 
c^{|\lambda|-|\mu|}
\tP_{\lambda/\mu} (x; q,t),
\qquad
c\in\bC ,
\label{eq:scaleTrans}
\ee
\be
\tP_{\lambda/\mu} \left(x; q^{-1},t^{-1}\right) 
=
\tP_{\lambda/\mu} (x; q,t),
\ee
\be
\tP_{\lambda^\vee/\mu^\vee} (vx; t,q) 
= 
{g_\lambda(q,t)\over g_\mu(q,t) }
\Endomega qt{} \tP_{\lambda/\mu} (x; q,t),
\label{eq:skewConjugate}
\qquad 
\Endomega qt{} (p_n) 
=
(-1)^{n-1}{1-q^n \over 1-t^n} p_n.
\ee
%
%
When $t=q$, the Schur function has the extra symmetries
\be
s_{\lambda^\vee}(x) = \iota s_\lambda(-x) = (-1)^{|\lambda|}\iota
s_\lambda(x).
\ee

The following Cauchy formula is especially important:
\ba
\sum_\lambda
g_\lambda(q,t)
\tP_\lambda(x;q,t) \tP_\lambda(y;q,t)
=
\Pi(v x,y)
&:=&
\exp\left\{
\sum_{n>0}{ v^n \over n}{1-t^n \over 1-q^n} p_n(x) p_n(y)
\right\}
\cr
&=&
\prod_{k\geq 0}
\prod_{i,j}
{1-tv x_i y_j q^k 
\over 
 1- v x_i y_j q^k },
\quad |q|<1.
\label{eq:AppCauchy}
\\
%
\sum_\lambda
\tP_\lambda(x;q,t) \tP_{\lambda^\vee} (y;t,q)
=
\Pi_0(x,y)
&:=&
\exp\left\{
\sum_{n>0}{(-1)^{n-1}\over n} p_n(x) p_n(y)
\right\}
\cr
&=&
\prod_{i,j}(1+ x_i y_j).
\label{eq:AppconjugateCauchy}
\ea
%
The Cauchy formulas for the skew Macdonald function are
 \ba
\sum_\lambda
{g_\lambda(q,t)\over g_\mu(q,t) }
 \tP_{\lambda/\mu} (x;q,t) \tP_{\lambda/\nu} (y;q,t)
&=&
\Pi(v x,y)
\sum_\lambda
\tP_{\mu/\lambda} (y;q,t) \tP_{\nu/\lambda} (x;q,t)
{g_\nu(q,t)\over g_\lambda(q,t)} ,
\cr
\sum_\lambda
\tP_{\lambda/\mu} (x;q,t) \tP_{\lambda^\vee/\nu^\vee} (y;t,q)
&=&
\Pi_0(x,y)
\sum_\lambda
\tP_{\mu^\vee/\lambda^\vee} (y;t,q) \tP_{\nu/\lambda} (x;q,t).
\label{eq:AppskewCauchy}
\ea
%
If we denote by $\Endomega qtx$ 
the endmorphism $\Endomega qt{}$ on variables $x$, then
\be
\Pi(v x,y;q,t)
=
\Pi(v^{-1} x,y;q^{-1},t^{-1})
=
\Endomega tqx \Endomega tqy\Pi(v^{-1} x,y;t,q).
\ee


\subsection{Specialization formulas} 


We denote
\ba
p_n(c q^\lambda t^\rho) 
&:=&
c^n\sum_{i=1}^\infty (q^{n\lambda_i}-1)t^{n(\ha-i)}
+ { c^n \over t^{n\over 2} - t^{-{n\over 2}} }, 
\qquad 
c\in\bC,
\cr
p_n(c q^\lambda t^\rho, c L t^{-\rho} ) 
&:=&
p_n(c q^\lambda t^\rho) + p_n(c L t^{-\rho} ), 
\cr 
&=&
c^n\sum_{i=1}^\infty (q^{n\lambda_i}-1)t^{n(\ha-i)}
+ c^n {1 - L^n \over t^{n\over 2} - t^{-{n\over 2}} }, 
\qquad 
c, L\in\bC.
\ea
%
Then by using (\ref{eq:skewConjugate}) and (\ref{eq:ACsumP}), we obtain
\be
\tP_{\mu^\vee/\nu^\vee}
\left(-t^{\lambda^\vee}q^{\rho},\ -L q^{-\rho};t,q\right)
=
{g_\mu(q,t)\over g_\nu(q,t)}
\tP_{\mu /\nu } 
\left(q^{-\lambda}t^{-\rho},\ L t^{\rho};q,t\right),
\label{eq:ACMac}
\ee
%
%

%
The Macdonald function in the power sums
$p_n = (1-L^n)/(t^{{n\over 2}}- t^{-{n\over 2}})$ is
\cite{Mac}(Ch.\ VI.6) 
\be
P_\lambda\left(t^\rho, L t^{-\rho};q,t\right) 
=
\prod_{s\in\lambda} 
(-1)t^\ha q^{a'(s)}
{
1-L q^{-a'(s)} t^{\ell'(s)}
\over 
1-q^{a(s)} t^{\ell(s)+1} 
},
\label{eq:Specialization}
\ee
for a generic $L\in\bC$.
By replacing $(q,t)$ and $\lambda$ 
with $(t,q)$ and $\lambda^\vee$, respectively,
\be 
P_{\lambda^\vee}\left(q^\rho, L q^{-\rho};t,q\right) 
=
\prod_{s\in\lambda}  q^{-\ha} q^{-a'(s)}
{
1-L q^{a'(s)} t^{-\ell'(s)}
\over 
1-q^{-a(s)-1} t^{-\ell(s)} 
}.
\ee
%
%
Then we have
\be
g_\lambda(q,t) 
{
\tP_\lambda(t^\rho,\Li t^{-\rho};q,t)
\over 
\tP_{\lambda^\vee}(q^\rho,\Lii q^{-\rho};t,q)}
=
\prod_{s\in\lambda}q^{a'(s)}t^{-\ell'(s)}
{
1-\Li q^{-a'(s)}t^{\ell'(s)}
\over 
1-\Lii q^{a'(s)}t^{-\ell'(s)}},
\qquad \Li, \Lii\in\bC.
\label{eq:DualFormula}
\ee
%
Note that
\be
\iota P_\lambda\left(t^{-\rho}, L t^{\rho};q,t\right) 
=
P_\lambda\left(t^{\rho}, L t^{-\rho};q,t\right), 
\qquad L\in\bC.
\ee


If $L=t^{-N}$ with $N\in\bN$, then
$
p_n(q^\lambda t^\rho, t^{-N-\rho} ) 
= \sum_{i=1}^N q^{n\lambda_i} t^{n(\ha-i)}
$
is the power sum symmetric polynomial in $N$ variables
$\{q^{\lambda_i}t^{\ha-i}\}_{1\leq i \leq N}$,
hence
$
P_\lambda\left(t^\rho, t^{-N-\rho};q,t\right) 
$ 
reduces to the Macdonald symmetric polynomial in $N$ variables.
Therefore
\be
P_\lambda\left(t^\rho, t^{-N-\rho};q,t\right) 
=
0,
\qquad
{\rm for}
\quad
\ell(\lambda) > N\in\bN.
\ee
Note that 
\be
{\cal W}_{\lambda,\mu} (q,t):= 
\tP_\lambda\left(      t^\rho,t^{-N-\rho};q,t\right)
\tP_\mu \left(q^\lambda t^\rho,t^{-N-\rho};q,t\right),
\qquad
N\in\bN,
\ee
has a nice symmetry \cite{Mac}(Ch.\ VI.6): 
\be
{\cal W}_{\lambda,\mu} (q,t)
= {\cal W}_{\mu,\lambda} (q,t).
\label{eq:Wsymm}
\ee




When  $L=0$ (the case of principal specialization),
\ba
\tP_\lambda\left(t^{\rho}; q,t\right)
\prod_{s\in\lambda} (-1)q^{-a(s)} t^{\ell(s)}
&=&
\tP_\lambda\left(t^{-\rho}; q,t\right)
=
\tP_{\lambda^\vee}\left(-q^{\rho}; t,q\right)
/g_\lambda(q,t)
\cr
&=&
\iota \tP_\lambda\left(t^{\rho}; q,t\right)
=
\iota\tP_{\lambda^\vee}\left(-q^{-\rho}; t,q\right)
/g_\lambda(q,t).
\label{eq:ACprincipalMac}
\ea


\section*{Appendix C : Refined BPS State Counting}

\renewcommand{\theequation}{C.\arabic{equation}}\setcounter{equation}{0}
\renewcommand{\thesubsection}{C.\arabic{subsection}}\setcounter{subsection}{0}


From the instanton expansion of Nekrasov's partition function,
\beq
Z_{Nek} = 1 + \sum_{k=1}^\infty \Lambda^k \Zk ktq{Q_\alpha}~,
\eeq
we can compute the refined Gopakumar-Vafa integer invariant $N_\beta^{(j_L,j_R)}$ 
as follows. We expect the following multicover structure of the partition function
\beq
Z_{Nek} = \exp \left( \sum_{n=1}^\infty \frac{\GV{}{}{t^n}{q^n}{Q_\alpha^n, Q_B^n}}{n} \right)~,
\eeq
from the argument of Gopakumar-Vafa type. 
Assuming the scale parameter $\Lambda$ is proportional to the K\"ahler
parameter $Q_B$ of the base space ${\bf P}^1$ of ALE fibration, we expand
\beq
\GV{}{}tq{Q_\alpha, Q_B} = \sum_{k=1}^\infty Q_B^k \GV k{}tq{Q_\alpha}~, \label{base-expansion}
\eeq
where
\beq
\GV k{}tq{Q_\alpha} = \sum_{\{\ell_\alpha\}} \sum_{(j_L,j_R)} 
\frac{N_{k, \{\ell_\alpha\}}^{(j_L, j_R)}}{(q^{1/2} - q^{-1/2})(t^{1/2} - t^{-1/2})} 
\chi_{j_L} (u) \chi_{j_R}(v) \prod_{\alpha=1}^{N-1} Q_\alpha^{\ell_\alpha}~, \label{fiber-expansion}
\eeq
and $\chi_j(x)$ is the irreducible character of $SU(2)$ with spin $j$. 
We have introduced the notations $u^2 = t\cdot q$ and $v^2 = q/t$. 
Comparing the coefficients of $\Lambda^k \sim Q_B^k$, up to $k=4$ we obtain
\beqa
\GV 1{}tq{Q_\alpha} &=& \Zk 1tq{Q_\alpha}, \CR
\GV 2{}tq{Q_\alpha} &=& \Zk 2tq{Q_\alpha} - \frac{1}{2} \left( \Zk 1tq{Q_\alpha} \right)^2
-\frac{1}{2} \Zk 1{t^2}{q^2}{Q_\alpha^2}, \CR
\GV 3{}tq{Q_\alpha} &=& \Zk 3tq{Q_\alpha} -\Zk 2tq{Q_\alpha} \Zk 1tq{Q_\alpha}
+ \frac{1}{3} \left( \Zk 1tq{Q_\alpha} \right)^3 -\frac{1}{3} \Zk 1{t^3}{q^3}{Q_\alpha^3}, \CR
\GV 4{}tq{Q_\alpha} &=& \Zk 4tq{Q_\alpha} - \Zk 3tq{Q_\alpha} \Zk 1tq{Q_\alpha}
-\frac{1}{2} \left( \Zk 2tq{Q_\alpha} \right)^2  \CR
& &~~~+ \Zk 2tq{Q_\alpha} \left( \Zk 1tq{Q_\alpha} \right)^2 
-\frac{1}{4} \left( \Zk 1tq{Q_\alpha} \right)^4 -\frac{1}{2}  \Zk 2{t^2}{q^2}{Q_\alpha^2} \CR
& &~~~~~+ \frac{1}{4} \left( \Zk 1{t^2}{q^2}{Q_\alpha^2} \right)^2.
\eeqa
There is a cancellation of $\Zk 1{t^4}{q^4}{Q_\alpha^4}$ in the computation of $\GV 4{}tq{Q_\alpha}$.

In \cite{AK} we reported some results for $SU(2)$ theory with no Chern-Simons coupling.
This corresponds to the refined GV invariants for the local Hirzebruch surface of ${\bf F}_0$.
For $SU(2)$ theory the expansion at instanton number $k$ becomes
\beq
\GV k{}tq{Q_F} = \sum_{n=0}^\infty \sum_{(j_L,j_R)} 
\frac{N_{kB + nF}^{(j_L, j_R)}}{(q^{1/2} - q^{-1/2})(t^{1/2} - t^{-1/2})} 
\chi_{j_L} (u) \chi_{j_R}(v) v^{2k} Q_F^{k+n}~,
\eeq
where $Q_F$ is the K\"ahler parameter of the fiber ${\bf P}^1$.
The analysis of the symmetry of Nekrasov's partition function made in section 2
instructs us to factor out $v^{2k} Q_F^k$ in computing $N_{kB + nF}^{(j_L, j_R)}$.
Our results are
\beq
N_{B + n F}^{(j_L, j_R)} = \delta_{j_L, 0} \delta_{j_R, n+\frac{1}{2}}~,
\eeq
for one instanton and 
\beq
\bigoplus_{(j_L, j_R)} N_{2B + nF}^{(j_L, j_R)} \left( j_L, j_R \right) =
\bigoplus_{\ell=1}^n \bigoplus_{m=1}^{n-\ell+1}  \left[\frac{m+1}{2} \right] 
\left( \frac{\ell -1}{2}, \frac{3\ell +2m}{2}\right)~, \label{2inst-spin}
\eeq
for two instantons.

We have computed the invariants of $SU(2)$ theory with the Chern-Simons coupling
$m=1,2$, which are expected to give the refined GV invariants for local ${\bf F}_1$ 
and ${\bf F}_2$. It has been known that the GV invariants of
${\bf F}_0$ and ${\bf F}_2$ are simply related by a \lq\lq shift\rq\rq\ of the K\"ahler
parameters. We have found this relation survives for the refined GV invariants up to 
instanton number $3$. To describe the result neatly, 
let $\GV k{(m)}tq{Q_F}$ be the coefficients of the
instanton expansion \eqref{base-expansion} for local ${\bf F}_m$. Then what we
have checked is 
\beq
\GV k{{(2)}}tq{Q_F} 
= Q_F^k \cdot 
\GV k{{(0)}}tq{Q_F}, \quad (1\leq k \leq 3)~,
\eeq
which implies $N_{kB + nF}^{(j_L, j_R)}$ for local ${\bf F}_0$ is the same as
$N_{kB + (n+k) F}^{(j_L, j_R)}$ for local ${\bf F}_2$. 
We would like to stress this is a somewhat surprising result, since the refined 
GV invariants are not BPS protected quantities and they may jump under the 
deformation of complex structures\footnote{However, for local CY the deformation
of complex structure may not be well-defined, because of noncompactness of the
total space.}. For the GV invariants which are BPS protected, the
agreement of the invariants may be explained by the fact that ${\bf F}_2$ is obtained 
from ${\bf F}_0$ by a deformation of complex structure\footnote{We thank 
Y. Konishi and S. Minabe for discussion on this issue.}. However, for BPS nonprotected quantities
it is not certain if the same argument applies. In any case
what we have found supports the expectation
that on noncompact Calabi-Yau manifold the refined  GV invariants are
actually invariant under the complex structure deformation, which is pointed
out in \cite{HIV}. 

For local ${\bf F}_1$ the invariants are qualitatively different from local ${\bf F}_0$
at one instanton. We have
\beq
N_{B + n F}^{(j_L, j_R)} = \delta_{j_L, 0} \delta_{j_R, n}~.
\eeq
For ${\bf F}_0$ and ${\bf F}_2$ the right spin $j_R$ at one instanton is always half-integer,
while for ${\bf F}_1$ it is integer. However, at two instanton our computation shows 
that the refined GV invariants of ${\bf F}_1$
are related to ${\bf F}_0$ quite similarly to the relation between ${\bf F}_0$ and ${\bf F}_2$. 
We have checked that
\beq
\GV {2k}{(1)}tq{Q_F} = Q_F^k \cdot \GV {2k}{(0)}tq{Q_F}, \quad (k=1)~. \label{F0F1}
\eeq
It has been pointed out that for even instanton
number one may expect the GV invariants of local ${\bf F}_0$ and local ${\bf F}_1$ are
related \cite{IK-P1}. 
It is tempting to conjecture that the above relation is valid for any $k$. 

For general values of the Chern-Simons coupling, our preliminary computation 
shows that the refined invariants have no simple relation to those of
local ${\bf F}_{0,1,2}$. Even wrong the structure of $Spin(4)$ character seems
lost in this region. This may be related to the fact that the five-dimensional theory
is physically not well-defined for these Chern-Simons couplings.

For $SU(3)$ case the computation of the refined invariants gets more involved.
The corresponding local toric Calabi-Yau geometry is the ALE fibration
of $A_2$ type over ${\bf P}^1$ and we have two K\"ahler parameters $Q_1 := e^{-t_{F_1}}$ 
and $Q_2 := e^{-t_{F_2}}$, for the fiber. The instanton expansion takes the form
\beq
\GV k{}tq{Q_i} = \sum_{n_1, n_2=0}^\infty \sum_{(j_L,j_R)} 
\frac{N_{\beta(n_1,n_2)}^{(j_L, j_R)}}{(q^{1/2} - q^{-1/2})(t^{1/2} - t^{-1/2})} 
\chi_{j_L} (u) \chi_{j_R}(v) v^{3k} Q_1^{k+n_1} Q_2^{k+n_2}~,
\eeq
where $\beta(n_1, n_2):=kB + n_1F_1+ n_2F_2$ represents two cycles wrapping $k$ times 
on the base space. The analysis of the symmetry of Nekrasov's partition function 
made in section 2 instructs us to factor out $v^{3k} (Q_1Q_2)^k$. At one instanton
we found that the spin contents for the homology class $B + n_1F_1+ n_2F_2$ are
\beq
(0, n_{\max}) \oplus (0, n_{\max}-1) \oplus \cdots \oplus (0, |n_1 - n_2|)~,
\eeq
where $n_{\max} := \max (n_1, n_2)$.
We note that the left spin always vanishes at one instanton.
When $n_1=0$ or $n_2=0$ the geometry reduces to local ${\bf F}_1$ and the 
above result is consistent with the refined GV invariants of local ${\bf F}_1$. 
At two instanton since we cannot find any simple rule for the refined GV invariants,
let us present a short list of our computation. When $n_1=0$ or $n_2=0$,
the result is again consistent with \eqref{2inst-spin} in view of the relation \eqref{F0F1}.

\vskip10mm

\begin{tabular}{| c | l |}
\hline
\rule[-12pt]{0pt}{32pt} 
$(n_1, n_2)$ 
&~ 
spin contents \\
\hline
\kern75pt{} 
&~\\
$ (1,0), (0,1), (1,1)$ 
&~ 
$\emptyset$ \\
&~\\
$ (2,0), (0,2)$ 
&~ 
$(0, \frac{5}{2})$ \\
&~\\
$ (2,1), (1,2)$ 
&~ $(0, \frac{5}{2}) \oplus (0,\frac{3}{2})$ \\
&~\\
$ (3,0), (0,3)$ 
&~ 
$(\frac{1}{2}, 4) \oplus (0, \frac{7}{2}) \oplus (0,\frac{5}{2})$ \\
&~\\
$ (2,2)$ 
&~ 
$(0, \frac{7}{2}) \oplus 2 (0,\frac{5}{2}) 
\oplus 2 (0, \frac{3}{2}) \oplus 2 (0,\frac{1}{2})$ \\
&~\\
$ (3,1), (1,3)$ 
&~  
$(\frac{1}{2}, 4) \oplus (\frac{1}{2}, 3) 
\oplus 2 (0, \frac{7}{2}) \oplus 3 (0,\frac{5}{2}) \oplus (0,\frac{3}{2})$ \\
&~\\
$ (4,0), (0,4)$ 
&~ 
$(1,\frac{11}{2})
\oplus (\frac{1}{2}, 5) \oplus (\frac{1}{2}, 4)
\oplus 2 (0, \frac{9}{2}) \oplus (0,\frac{7}{2}) \oplus (0,\frac{5}{2})$ \\
&~\\
$ (3,2), (2,3)$ 
&~ 
$(\frac{1}{2}, 4) \oplus (\frac{1}{2}, 3)  \oplus (\frac{1}{2}, 2)$ \\
&~ 
$\oplus (0, \frac{9}{2}) \oplus 3 (0,\frac{7}{2}) \oplus 5 (0, \frac{5}{2}) 
\oplus 4 (0, \frac{3}{2}) \oplus 2 (0,\frac{1}{2})$ \\
&~\\
$ (4,1), (1,4)$ 
&~ 
$(1, \frac{11}{2}) \oplus (1, \frac{9}{2}) 
\oplus (\frac{1}{2}, 5) \oplus 3 (\frac{1}{2}, 4)  \oplus (\frac{1}{2}, 3)$  \\
&~
$\oplus 3 (0, \frac{9}{2}) \oplus 5 (0,\frac{7}{2}) \oplus 3 (0, \frac{5}{2}) \oplus (0,\frac{3}{2})$ \\
&~\\
$ (5,0), (0,5)$ 
&~ 
$(\frac{3}{2}, 7) \oplus (1, \frac{13}{2}) \oplus (1, \frac{11}{2})
\oplus 2 (\frac{1}{2}, 6) \oplus (\frac{1}{2}, 5)  \oplus (\frac{1}{2}, 4)$ \\
&~
$\oplus 2 (0, \frac{11}{2})  \oplus 2 (0, \frac{9}{2}) \oplus (0, \frac{7}{2}) \oplus (0, \frac{5}{2})$  \\
&~\\
$ (3,3)$ 
&~ 
$(\frac{1}{2}, 5) \oplus 2 (\frac{1}{2}, 4)  \oplus 2 (\frac{1}{2}, 3) 
\oplus 2 (\frac{1}{2}, 2) \oplus 2 (\frac{1}{2}, 1)$ \\
&~
$\oplus (0, \frac{11}{2}) \oplus 3 (0,\frac{9}{2}) \oplus 6 (0, \frac{7}{2}) 
\oplus 8 (0, \frac{5}{2}) \oplus 8 (0,\frac{3}{2})  \oplus 6 (0, \frac{1}{2})$ \\
&~\\
\hline
\end{tabular}

\newpage

\Section*{Appendix D : $q$-Dunkl Operator Realization for the Refined Topological Vertex}


\renewcommand{\theequation}{D.\arabic{equation}}\setcounter{equation}{0}
\renewcommand{\thesubsection}{D.\arabic{subsection}}\setcounter{subsection}{0}


In this appendix, 
we use the Macdonald polynomials $P^N_\lambda(x;q,t)$ in the finite number of variables 
$x=(x_1,x_2,\cdots, x_N)$
with setting $x_{N+1}=x_{N+2}=\cdots=0$.
Here we assume that 
$|q|$, $|t| > 1 $, 
and define the following refined topological vertex (without framing factor)
\ba
V_{\mu \lambda}{}^\nu 
&:=& 
\lim_{N\rightarrow \infty} 
\sum_\sigma\iota \tP^N_{\mu ^\vee /\sigma^\vee}(-t^{\lambda^\vee} q^{\rho};t,q) \ 
\tP^N_{\nu /\sigma}(q^{\lambda} t^{\rho};q,t)
\tP^N_\lambda(t^\rho;q,t)
v^{|\sigma|},
\cr
V_{\mu }{}^{\lambda\nu} 
&:=& 
\lim_{N\rightarrow \infty} 
\sum_\sigma
\tP^N_{\mu ^\vee/\sigma^\vee}(-t^{\lambda^\vee} q^{\rho};t,q) \
\iota \tP^N_{\nu /\sigma}(q^{\lambda} t^{\rho};q,t)
\tP^N_{\lambda^\vee}(-q^{\rho};t,q)
v^{|\sigma|}
=\iota V_{\mu \lambda}{}^\nu .
\label{eq:appRTV}
\ea
These also reproduce Nekrasov's partition function.

Let $Y_i$, ($i=1,\cdots,N$) 
be the $q$-Dunkl operator 
\cite{rf:Cherednik}
\cite{rf:KirillovNoumi}
acting on the variables
$x_i$, ($i=1,\cdots,N$);
\ba
Y_i(x) &=& t^{-{N\over 2}}
T_i T_{i+1} \cdots T_{N-1} 
\omega
T_1^{-1} \cdots T_{i-1}^{-1},
\cr
T_i &=& t^\ha + t^{-\ha} {1 - t x_i/x_{i+1} \over 1 -
x_i/x_{i+1}} (s_i - 1),
\ea
where
\be
s_i = (i,i+1),
\qquad
\omega = \tau_N s_{N-1}\cdots s_1,
\qquad
\tau_N(x_i) = q^{\delta_{i,N}} x_i .
\ee
They commute with each other, 
\be
[Y_i(x),Y_j(x)]=0,
\ee
and the Macdonald polynomials are 
eigenfunctions of any symmetric operator $f$ in them:
\be
f(Y_1(x),\cdots,Y_N(x)) P^N_\lambda(x;q,t) 
= f(q^{\lambda_1} t^{\ha -1},\cdots,q^{\lambda_N} t^{\ha -N})
P^N_\lambda(x;q,t). 
\ee
Let $\tY_i(x)$ be the dual $q$-Dunkl operator 
which is given by replacing $q$ with $t$ in $Y_i(x)$,
i.e.
\be
f(\tY_1(x),\cdots,\tY_N(x)) P^N_\lambda(x;t,q) 
= f(t^{\lambda_1} q^{\ha -1},\cdots,t^{\lambda_N} q^{\ha -N})
P^N_\lambda(x;t,q). 
\ee
Note that
$Y_i(x)$ and $\tY_i(x)$ 
may not commute with each other.

Using these (dual) $q$-Dunkl operators,
our vertices in (\ref{eq:appRTV})
are written as follows:
\ba
V_{\mu\lambda}{}^\nu 
&=& 
\lim_{N\rightarrow \infty} \sum_\sigma 
v^{|\sigma|}
\tP^N_{\nu /\sigma}(-Y(x);q,t) \ 
\Endomega qt{}
\left(\iota \tP^N_{\mu^\vee /\sigma^\vee}(\tY(x);t,q)
\tP^N_{\lambda^\vee}(x;t,q)\right)
\vert_{x=t^{\rho}},
\cr
V_{\mu}{}^{\lambda\nu} 
&=& 
\lim_{N\rightarrow \infty} \sum_\sigma 
v^{|\sigma|}
\tP^N_{\mu^\vee /\sigma^\vee}(-\tY(x);t,q) \ 
\Endomega qt{}
\left(\iota \tP^N_{\nu /\sigma}(Y(x);q,t)
\tP^N_{\lambda}(-x;q,t)\right)
\vert_{x=q^{\rho}}.
\ea
Here $\Endomega qt{}$ is the involution in 
(\ref{eq:skewConjugate}).


Therefore the summation in the Young diagrams in Nekrasov's formula
is formally performed by using these $q$-Dunkl operators.
For example, the $SU(2)$ partition function 
in (\ref{eq:suiiZ}) is
\ba
\tZm{}{}
&=&
\lim_{N\rightarrow \infty} 
\tPi\left(-Q_1 Y(x),\tY(z)\right)^{-1} \
\tPi\left(-Q_2 Y(w),\tY(y)\right)^{-1}
\cr
&&\qquad 
\times
\tPi(-\Lambda x,y) \
\tPi(-\Lambda z,w)
\vert_{x=z=t^{\rho}, y=w=q^{\rho}}.
\ea

In the $SU(\nn)$ case,
let
\ba
D_0
&:=&
\prod_{\a =1}^\nn \tPi\left(-\Lambda x^\a ,y^\a \right),
\cr
D_\a 
&:=&
\prod_{\b =\a +1}^\nn 
\tPi\left(-Q_{\a ,\b } Y(x^\a ),\tY(x^\b )\right)^{-1}
\tPi\left(-Q_{\b ,\a } Y(y^\b ),\tY(y^\a )\right)^{-1},
\quad
0<\a <\nn,~~~~~~
\ea
and 
$
D'_\a := D_\a  
\Endomega tq{x^\a }\Endomega qt{y^\a }
$,
then the $SU(\nn)$ partition function in (\ref{eq:suNZ}) with $m=0$ is
\ba
\tZm{}{0} 
&=& 
\sum_{\Ya\lambda \a  ,\Ya\mu \a  ,\Ya\nu \a }
V_{\bullet \Ya\lambda 1}{}^{\Ya\mu 1} 
V_{\Ya\mu 1  \Ya\lambda 2}{}^{\Ya\mu 2} 
\cdots
V_{\Ya\mu {\nn-2} \Ya\lambda {\nn-1}}{}^{\Ya\mu {\nn-1}} 
V_{\Ya\mu {\nn-1} \Ya\lambda {\nn   }}{}^{\bullet}
\cr
&&\hskip24pt\times
V_{\bullet       }{}^{\Ya\lambda {\nn  }\Ya\nu {\nn-1}}
V_{\Ya\nu {\nn-1}}{}^{\Ya\lambda {\nn-1}\Ya\nu {\nn-2}}
\cdots
V_{\Ya\nu 2}{}^{\Ya\lambda 2\Ya\nu 1}
V_{\Ya\nu 1}{}^{\Ya\lambda 1\bullet}
\cr
&&\hskip24pt\times
\prod_{\a =1}^\nn 
Q_{B^\a }^{|\Ya\lambda \a  |}
\prod_{\a =1}^{\nn-1} 
v^{-|\Ya\mu \a |-|\Ya\nu \a |}
Q_{\a ,\a +1}^{|\Ya\mu \a  |}
Q_{\a +1,\a }^{|\Ya\nu \a  |}
\cr
&=& 
D'_{\nn-1}\cdots 
D'_2
D_1 D_0
\vert_{x^\a =t^{\rho}, y^\a =q^{\rho}}.
\ea
Since 
$
\Endomega tqx\Endomega qty
\tPi(x,y)
=
\tPi(x,y)
$,
we have the following $q$-Dunkl operator realization for Nekrasov's formula
\be
\tZm{}{0} 
= D_{\nn-1} \cdots D_2 D_1 D_0
\vert_{x^\a =t^{\rho}, y^\a =q^{\rho}}.
\ee


\Section*{Appendix E : Notations and identities for Partitions}


\renewcommand{\theequation}{E.\arabic{equation}}\setcounter{equation}{0}
\renewcommand{\thesubsection}{E.\arabic{subsection}}\setcounter{subsection}{0}




For each square  $s=(i,j)$ in the Young diagram of 
a partition $\lambda = (\lambda_1,\lambda_2,\cdots)$, we define
\beq
a_\lambda(s) := \lambda_i - j, \quad \ell_\lambda(s) := \lambda_j^\vee - i, 
\quad a'(s) := j -1, \quad \ell'(s) := i -1~,
\eeq
where $\lambda_j^\vee$ denotes the conjugate (dual) diagram. 
They are called arm length, leg length, arm colength and leg colength,
respectively.
The hook length $h_\lambda(s)$ and the content $c(s)$ at $s$ are given by
\beq
h_\lambda(s) := a_\lambda(s) + \ell_\lambda(s) + 1~, \quad 
c(s) := a'(s) - \ell'(s)~.
\eeq
The weight $|\lambda |$ and $||\lambda ||^2$ are 
\be
|\lambda |:=\sum_i \lambda_i,
\qquad
||\lambda ||^2:=\sum_i \lambda_i{}^2 =2 \sum_{s\in\lambda}(a(s)+\ha).
\ee
We also need the following integer
\beq
n(\lambda)  := \sum_{s \in \lambda} \ell'(s) = \sum_{i=1}^\infty (i-1) \lambda_i
= \frac{1}{2} \sum_{i=1}^\infty \lambda_i^\vee ( \lambda_i^\vee -1) 
= \sum_{s \in \lambda} \ell_\lambda(s)~.
\eeq
Similarly, we have
\beq
n(\lambda^\vee)  := \sum_{s \in \lambda} a'(s) = \sum_{s \in \lambda} a_\lambda(s)~.
\eeq
They are related to the integer $\kappa(\lambda)$ as follows:
\beq
\kappa(\lambda) := 2 \sum_{s \in \lambda} (j-i) = 2(n(\lambda^\vee) - n(\lambda))
= |\lambda| + \sum_{i=1}^\infty \lambda_i (\lambda_i -2i)~.
\eeq

Note that, since
$
\{\lambda_i-j\}_{j=1}^{\lambda_i}
=
\{j-1\}_{j=1}^{\lambda_i},
$
\be
\sum_{(i,j)\in\lambda}
f(\lambda_i-j , \ i)
=
\sum_{(i,j)\in\lambda}
f(j-1 ,\ i),
\label{eq:appsetFormulaI}
\ee
for any function $f$.
Also since
\be
\{\lambda_i-j\}_{j=1}^{\mu_i}
\cup
\{-\mu_i+j-1\}_{j=1}^{\lambda_i}
=
\{j\}_{j=-\mu_i}^{\lambda_i-1}
=
\{\lambda_i-j\}_{j=1}^{\lambda_i}
\cup
\{-\mu_i+j-1\}_{j=1}^{\mu_i},
\ee
we have
\be
\sum_{(i,j)\in\mu}(\lambda_i-j)-
\sum_{(i,j)\in\lambda}(\mu_i-j+1)
=
\sum_{(i,j)\in\lambda}(\lambda_i-j)-
\sum_{(i,j)\in\mu}(\mu_i-j+1).
\label{eq:appsetFormulaII}
\ee


We list their relations:

$$
\begin{array}{c c c c c c c}

\hline
%
\rule[-17pt]{0pt}{42pt} 
|\lambda|,||\lambda||^2, n, \kappa
&&
\sum_i  
&& 
\sum_{(i,j)\in\lambda}
&& 
\sum_{s\in\lambda} 
\cr

\hline
&&&&&&\cr

|\lambda| 
&=&
\sum_i \lambda_i
&=&
\sum_{(i,j)\in\lambda} 1
&=&
\sum_{s\in\lambda} 1,
\cr

||&&&&&&\cr

|\lambda^\vee|
&=&
\sum_j \lambda^\vee_j.
&&
&&\cr

&&&&&&\cr

\ha||\lambda||^2
&=&
\ha\sum_i {\lambda_i}^2
&=&
\sum_{(i,j)\in\lambda} (\lambda_i-j+\ha)
&=&
\sum_{s\in\lambda} \left(a(s)+\ha\right).
\cr

&&&&&&\cr

\ha||\lambda^\vee||^2
&=&
\ha\sum_j \lambda_j^{\vee 2}
&=&
\sum_{(i,j)\in\lambda} (\lambda^\vee_j-i+\ha)
&=&
\sum_{s\in\lambda} \left(\ell(s)+\ha\right),
\cr

&&&&&&\cr
\hline
&&&&&&\cr

n(\lambda)
&=&
\sum_i (i-1){\lambda_i}
&=&
\sum_{(i,j)\in\lambda}(i-1)
&=& 
\sum_{s\in\lambda} \ell'(s),
\cr

||&&&&&&\cr

\ha( || \lambda^\vee||^2- |\lambda^\vee|)
&=&
\ha\sum_j \lambda^\vee_j(\lambda^\vee_j -1)
&=&
\sum_{(i,j)\in\lambda}( \lambda^\vee_j-i)
&=& 
\sum_{s\in\lambda} \ell(s).
\cr

&&&&&&\cr

n(\lambda^\vee)
&=&
\sum_j (j-1){\lambda^\vee_j}
&=&
\sum_{(i,j)\in\lambda}(j-1)
&=& 
\sum_{s\in\lambda} a'(s),
\cr

||&&&&&&\cr

\ha( || \lambda||^2- |\lambda|)
&=&
\ha\sum_i \lambda_i (\lambda_i -1)
&=&
\sum_{(i,j)\in\lambda}( \lambda_i-j)
&=& 
\sum_{s\in\lambda} a(s).
\cr

&&&&&&\cr
\hline
&&&&&&\cr

\ha\kappa(\lambda) 
&=&
\ha\sum_i \lambda_i( \lambda_i+1-2i),
&&
&&
\sum_{s\in\lambda} c(s),
\cr

||&&&&&&||\cr

n(\lambda^\vee)-n(\lambda)
&=&
\sum_i i\left({\lambda^\vee_i}-{\lambda_i}\right)
&=&
\sum_{(i,j)\in\lambda} (j-i)
&=&
\sum_{s\in\lambda} (a'(s)-\ell'(s)),
\cr

||&&&&&&\cr

\ha( || \lambda||^2 - ||\lambda^\vee||^2)
&=&
\ha\sum_i \left({\lambda_i}^2 - \lambda_i^{\vee 2}\right)
&=&
\sum_{(i,j)\in\lambda} (\lambda_i-\lambda^\vee_j+i-j)
&=&
\sum_{s\in\lambda} (a(s)-\ell(s)).
\cr

&&&&&&\cr

n(\lambda^\vee)+n(\lambda)+|\lambda|
&=&
\sum_i \left(i-\ha\right)\left({\lambda_i}+{\lambda^\vee_i}\right)
&=&
\sum_{(i,j)\in\lambda} (i+j-1)
&=&
\ds{\sum_{s\in\lambda}} (a'(s)+\ell'(s)+1),
\cr

||&&&&&&\cr

\ha( || \lambda||^2 + ||\lambda^\vee||^2)
&=&
\ha\sum_i \left({\lambda_i}^2 +\lambda_i^{\vee 2}\right)
&=&
\ds{\sum_{(i,j)\in\lambda}} (\lambda_i+\lambda^\vee_j-i-j+1)
&=&
\ds{\sum_{s\in\lambda}} (a(s)+\ell(s)+1),
\cr

&&&&&&||\cr

&=&
\ha\sum_i \lambda_i( \lambda_i-1+2i),
&&&&
\sum_{s\in\lambda} h(s).
\cr

&&&&&&\cr
\hline

\end{array}
$$




\begin{thebibliography}{99}

\bibitem{Nek} N.~Nekrasov, ``Seiberg-Witten Prepotential from Instanton Counting,'' 
{\it Adv. Theor. Math. Phys. } {\bf 7} (2003) 831, {\tt arXiv:hep-th/0206161}.

\bibitem{AMV} M.~Aganagic,  M.~Mari\~no  and C.~Vafa, 
``All Loop Topological String Amplitudes From Chern-Simons Theory,''
{\it Commun. Math. Phys.} {\bf 247} (2004) 467, {\tt arXiv:hep-th/0206164}.

\bibitem{AKMV} M.~Aganagic,  A.~Klemm, M.~Mari\~no and C.~Vafa, ``The Topological Vertex,''
{\it Commun. Math. Phys. } {\bf 254} (2005) 425, {\tt arXiv:hep-th/0305132}.

\bibitem{KKV} S.~Katz, A.~Klemm and C.~Vafa,
``Geometric Engineering of Quantum Field Theory,''
{\it Nucl. Phys.} {\bf B 497} (1997) 173, {\tt arXiv:hep-th/9609239}.

\bibitem{KMV} S.~Katz, P.~Mayr and C.~Vafa,
``Mirror Symmetry and Exact Solution of $4D$ ${\cal N}=2$ Gauge Theories I ,''
{\it Adv. Theor. Math. Phys.} {\bf 1} (1998) 53, {\tt arXiv:hep-th/9706110}.


\bibitem{rf:NakajimaYoshioka}
H.~Nakajima and K.~Yoshioka,
``Instanton Counting on Blowup I,''  {\it Invent. Math.} {\bf 162} (2005) 313,
{\tt arXiv:math.AG/0306198}.

\bibitem{IK-P1} A.~Iqbal and A.-K. Kashani-Poor,
``Instanton Counting and Chern-Simons Theory,''
{\it Adv. Theor. Math. Phys.} {\bf 7} (2003) 457, {\tt arXiv:hep-th/0212279}.

\bibitem{IK-P2} A.~Iqbal and A.-K. Kashani-Poor,
``$SU(N)$ Geometries and Topological String Amplitudes,''
{\it Adv. Theor. Math. Phys.} {\bf 10} (2006) 1, {\tt arXiv:hep-th/0306032}.

\bibitem{EK1} T.~Eguchi and H.~Kanno,
``Topological Strings and Nekrasov's Formulas,"
{\it JHEP} {\bf 0312}  (2003) 006, {\tt arXiv:hep-th/0310235}.

\bibitem{EK2} T.~Eguchi and H.~Kanno,
``Geometric transitions, Chern-Simons theory and Veneziano type amplitudes,''
{\it Phys. Lett.} {\bf B585}  (2004) 163, {\tt arXiv:hep-th/0312234}.

\bibitem{HIV} T.~Hollowood, A.~Iqbal and C.~Vafa,
``Matrix Models, Geometric Engineering and Elliptic Genera,"
{\it JHEP} {\bf 0803}  (2008) 069, {\tt arXiv:hep-th/0310272}.

\bibitem{Zhou} J.~Zhou, 
``Curve Counting and Instanton Counting,'' {\tt arXiv:math.AG/0311237}.

\bibitem{IKV} A.~Iqbal, C.~Kozcaz and C. Vafa,
``The Refined Topological Vertex,'' {\tt arXiv:hep-th/0701156}.

\bibitem{AK}
H.~Awata and H.~Kanno, 
``Instanton counting, Macdonald function and the moduli space of $D$-branes,''
{\it JHEP}  {\bf 0505} (2005) 039, {\tt arXiv:hep-th/0502061}. 

\bibitem{Mac} I.G.~Macdonald, 
{\it Symmetric functions and Hall polynomials}, Second Edition, 
Oxford University Press, 1995. 


\bibitem{Taki} M.~Taki,
``Refined Topological Vertex and Instanton Counting,''
{\it JHEP}  {\bf 0803} (2008) 048, 
{\tt arXiv:0710.1776[hep-th]}.


\bibitem{ORV} A.~Okounkov, N.~Reshetikhin and C.~Vafa,
``Quantum Calabi-Yau and Classical Crystals,'' {\tt arXiv:hep-th/0309208}.

\bibitem{IKS}
A.~Iqbal, C.~Kozcaz and K. Shabbir,
``Refined Topological Vertex, Cylindric Partitions and $U(1)$ Adjoint Theory,''
{\tt arXiv:0803.2260[hep-th]}.

\bibitem{GV1} R.~Gopakumar and C.~Vafa,
``M theory and Topological Strings I,''
{\tt arXiv:hep-th/9809187}.

\bibitem{GV2} R.~Gopakumar and C.~Vafa,
``M theory and Topological Strings II,''
{\tt arXiv:hep-th/9812127}.

\bibitem{KKV-M}
S.~Katz, A.~Klemm and C.~Vafa, 
``M-theory, topological strings and spinning black holes,''
{\it Adv. Theor. Math. Phys. } {\bf 3} (1999) 1445, {\tt arXiv:hep-th/9910181}.

\bibitem{Peng}
P.~Peng, 
``A Simple Proof of Gopakumar-Vafa Conjecture for
Local Toric Calabi-Yau Manifolds,''
{\tt arXiv:math.AG/0410540}.

\bibitem{Kon1}
Y. Konishi,
``Pole Structure of Topological String Free Energy,''
{\it Publ. Res. Inst. Math. Sci.} {\bf 42} (2006) 173, 
{\tt arXiv:math.AG/0411357}.

\bibitem{Kon2}
Y. Konishi,
``Integrality of Gopakumar-Vafa Invariants of
Local Toric Calabi-Yau Threefolds,''
{\it Publ. Res. Inst. Math. Sci.} {\bf 42} (2006) 605,
{\tt arXiv:math.AG/0504188}.

\bibitem{rf:AKOS}
H. Awata, H. Kubo, S. Odake and J. Shiraishi,
``Quantum $W_N$ Algebras and Macdonald Polynomials,"
 {\it Commun. Math. Phys.} {\bf 179}  (1996) 401,
{\tt arXiv:q-alg/9508011}.

\bibitem{NY2} H.~Nakajima and K.~Yoshioka,
``Instanton Counting on Blowup II. $K$-theoretic Partition Function,''
{\tt arXiv:math.AG/0505553}.

\bibitem{GNY} L. G\"ottsche, H.~Nakajima and K.~Yoshioka,
``$K$-theoretic Donaldson Invariants via Instanton Counting,''
{\tt arXiv:math.AG/0611945}.


\bibitem{GSV}
S.~Gukov, A.~Schwarz and C.~Vafa,
``Khovanov-Rozansky Homology and Topological Strings,''
{\it Lett. Math. Phys.} {\bf 74} (2005) 53, 
{\tt arXiv:hep-th/0412243}.

\bibitem{GIKV} S.~Gukov, A.~Iqbal, C.~Kozcaz and C. Vafa,
``Link Homologies and the Refined Topological Veretx,''
{\tt arXiv:0705.1368[hep-th]}.


\bibitem{Tac}
Y.~Tachikawa,
``Five-dimensional Chern-Simons terms and Nekrasov's instanton counting,
{\it JHEP} {\bf 0402} (2004) 050, {\tt arXiv:hep-th/0401184}.


\bibitem{rf:NekrasovOkounkov}
N.~Nekrasov and A.~Okounkov, 
``Seiberg-Witten Theory and Random Partitions,'' 
{\it The unity of mathematics,} Progr. Math. Vol. 244, Birkh\"auser, Boston, 2006, p525,
{\tt arXiv:hep-th/0306238}.


\bibitem{IK-P3} A.~Iqbal and A.-K. Kashani-Poor,
``The Vertex on a Strip,'' 
{\it Adv. Theor. Math. Phys.} {\bf 10} (2006) 317, {\tt arXiv:hep-th/0410174}.


\bibitem{rf:AOS}
  H.~Awata, S.~Odake and J.~Shiraishi,
  ``Integral Representations of the Macdonald Symmetric Polynomials,"
   {\it Commun. Math. Phys.} {\bf 179} (1996) 647,
  {\tt arXiv:q-alg/9506006}.


\bibitem{KM}
Y. Konishi and S. Minabe,
``Flop Invariance of the Topological Vertex,''
{\it Internat. J. Math.} {\bf 19} (2008) 27, 
{\tt arXiv:math.AG/0601352}.

\bibitem{rf:LiLiuZhou}
J.~Li, K.~Liu and J.~Zhou, 
``Topological String Partition Functions as Equivariant Indices,''
{\tt arXiv:math.AG/0412089}.

\bibitem{PS} R.~Poghossian and M.~Samsonyan,
``Instantons and the 5D U(1) gauge theory with extra adjoint,"
{\tt arXiv:0804.3564 [hep-th]}.


\bibitem{rf:Cherednik}
I.~Cherednik, 
``Double affine Hecke algebras and Macdonald conjectures,''
{\it Annals of Math.} {\bf 141} (1995) 191.

\bibitem{rf:KirillovNoumi}
A.N.~Kirillov and M.~Noumi, 
``Affine Hecke algebras and raising operators for Macdonald polynomials,''
{\tt arXiv:q-alg/9605004}.


\bibitem{Taki2} M.~Taki,
``Flop Invariance of Refined Topological Vertex and Link Homologies,''
{\tt arXiv:0805.0336[hep-th]}.




\end{thebibliography}
\end{document}